\documentclass[a4paper,11pt]{article}
\usepackage{jheppub} 
\usepackage{lineno}

 \usepackage{physics}
 \usepackage{indentfirst}
\usepackage{suffix}
\usepackage{bm}
\usepackage{subcaption}

\makeatother

\newcommand{\half}{\frac{1}{2}}
\newcommand{\third}{\frac{1}{3}}
\newcommand{\quarter}{\frac{1}{4}}
\newcommand{\ve}{\varepsilon}
\newcommand{\expect}[1]{\left\langle {#1} \right\rangle}

\title{\boldmath Open-Closed String Duality, Branes, and Topological Recursion}

\author{Ashton Lowenstein}
\affiliation{Department of Physics and Astronomy, University of Southern California,\\
Los Angeles, CA 90089, U.S.A.}

\emailAdd{alowenst@usc.edu}

\abstract{We consider matrix models exhibiting open-closed string duality in two-dimensional string theories with various amounts of supersymmetry. In particular, a relationship between matrix models in the $\beta = 2$ Wigner-Dyson class and models in the $(1 + 2\Gamma, 2)$ Altland-Zirnbauer class relates the perturbative solutions of the two systems' string equations. Point-like operator insertions in the closed string theory are mapped to the topological expansion of the free energy in the open string theory. We compute correlation functions of macroscopic loop operators and FZZT branes in a general topological gravity background. The relationship between the topological recursion of moduli space volumes and branes is discussed by analyzing the Virasoro conditions in the matrix models.}

\begin{document}

\maketitle
\flushbottom

\section{Introduction}

The last 10 years has seen a resurgence in the study of two-dimensional quantum gravity, and methods like random matrix theory (RMT), intersection theory and topological gravity, and non-critical string theory are once again in vogue. Developments in the $0+1$-dimensional SYK theory \cite{KitaevSYK, Maldacena:2016hyu, Kitaev:2017awl} led to an interest in Jackiw-Teitelboim (JT) gravity \cite{JACKIW1985343, TEITELBOIM198341, Almheiri:2014cka}, because both theories are described by Schwarzian dynamics in a certain regime. The landmark discovery made by Mirzakhani regarding volumes of moduli spaces of bordered hyperbolic Riemann surfaces \cite{Mirzakhani:2006fta} and the subsequent connection to topological recursion in RMT made by Eynard and Orantin \cite{Eynard:2007fi} provided some of the tools necessary for Saad, Shenker, and Stanford to show that JT gravity is described by a double scaled matrix model \cite{Saad:2019lba}. Such results reignited an interest in matrix models and topological gravity that started nearly 40 years ago, as borne out by applications to non-perturbative quantum gravity \cite{Johnson:2019eik, Johnson:2021tnl}, flat space holography \cite{Kar:2022sdc, Rosso:2022tsv}, and gauge-string duality \cite{Gopakumar:2022djw}.

The use of random matrix techniques in quantum field theory began in the late 70s ({\it e.g.} \cite{Brezin:1977sv}) and saw its first golden age for applications to quantum gravity in the late 80s and early 90s with works like \cite{Kazakov:1989bc, BREZIN1990144, DOUGLAS1990635,  PhysRevLett.64.127,  GROSS1990333, BANKS1990279, Dijkgraaf:1990rs, MOORE1991665, Dalley:1992br, Dalley:1991vr, Dalley:1991qg, Dalley:1991yi}. A review and summary of the state of low-dimensional string theory and RMT at that time can be found in \cite{ginsparg1993lectures}. A matrix model is defined by a choice of the class of $N \times N$ matrices $M$ and a potential $V(M)$, usually a polynomial defined by a set of coupling constants
	\begin{equation}
		V(M) = \half M^2 +  \sum_{i = 3}^\infty g_i M^i.
	\end{equation}
Many 2d quantum gravity theories are described by matrix models of $N \times N$ Hermitian matrices with a polynomial potential. The matrix integral $Z$ is defined by
	\begin{equation}
		Z = \int dM e^{-N \tr V(M) },
	\end{equation}
where $dM$ is the flat, translation invariant measure on the space of matrix elements $M_{ij}$. Observables in the theory are scalar functions of $M$ and their expectation values are computed by inserting them into the matrix integral. Such quantities have an expansion in powers of $1/N$, \`{a} la 't Hooft \cite{HOOFT1974461}. This expansion applied to the matrix integral itself has the interpretation as an enumeration of tessellations\footnote{The numbers of sides that a face is allowed to have is determined by the rank of $V$.} of closed surfaces \cite{Brezin:1977sv, BESSIS1980109}. Thinking of this like a sum over geometries allows one to interpret $Z$ like a sort of discrete quantum gravity path integral.

The study of continuum physics via RMT often involves the use of the double scaling limit\footnote{There are interesting cases where the finite or large-$N$ matrix model has an interpretation in terms of string theory constructions, see for example \cite{Gopakumar:2022djw}.}, a process in which the size of the matrix $N$ is taken to infinity while tuning the parameters $g_i$ of $V$ to their critical values \cite{KAZAKOV1985295, DOUGLAS1990635, BREZIN1990144, GROSS1990333}. This can be interpreted geometrically as taking the number of faces in a tessellation to infinty while sending the average size of a face to 0, producing a smooth closed surface with non-vanishing area. The expansion in powers of $1/N$ becomes an expansion in a new small parameter $\hbar$, which is a renormalized version of the former. Surfaces with $g$ handles come weighted by a factor $\hbar^{2g-2}$, and the matrix integral $Z$ can be written as the (asymptotic) sum
	\begin{equation}
		Z = \sum_{g = 0}^\infty \hbar^{2g-2} Z_g.
	\end{equation}
It is in this sense that expansions in powers of $\hbar$ are deemed topological expansions.

A convenient way to take the double scaling limit involves the method of orthogonal polynomials. This approach introduces two interlinked and powerful technologies. The first, often referred to as the Dyson gas or many-body formalism, is an auxiliary quantum mechanics governed by the Schrodinger equation with Hamiltonian $\mathcal{H} = -\hbar^2 \partial_x^2 + u(x)$. The wavefunctions $\psi(x,E)$ of this theory are essential to studying the statistics of the double scaled theory, including the spectral density and the spectral form factor \cite{Johnson:2022wsr}. The Schrodinger potential $u(x)$ links us to the second technology. The function $u$ obeys a differential equation\footnote{For historical reasons $u$ is called the string susceptibility and its equation of motion is called the string equation.} defined by a set of coupling constants $t_k$ 
	\begin{equation}
		\sum_{k = 1}^\infty t_k R_k + x = 0, 
	\end{equation}
where $R_k$ is the $k^{\text{th}}$ Gelfand-Dikii differential polynomial in $u$ and its derivatives \cite{IMGelfand_1975}. Moreover, $u$ obeys the KdV flow equations
	\begin{equation}
		\frac{\partial u}{\partial t_k} = R_{k +1}',
	\end{equation}
where the couplings $t_k$ are interpreted as KdV times.

In \cite{Kazakov:1989bc} Kazakov made the connection between a subset of double scaled matrix models  and Liouville theory coupled to minimal CFT matter. When the matrix potential is the polynomial $V(M) = \half M^2 + g_{2p}M^{2p}$ the double scaled version is referred to as the $p^\text{th}$ multicritical model and is dual to Liouville plus the $(2,2p-1)$ minimal model. When multiple of the $g_i$ are turned on and sent to their critical values, the double scaled model is referred to as a massive interpolation between the multicritical models.

An alternative interpretation of the matrix model \cite{Maldacena:2004sn} was later afforded by the discovery of ZZ and FZZT branes in two-dimensional string theory \cite{Fateev:2000ik, Teschner:2001rv, Zamolodchikov:2001ah, Seiberg:2003nm}. The index $i = 1, \dots, N$ on the matrix labels a configuration of $N$ ZZ branes, and the matrix element $M_{ij}$ represent stringy degrees of freedom stretching between the $i^\text{th}$ and $j^\text{th}$ branes. The matrix model naturally lives on ZZ branes because of the identification between the eigenvalues $\lambda$ of $M$ and the associated Liouville direction $\varphi$ via $\varphi \sim \log \lambda$. Since ZZ branes represent Dirichlet boundary conditions on $\varphi$, they should be associated directly with the eigenvalues $\lambda$. Some number of FZZT branes can be inserted into the theory via determinant operators $\det(M + \mu_i)$, where $\mu_i$ corresponds to the cosmological constant on each brane. The FZZT branes act as probes of the ZZ brane spacetime. By integrating in and out auxiliary degrees of freedom, one can equivalently interpret the matrix model with determinant operator insertions as describing strings stretching between the FZZT and ZZ branes, as well as strings stretching just between the FZZT branes.

The notion of open-closed duality in two-dimensional string theory has existed since at least the early 90s, but was vastly overshadowed by the end of the decade by another open-closed duality, the AdS/CFT correspondence \cite{Maldacena:1997re}. There is a simple connection between matrix models dual to closed string theories and matrix models dual to open string theories which is facilitated by the differential equation (see e.g. \cite{Johnson:1992wr})
	\begin{equation}
		(u + s)\mathcal{R} - \frac{\hbar^2}{2}\mathcal{R}\mathcal{R}'' + \frac{\hbar^2}{4}(\mathcal{R}')^2 = \hbar^2 \Gamma^2, \label{eqn:bigstringeqfirstappearance}
	\end{equation}
where $\mathcal{R} = \sum_k t_k R_k + x$. Solutions to this equation for $\Gamma \neq 0$ can be interpreted as describing open string physics. Since this equation can be obtained from the closed string equation by a $\rho$-dependent shift of the coupling constants $t_k$, we have an apparent duality transformation between the two sectors of the string theory \cite{Johnson:1993vk}. This duality means that observables computed in the closed string theory involving geodesic loops are equivalently computed by the $\Gamma \neq 0$ solutions in the open string theory.

FZZT brane insertions are deeply connected to macroscopic loop insertions in the matrix model. For example, prior to double scaling
	 \begin{equation}
        \det(M + \mu) = \sum_{n = 0}^\infty \frac{(-1)^n}{n!}\prod_{i = 1}^n \int_0^\infty \frac{d\beta_i}{\beta_i}e^{-\mu\beta_i} \tr e^{-\beta_i M}.
       	\end{equation}
From a string worldsheet perspective, an FZZT brane is a substrate on which the worldsheet can form a geodesic boundary of any length. This is captured by the above formlua where the determinant. The expectation value of the double scaled macroscopic loop operator $\langle e^{-\beta \mathcal{H}}\rangle$ , where $\mathcal{H}$ is the Hamiltonian of the auxiliary quantum mechanics, has the interpretation of a gravity path integral on a surface with an asymptotic boundary of renormalized length $\beta$ \cite{BANKS1990279}. Correlation functions of the macroscopic loop operator have a topological expansion similar to $Z$
	\begin{equation}
		\langle e^{-\beta_1\mathcal{H}} \cdots e^{-\beta_n\mathcal{H}} \rangle = \sum_{g = 0}^\infty \hbar^{2g-2+n}\mathcal{Z}_{g,n}(\beta_1,\dots,\beta_n),
	\end{equation}
where $\mathcal{Z}_{g,n}$ is the path integral over surfaces with genus $g$ and $n$ asymptotic boundaries.

Many quantities of interest in double scaled matrix models have a perturbative expansion in $\hbar$, including the wavefunctions $\psi$, the potential $u$, and observables like macroscopic loops. Another quantitiy of particular interest in both the statistical interpretation of the matrix models as well as in topological recursion is the resolvent $R(\lambda) = (M - \lambda)^{-1}$. It's correlation functions are expanded in powers of $\hbar$ as
	\begin{equation*}
		\left\langle \frac{1}{M-\lambda_1} \cdots \frac{1}{M-\lambda_n} \right \rangle = \sum_{g =0}^\infty \hbar^{2g} R_{g,n}(\lambda_1,\dots,\lambda_n).
	\end{equation*}
In the double scaling limit it is convenient to introduce a complex uniformizing coordinate $z$ to replace the scaling part of the eigenvalue $\lambda$. One then considers a slightly altered form for the $\hbar$-corrections to the resolvent $n$-point function
	\begin{equation*}
		W_{g,n}(z_1,\dots,z_n) = (-1)^n z_1 \cdots z_n R_{g,n}(-z_1^2,\dots,-z_n^2).
	\end{equation*}
The functions $W_{g,n}$ satisfy the famous topological recursion relations \cite{Eynard:2007fi}
	\begin{equation}
	\begin{aligned}
		W_{g,n}(z_1,\dots,z_n) &= \underset{z \to 0}{\text{Res}}\Bigg\{ \frac{1}{z_1^2 - z^2} \frac{1}{4y(z)}\bigg[W_{g-1,n+1}(z,-z,J) \\
		&+\sum'_{\underset{g_1+g_2=g}{I \cup I' = J}}W_{g_1,|I|+1}(z,I)W_{g_2,|I'|+1}(-z,I)\bigg]\Bigg\}, \label{eqn:Wrecursion}
	\end{aligned}
	\end{equation}
where $y(z)$ is the double scaled spectral curve of the matrix model, $J = \{z_2,\dots,z_n\}$, and $\sum'$ denotes a sum over stable configurations.

In the special case that the matrix model has $y(z) = \frac{\sin(2\pi z)}{4\pi}$, the resolvent functions $W_{g,n}$ are related to the Weil-Petersson volume $V_{g,n}$ of the moduli space of Riemann surfaces with genus $g$ and $n$ borders. The recursion relation \eqref{eqn:Wrecursion} reproduces Myrzakhani's recursion relation for $V_{g,n}$ \cite{Mirzakhani:2006fta}. In hindsight it is not so surprising that this connection exists, given the connection between Myrzakhani's own work and the Witten-Kontsevich theorem, which relates the intersection numbers of certain cohomology classes on the moduli space $\mathcal{M}_{g,n}$ of Riemann surfaces with $g$ handles and $n$ marked points, to the free energy of a general class of matrix model \cite{Witten:1990hr, cmp/1104250524}. Such a matrix model with that spectral curve was identified as being dual to JT gravity \cite{Saad:2019lba}, in part using the fact that the volumes $V_{g,n}$ are used the compute the gravity path integral. A natural question to ask in light of that is whether other two-dimensional gravity path integrals can be computed in terms of corresponding volumes, and if so, is there a corresponding matrix model dual? In \cite{Mertens:2020hbs} minimal string theory was compared to its matrix model dual, and moduli space volumes were identified on the field theory side. A model of three-dimensional quantum gravity, the Virasoro Minimal String, was explored recently in \cite{Collier:2023cyw} and linked to the moduli spaces of Riemann surfaces on the gravity side. In \cite{Okuyama:2021eju} the authors start on the matrix model side and consider a generalization of $V_{g,n}$ in any topological gravity background.

While the moduli space volumes for ordinary Riemann surfaces is clear, the picture becomes more complicated when supersymmetry is included. The jump to $\mathcal{N} = 1$ super-Riemann surfaces is fairly straightforward, but as the supersymmetry is extended further things become more difficult \cite{Stanford:2019vob, turiaci2023mathcaln2}. Another remarkable feature of the differential equation \eqref{eqn:bigstringeqfirstappearance} is that it has the capacity to describe both $\mathcal{N} = 1$ (i.e. the 0A theories) \cite{Johnson:2003hy, Klebanov:2003wg} and $\mathcal{N} = 2$ superstrings \cite{Johnson:2023ofr}. This can be leveraged to make predictions not only about the physics of asymptotic boundaries in the theories but geodesic ones, and hence moduli space volumes, as well.

The organization of this paper is as follows. In {\bf section \ref{scn:doublescaledmatrixmodels}} we review the matrix models of interest, including their double scaling limits. In {\bf section \ref{scn:stringequationperturbationtheory}} we begin to pursue perturbation theory, starting with the string equations of our various models. In {\bf section \ref{scn:macroscopicloopperturbationtheory}} we develop perturbation theory of macroscopic loops in the theories, with a focus on applying the techniques from the string equations. In {\bf section \ref{scn:geodesicloopperturbationtheory}} we examine a unique feature of the open-closed string system, namely its ability to describe geodesic boundaries. In {\bf section \ref{scn:generalizedweilpeterssonvolumes}} we consider a generalization of the Weil-Petersson volumes and how they fit into the open-closed string theory. In {\bf section \ref{scn:topologicalrecursion}} we discuss the presence of topological recursion in these models.


\section{Double scaled matrix models} \label{scn:doublescaledmatrixmodels}

\subsection{Wigner-Dyson ensembles}

There are two classes of matrix models that will be of use to us. First are the multicritical Wigner-Dyson $\beta$-models. Such a model is defined in the finite-$N$ regime by the matrix integral
	\begin{equation}
		Z = \int \prod_{i = 1}^N d\lambda_i \Delta(\bm{\lambda})^\beta e^{ -N\sum_{i = 1}^N V(\lambda_i)}, \label{eqn:WDmatrixintegral}
	\end{equation}
where $\lambda_i$ are the eigenvalues of the random matrix $M$, $\Delta(\bm{\lambda}) = \prod_{i<j}(\lambda_j - \lambda_i)$ is the Vandermonde determinant, and $V(\lambda) = \sum_{i = 1}^{k} g_i \lambda^{2i}$. The cases $\beta = 1,2,4$ arise from considering identifiable classes of matrices, namely real symmetric, hermitian, and quaternion symmetric, respectively. The Vandermonde determinant appears as the Jacobian for the transformation that diagonalizes $M$. A numerical factor coming from integrating out the symmetry group is usually discarded. When the potential $V$ is quadratic, the models are commonly referred to as the Gaussian Orthogonal Ensemble (GOE), Gaussian Unitary Ensemble (GUE), and the Gaussian Symplectic Ensemble (GSE).

In all cases, it is conventional to order the eigenvalues by increasing label $i$, fixing the domain of integration to respect this ordering. When $\beta$ is even it is convenient to enlarge the domain of integration to $\mathbb{R}^N$ at the cost of a combinatorial factor\footnote{The choice is made here to scale the potential by a factor of $N$. This is accomplished by rescaling the eigenvalues to $\lambda/\sqrt{N}$. In the large-$N$ this confines the eigenvalues to be in a finite window $[-a,a]$.}. The surfaces tessellated in the $1/N$-expansion are orientable for $\beta = 2$, while the $\beta = 1,4$ theories include both oriented and unoriented surfaces. 

Henceforth we will specialize to the $\beta = 2$ theories. In order to double scale a model, introduce the set of polynomials $p_i(\lambda)$ that are orthogonal with respect to the measure $e^{-NV}$
	\begin{equation}
		\int_{-\infty}^\infty d\lambda e^{-NV(\lambda)}p_i(\lambda)p_j(\lambda) = h_i \delta_{ij}.
	\end{equation}
These polynomials are normalized so that $p_i(\lambda) = \lambda^i + \cdots$. They obey the recursion relation
	\begin{equation}
		\lambda p_i(\lambda) = p_{i+1}(\lambda) + R_i p_{i-1}(\lambda). \label{eqn:orthogonalpolynomialrecursion}
	\end{equation}
The recursion coefficients $R_i$ are related to one another through their own recursion relation, which can be worked out by studying the action of the derivative $\frac{d}{d\lambda}$ on the polynomials. After integrating by parts
	\begin{equation}
		N\int_{-\infty}^\infty d\lambda e^{-NV(\lambda)} V'(\lambda) p_n(\lambda)p_{n-1}(\lambda) = nh_n. \label{eqn:recursioncoefficientsource}
	\end{equation}
By invoking the recursion relation \eqref{eqn:orthogonalpolynomialrecursion} to simplify, one arrives at an equation for the $R_i$.

The Vandermonde determinant $\Delta$ is expressible as the determinant of a matrix $\Delta = \det(\lambda_i^{j-1})$ where $i,j = 1,\dots,N$. By taking linear combinations of the rows and columns of that matrix, it is equivalently expressed as $\Delta = \det(p_{j-1}(\lambda_i))$, with $i,j = 1,\dots, N$ again. While the family of orthogonal polynomials $p_i$ is infinte, the matrix model only utilizes the first $N$ of them. This can be thought of in a many-body setting as saying the matrix model describes a system of fermions, making the dependence on only the first $N$ polynomials akin to having a Fermi sea.

It is advantageous for several reasons to think of the matrix integral $Z$ as being the exponential of a free energy function $F$. First, this fits nicely with the physical interpretation of the model describing some sort of statistical ensemble of Hamiltonians. Second, it makes direct contact with the study of integrable systems and topological gravity. The Witten-Kontsevich theorem casts the free energy $F$ of a matrix model as the generating function of intersection numbers on the moduli space of Riemann surfaces \cite{Witten:1990hr, cmp/1104250524}, and asserts that the matrix integral $Z$ is related to a tau function of the KdV hierarchy. Third, the critical behavior of the free energy provides guidance for how to implement the double scaling limit and extract continuum physics.

In the large-$N$ limit, the label $i$ on the eigenvalues and polynomials can be turned into a continuous parameter $i/N = X \in [0,1)$. The large-$N$ free energy $F$ of the theory is misbehaved at $X = 1$
	\begin{equation}
		F = -N^2 \int_0^1 dX (1-X)\log R(X). \label{eqn:largeNF}
	\end{equation}
Investigation of this region is what yields the double scaling limit. For the $k^\text{th}$ model, scale away from $X = 1$ via $X = 1 + (x-\mu)\delta^{2k}$, where $\delta \to 0$ and $x \in (-\infty,\mu]$. The parameter $\mu$ is referred to as the Fermi level in the many-body formalism. Since the eigenvalues are ordered, the point $X = 1$ corresponds to the largest eigenvalue. Therefore we scale away from that point via $\lambda = a - E \delta^{2}.$ The recursion coefficient $R(X)$ takes the critical value $R(1) = 1$ at the edge of the spectrum. Its scaling form is $R(X) = 1 - u(x)\delta^2$. Finally, we define $N^{-1} = \hbar \delta^{2k+1}$.

The double scaling limit yields two important outcomes. First, define the double scaling limit of the orthogonal polynomials via
	\begin{equation}
		\frac{1}{\sqrt{h_i}}e^{-NV(\lambda)/2}p_i(\lambda) \xrightarrow{\text{double scale}} \psi(x,E).
	\end{equation}
Thiewavefunction $\psi$ is related to the Baker-Akheizer function of the underlying integrable hierarchy. By double scaling the recursion relation \eqref{eqn:orthogonalpolynomialrecursion}, one finds
	\begin{equation}
		\Big(-\hbar^2 \partial^2 + u(x)\Big)\psi(x,E) = E\psi(x,E)
	\end{equation}
where $\partial \equiv \partial/\partial x$. The double scaling limit of the orthogonal polynomials obey the Schrodinger equation with potential $u(x)$. This auxiliary quantum mechanics is useful for several reasons: the wavefunctions can be used to construct the self-reproducing kernel $K(E,E')$, which is essential to statistical interpretations of the double scaled model, and the wavefunctions can also be interpreted as the partition function of an FZZT brane probing the ZZ brane background \cite{Carlisle:2005wa}.

The second important outcome determines the function $u(x)$\footnote{Strictly speaking this is true for models where the double scaled spectrum is one connected part of $\mathbb{R}$. These are called single-cut matrix models. A slightly different approach is needed to describe the double-cut models, for instance.}. By double scaling the outcome of \eqref{eqn:recursioncoefficientsource} one arrives at a differential equation for the string susceptibility. For the $k^\text{th}$ model the differential equation is $R_k + x = 0$. For example, the equation governing the $k = 2$ model is
	\begin{equation*}
		-\frac{\hbar^2}{3} u'' + u^2 + x = 0,
	\end{equation*}
which is the Painleve I equation. One can construct more complicated models in the double scaling limit by taking linear combinations of the Gelfand-Dikii polynomials
	\begin{equation}
		\sum_k t_k R_k + x = 0. \label{eqn:closedstringeq}
	\end{equation}
Such models are referred to as massive interpolations, and can be thought of as RG flows between the different multicritical models. Since each multicritical model is dual to Liouville theory coupled to minimal CFT matter, this is like flowing between different gravity + matter theories. For historical reasons, equations of the form \eqref{eqn:closedstringeq} are called string equations.

The numbers $t_k$ that define the string equation of the model have an interpretation in the string theory language as being the coupling constants for gravitationally dressed CFT primaries $\sigma_k$, or closed string operators \cite{GROSS1990333}. Written in terms of the double scaled matrix model, their expectation values are \cite{BANKS1990279}
	\begin{equation}
		\langle \sigma_k \rangle = \int_{-\infty}^\mu dx \mel{x}{\mathcal{H}^{k + \half}}{x}.
	\end{equation}
By applying the double scaling ansatz to \eqref{eqn:largeNF}, the double scaled free energy $F$ is expressed in terms of the function $u$ by
	\begin{equation}
		F(\mu) = \frac{1}{\hbar^2} \int_{-\infty}^\mu dx (x-\mu)u(x),
	\end{equation}
which can be inverted to give
	\begin{equation}
		u(x) = \hbar^2\frac{\partial^2 F}{\partial x^2}. \label{eqn:utof}
	\end{equation}
Since $\langle \sigma_k \rangle = \partial_{t_k} F$ \cite{GROSS1990333, Witten:1990hr} the relationship between $\sigma_k$ and the Schrodinger Hamiltonian $\mathcal{H}$ can be recast \cite{BANKS1990279} as
	\begin{equation}
		\frac{\partial u}{\partial t_k} = R'_{k+1}, \label{eqn:KdVflows}
	\end{equation}
which are the generalized KdV flows. The flow equations organize the operator content of the double scaled matrix model and control the perturbation theory of interesting observables.

The KdV flows are generated by an infinite set of commuting vector fields $\xi_k$ defined by
	\begin{equation}
		\xi_k = \sum_{l = 0}^\infty R_k^{(l+1)} \frac{\delta}{\delta u^{(l)}},
	\end{equation}
where here the superscripts denote derivatives with respect to $x$. Indeed one can easily confirm that
	\begin{equation}
		\frac{\partial u}{\partial t_k} = \xi_{k+1}\cdot u.
	\end{equation}
These vector fields are related to the point-like operators $\sigma_k$ through connected correlation functions \cite{BANKS1990279}
	\begin{equation}
		 \hbar^2\frac{\partial^2}{\partial \mu^2} \langle \sigma_{k_1}\cdots \sigma_{k_n}\rangle_\mu = \xi_{k_n + 1} \cdots \xi_{k_1+1} \cdot u(\mu), \label{eqn:sigmatoxi}
	\end{equation}
where the $\mu$ subscript on the correlation function denotes that it is evaluated at $x = \mu$.

Double scaled matrix models with a string equation of the form \eqref{eqn:closedstringeq} make up a subset of the class of models that describe closed string physics. One way of interpreting this statement is that the surfaces described by the model are closed and hence look like closed string worldsheets. However, it is possible to introduce asymptotic boundaries to these theories using macroscopic loop operators. In the finite-$N$ regime, these operators are represented by insertions of thermal partition functions $\tr e^{- l M}$ into the matrix integral. In a holographic interpretation, this is like saying there is a dual quantum system with this partition function living on the boundary \cite{Saad:2019lba}.


\subsection{Altland-Zirnbauer ensembles}

The second class of matrix model useful here are the $(\alpha,\beta)$ Altland-Zirnbauer (AZ) models, defined by the matrix integral
	\begin{equation}
		Z = \int \prod_{i = 1}^N d\lambda_i \lambda_i^\alpha \prod_{j < k}|\lambda_k^2 - \lambda_j^2|^\beta e^{-N\sum_{i = 1}^N V(\lambda_i)}. \label{eqn:AZmatrixintegral}
	\end{equation}
Of particular interest here will be the cases $(1,2)$ and $(1 + 2\Gamma, \beta)$ for some number $\Gamma$. The former is related to choosing the matrices to be positive and hermitian, {\it i.e.} choosing matrices of the form $M = H^\dagger H$.  The eigenvalues of $M$ are non-negative by construction, and so the matrix integral $Z$ is naturally defined by restricting the $\beta = 2$ Wigner-Dyson matrix integral \eqref{eqn:WDmatrixintegral} to $\mathbb{R}_+^N$. By changing coordinates $\lambda_i = y_i^2$ the matrix integral is of the form \eqref{eqn:AZmatrixintegral} with $\alpha = 1$ and $\beta = 2$. If $H$ is rectangular with size $(N + \Gamma) \times N$ the resulting matrix integral is of the form \eqref{eqn:AZmatrixintegral} with $\alpha = 1 + 2\Gamma$ and $\beta = 2$.

The double scaling procedure for these models goes through similarly to how it does in the Wigner-Dyson models, and all of the important objects defined for those models, e.g. the free energy, are important here. In particular one can still introduce orthogonal polynomials, whose recursion relation under multiplication by $\lambda$ double scales to a Schrodinger equation. Moreover, the double scaled version of the recursion coefficients, still referred to here as $u$, will still satisfy a differential equation, albeit a different one. Multicritical theories can still obtained for this class of matrix models by tuning the coupling constants to critical values during the double scaling procedure. For the $k^\text{th}$ multicritical model \cite{Dalley:1991qg}
	\begin{equation*}
		u \mathcal{R}_k^2 - \frac{\hbar^2}{2}\mathcal{R}_k\mathcal{R}_k'' + \frac{\hbar^2}{4}(\mathcal{R}_k')^2 = 0,
	\end{equation*}
where $\mathcal{R}_k \equiv R_k + x$. This is the most general equation of motion for the function $u$ that is both consistent with the KdV flows and which reproduces perturbative closed string physics -- the regime in which that perturbation theory is recovered will be discussed in the following section. The $k^{\text{th}}$ multicritical model is dual to $\mathcal{N}  = 1$ super-Liouville coupled to the $(2,4k)$ super minimal modelt. Just as with the Wigner-Dyson models, one can consider the massive interpolation between models
	\begin{equation*}
		u \mathcal{R}^2 - \frac{\hbar^2}{2}\mathcal{R}\mathcal{R}'' + \frac{\hbar^2}{4}(\mathcal{R}')^2 = 0,
	\end{equation*}
where now the more general $\mathcal{R} \equiv \sum_k t_k R_k + x$ is defined.

The string equation for the double scaled $(1,2)$ AZ model has two generalizations. The first is obtained by noticing that the function $u$ together with the coupling constants $t_k$ have a scaling symmetry \cite{Dalley:1991vr}
	\begin{equation}
		\sum_{k = 0}^\infty \left(k + \half \right) \frac{\partial u}{\partial t_k} + \half x \frac{\partial u}{\partial x} + u = 0. \label{eqn:CallanSymanzik}
	\end{equation}
Here we have absorbed $\hbar$ into the coupling constants and assigned the following mass dimensions
	\begin{equation}
		[u] = 1,\quad [x] = -\half,\quad [t_k] = -\left(k + \half\right).
	\end{equation}
The scaling relation \eqref{eqn:CallanSymanzik} is just a Callan-Symanzik equation for the model. Assuming that $u$ satisfies the KdV flow equations \eqref{eqn:KdVflows}, using the recursion relation for the Gelfand-Dikii polynomials, and integrating once, \eqref{eqn:CallanSymanzik} becomes \cite{Dalley:1991qg}
	\begin{equation*}
		u \mathcal{R}^2 - \frac{\hbar^2}{2}\mathcal{R}\mathcal{R}'' + \frac{\hbar^2}{4}(\mathcal{R}')^2 = \hbar^2\Gamma^2,
	\end{equation*}
which is the correct string equation for the double scaled $(1 + 2\Gamma, 2)$ AZ model. There is an apparent tension between the definition of $\mathcal{R}$ provided above and what one would actually obtain by following the procedure described to turn the scaling relation into the string equation. The choice to include the explicit factor of $k + \half$ is conceptually necessary to call \eqref{eqn:CallanSymanzik} a scaling symmetry. In the formalism used throughout this paper, unless noted otherwise, we choose to absorb any extra factors into the coupling constants $t_k$.

The second generalization comes from slightly altering the model starting at the level of the matrix integral. Recall that the $(1,2)$ and $(1 + 2\Gamma,2)$ models involved non-negative hermitian matrices, which is to say that the eigenvalues satisfied $\lambda \geq 0$. One can instead consider an ensemble of matrices whose eigenvalues satisfy $E \geq -s$ in the double scaling limit for some $s \in \mathbb{R}$, yielding the double scaled string equation \cite{Dalley:1991yi}
	\begin{equation*}
		(u+s) \mathcal{R}^2 - \frac{\hbar^2}{2}\mathcal{R}\mathcal{R}'' + \frac{\hbar^2}{4}(\mathcal{R}')^2 = 0.
	\end{equation*}
	
Both generalizations of the double scaled $(1,2)$ AZ model can be combined into one model, with the string equation
	\begin{equation}
		(u+s) \mathcal{R}^2 - \frac{\hbar^2}{2}\mathcal{R}\mathcal{R}'' + \frac{\hbar^2}{4}(\mathcal{R}')^2 = \hbar^2 \Gamma^2. \label{eqn:bigstringequation}
	\end{equation}
From the point of view of the differential equation $\Gamma$ need not be an integer, despite the fact that from the matrix model perspective it does not make sense to consider a matrix with non-integer size.  This generality remains true in the physical interpretation of the equation of motion: there will be some cases where $\Gamma$ should naturally be a non-negative integer, cases where the exact value of $\Gamma$ is not important, and even cases where $\Gamma = -\half$. The full differential equation \eqref{eqn:bigstringequation} will be called the DJM equation henceforth.

The DJM equation \eqref{eqn:bigstringequation} exhibits a somewhat miraculous universality. It was originally derived as the string equation describing double scaled multicritical complex matrix models \cite{Dalley:1991qg,Dalley:1992br}. It was subsequently argued that the DJM equation provides a consistent non-perturbative formulation of the multicritical bosonic closed string theory \cite{Dalley:1991vr,Dalley:1991yi}. It was later discovered that \eqref{eqn:bigstringequation} also has the capacity to describe type 0A and 0B superstrings \cite{Klebanov:2003wg, Johnson:2003hy}, with further explorations in a modern context in \cite{Johnson:2020heh, Johnson:2021owr}. Different models are defined by the different types of perturbative solutions \eqref{eqn:bigstringequation} allows. This is discussed in depth below.

In \cite{KOSTOV1990181} Kostov showed that a matrix model describing both open and closed string sectors simultaneously in the double scaling limit is obtained from just the closed string matrix model by including the deformation of the matrix potential
	\begin{equation}
		V_{\text{open}}(M) = \frac{\gamma}{N} \tr \log(1 - m^2 M^2), \label{eqn:Vopen}
	\end{equation}
which can be thought of as the insertion of a determinant operator in the matrix integral. The details concerning the double scaling limit of the pure gravity model are contained therein, as well as the beginnings of the generalization to arbitrary multicritical models. The story was made more complete in \cite{Dalley:1992br, Johnson:1992wr, Johnson:1993vk, Johnson:2004ut}, where the connections to the KdV hierarchy in the double scaling limit were established in generality. The following is a review of \cite{Johnson:1993vk}.

In the normalization of \cite{IMGelfand_1975} the first three Gelfand-Dikii polynomials are
	\begin{equation}
		\tilde{R}_0 = \half,\quad \tilde{R}_1 = -\quarter u,\quad \tilde{R}_2 = \frac{1}{16}(3u^2 - u''),
	\end{equation}
and the higher order ones are determined by the recursion relation
	\begin{equation}
		\tilde{R}_{k+1}' = \quarter \tilde{R}_k''' - u \tilde{R}_k' - \half u'\tilde{R}_k.
	\end{equation}
The closed string equation is written
	\begin{equation}
		\tilde{\mathcal{R}} \equiv \sum_{k = 0}^\infty \left(k + \half \right) t_k \tilde{R_k} = 0.
	\end{equation}

The string equation describing the open-closed string system is similar to the closed string equation, but involves a new object $\hat{R}$
	\begin{equation}
		\tilde{\mathcal{R}} + 2\hbar \Gamma \hat{R}(x,s) = 0.\label{eqn:Kostoveq}
	\end{equation}
The new function $\hat{R}$ is the Gelfand-Dikii resolvent
	\begin{equation}
		\hat{R}(x,s) = \mel**{x}{\frac{1}{ -\hbar^2 \partial_x^2 + u(x) - s}}{x} = \sum_{k = 0}^\infty \frac{\tilde{R}_k}{(-s)^{k + \half}}.
	\end{equation}
In the above, $\hbar$ is the closed string coupling constant, $\Gamma$ is the scaling part of $\gamma$ (the ratio of the open to closed string coupling), and $\rho$ is the scaling part of $m^2$ (the mass of the ends of the open strings). The variable $x$ is related to the KdV time $t_0$ by $t_0 = -4x$. The resolvent $\hat{R}$ solves the Gelfand-Dikii differential equation
	\begin{equation}
		4(u - s)\hat{R}^2 - 2\hat{R}\hat{R}'' + (\hat{R}')^2 = 1, \label{eqn:GDeqn}
	\end{equation}
where primes denote derivatives with respect to $x$. By solving (\ref{eqn:Kostoveq}) for $\hat{R}$ and substituting that into the GD equation, one obtains the (open-closed) string equation
	\begin{equation}
		(u - s) \tilde{\mathcal{R}}^2 - \half \tilde{\mathcal{R}}\tilde{\mathcal{R}}'' + \quarter (\tilde{\mathcal{R}}')^2 = \hbar^2 \Gamma^2, \label{eqn:clopenstringeq}
	\end{equation}
which we can plainly see is the DJM equation. In addition to the models described previously that share this string equation, it also describes open string physics.

The open-closed string equation (\ref{eqn:clopenstringeq}) can be obtained from the closed string equation (\ref{eqn:closedstringeq}) by a shift in the coupling constants $t_k$. Define the shifted variables
	\begin{equation}
	\begin{aligned}
		t_k &\to t_k + \frac{2\hbar\Gamma}{k + \half}(-s)^{-(k+\half)},\\
		x &\to x - \hbar\Gamma (-s)^{-\half}.
	\end{aligned}
	\end{equation}
By substituting these into the closed string equation (\ref{eqn:closedstringeq}) and invoking the expansion of the resolvent $\hat{R}$, one gets Kostov's string equation (\ref{eqn:Kostoveq}).

We pause here to clarify some terminology, since the same objects are used to describe different physics. The work of \cite{KOSTOV1990181,Johnson:1993vk, Johnson:2004ut} demonstrated a duality between purely closed strings and an interacting open-closed string system. In the ``KdV frame'' both matrix models are organized by the KdV flows and a string equation, each of which is dependent on a set of KdV times, or coupling constants. Before establishing any equivalence between the two models, the closed string matrix model is described by the string equation (\ref{eqn:closedstringeq}), while the open-closed system is described by the string equation (\ref{eqn:clopenstringeq}). The open-closed system can be described by the closed string equation (\ref{eqn:closedstringeq}) once one makes the change of coordinates
	\begin{equation*}
	\begin{aligned}
		t^{(\text{closed})}_k &= t^{(\text{open})}_k + \frac{2\hbar\Gamma}{k + \half}(-s)^{-(k+\half)},\\
		x^{(\text{closed})} &= x^{(\text{open})} - \hbar\Gamma (-s)^{-\half}.
	\end{aligned}
	\end{equation*}


\subsection{Resolvents and the eigenvalue density}

The matrix resolvent is defined by $R(\lambda) = (M - \lambda)^{-1}$, and is analytic everywhere except a cut along the eigenvalue distribution. It's expectation value has a discontinuity across the real axis, which is interpreted as the one-eigenvalue correlator, or density. General correlation functions of $R$ have a topological expansion in $1/N$ prior to double scaling. Examination of the perturbative contributions to these correlators yields the loop equations of the matrix model.

After diagonalizing the matrix to work in terms of the eigenvalues it becomes convenient to work in terms of the eigenvalue density $\rho(\lambda)$, which is the discontinuity of the resolvent across the cut along the spectrum: $R(\lambda + i\epsilon) - R(\lambda - i\epsilon) = i\pi \rho(\lambda).$
Equivalently, in the orthogonal polynomial formalism it is given by $\rho(\lambda) = \sum_{i = 0}^{N-1} \psi_i(\lambda)^2$. A similar, more powerful quantity constructed out of the orthogonal polynomials is the self reproducing kernel
	\begin{equation}
		K(\lambda,\lambda') = \sum_{i = 0}^{N-1} \psi_i(\lambda) \psi_i(\lambda'),
	\end{equation}
which can be used to probe more complicated statistical questions about the matrix model.

The resolvent, eigenvalue density, and kernel continue to be important in the double scaling limit, although out of the three we will primarily be concerned with the resolvent. However, through open-closed duality we will see that the kernel has an direct connection to the two-resolvent correlator that is mediated by the Gelfand-Dikii polynomials, and not the eigenvalue density.


\section{String Equation Perturbation Theory} \label{scn:stringequationperturbationtheory}

We now take some time to develop the perturbation theory of the string equations defined in the previous section. This section has two purposes: to display the general techniques for solving these differential equations and to collect results that will be useful later. In particular, we begin with a discussion of the closed and open string solutions of non-supersymmetric models. Using the open string sector as a bridge, we then study the closed string solutions in $0A$ and $\mathcal{N}  = 2$ supersymmetric theories. Notable examples are provided in each case.

The DJM equation naturally incorporates the parameters $z$ and $\Gamma$, though they have context-dependent interpretations. One of the key differences is that the parameters exist in both the open and closed sectors of the supersymmetric theories. In the open string setting, $z$ is related to the open string endpoint mass (or equivalently the cosmological constant of the brane) and $\Gamma$ related to the open-closed string coupling, and neither parameter has an interpretation in the closed string sector. In closed string sector of 0A theories, the parameter $\Gamma$ counts background RR flux on the worldsheet. 

Another key difference is the interpretation of the variable $x$. In non-supersymmetric systems, both the open and closed physics is described in the region $x < 0$. On the other hand, in the $0A$ and $\mathcal{N} = 2$ systems the open string physics is still in the $x < 0 $ region, but the closed string physics is captured by the $ x> 0$ part of the solution. The comparisons are illustrated in the following table.

\begin{table}[htp]
\caption{Locations of open and closed string physics}
\begin{center}
\begin{tabular}{c|c|c}
	&Open string & Closed string \\
	\hline
	$\mathcal{N} = 0$ & $x < 0$ & $x < 0$ \\
	\hline
	$\mathcal{N} = 1$ & $x < 0$ & $x > 0$ \\
	\hline
	$\mathcal{N} = 2$ &$ x < 0$ & $x > 0$.
\end{tabular}
\end{center}
\label{tbl:openvsclosed}
\end{table}%

We also note here that compared to \cite{IMGelfand_1975}, a different normalization for the Gelfand-Dikii polynomials will be utilized from here on, unless noted otherwise. The first several $R_k$ are given by
	\begin{equation}
	\begin{aligned}
		R_0 &= 1, \\
		R_1 &= u,\\
		R_2 &= u^2 - \frac{\hbar^2}{3}u'',\\
		R_3 &= u^3 - \frac{\hbar^2}{2}(u')^2 - \hbar^2 uu'' + \frac{\hbar^4}{10}u^{(4)}.
	\end{aligned}
	\end{equation}
The normalization used here fixes $R_k = u^k + \cdots$. The Gelfand-Dikii polynomials satisfy the recursion relation
	\begin{equation}
		R_{k + 1} = \frac{2k + 2}{2k+1} \left[uR_k - \frac{\hbar^2}{4}R_k'' - \half \int^x d\overline{x} u'(\overline{x})R_k\right]. \label{eqn:GDrecursion}
	\end{equation}
They also have an organization in powers of $\hbar$ (see \cite{Johnson:2021owr}, for example)
	\begin{equation}
	\begin{aligned}
		R_k &= r^{(0)}_k + \hbar^2 r^{(2)}_k + \cdots ,\quad\\
		r^{(0)}_k &= u^k,\quad \quad r^{(2)}_k  = - \frac{k(k-1)}{12}\Big[(k-2)(u')^2 + 2uu''\Big]u^{k-3}. \label{eqn:GDhbarexpansion}
	\end{aligned}
	\end{equation}

\subsection{Non-supersymmetric closed string equation}

The equation of motion for a general double scaled one-matrix $\beta = 2$ Wigner-Dyson model is
	\begin{equation}
		\sum_{k = 1}^\infty t_k R_k + x = 0, \label{eqn:Fullclosedstringequation}
	\end{equation}
A solution to this differential equaiton can be obtained in an asymptotic series in small $\hbar/|x|$
	\begin{equation}
		u(x) = \sum_{g = 0}^\infty \hbar^{2g}u_g(x).
	\end{equation}
The leading order contribution $u_0$ satisfies 
	\begin{equation}
		f(u_0) + x = 0, \quad \quad f(u_0) \equiv \sum_{k = 1}^\infty t_k u_0^k. \label{eqn:closedstringdiskequation}
	\end{equation}
Equation \eqref{eqn:closedstringdiskequation} is commonly referred to as the disk level string equation. When only the $p^\text{th}$ coupling constant $t_p$ is non-zero, the solution is $u_0 = (-x/t_p)^{1/p}$. Typically in these models the Fermi level is located at $\mu = 0$. A feature of the individual multicritical models is that $u_0(\mu) = 0$, which generalizes to massive interpolations: the disk level equation at $x = 0$ is $f(u_0(0)) = 0$, which is always solved by $u_0(0) = 0$ since $f$ has no constant term.

The derivatives of $u_0$ with respect to $x$ can be expressed in terms of the solution $u_0$ and the function $f$. We collect the first several with their values at the Fermi surface $x = \mu$ here:
	\begin{equation}
    \begin{aligned}
        u_0'[u_0] &= -\frac{1}{\dot{f}},\quad u_0'(\mu) = -\frac{1}{t_1}\\
        u_0''[u_0] &= -\frac{\ddot{f}}{\dot{f}^3},\quad u_0''(\mu) = -\frac{2t_2}{t_1^3},\\
        u_0'''[u_0] &= \frac{1}{\dot{f}^3}\left[\frac{\dddot{f}}{\dot{f}} - 3\left(\frac{\ddot{f}}{\dot{f}}\right)^2\right], \quad  u_0'''(\mu) = \frac{6t_1t_3 - 12 t_2^2}{t_1^5}, \label{eqn:nonsusy_fermieval}
    \end{aligned}
    \end{equation}
where a dot denotes a partial derivative with respect to $u_0$. 

The subsequent perturbative contributions to $u(x)$ are determined by the Gelfand-Dikii polynomials and the disk level string equation. The $g = 1$ equation is
	\begin{equation}
		\dot{f} u_1 - \frac{1}{12} \Big[\dddot{f}(u_0')^2 + 2\ddot{f}u_0''\Big] = 0,
	\end{equation}
The solution and its value at $x = \mu$, expressed in terms of the function $f$ and the coupling constants, are
	\begin{equation}
        u_1 = \frac{1}{12(f')^2}\left[\frac{\dddot{f}}{\dot{f}} - 2\left(\frac{\ddot{f}}{\dot{f}}\right)^2\right], \quad  u_1(\mu) = \frac{3t_1t_3 - 4t_2^2}{6t_1^4}.
    \end{equation}
   
It will prove convenient to consider the relationship between the perturbative corrections $u_g$ and the derivatives of the leading order solution $u_0$. For example, the second derivative $u_0''$ is related to the first derivative $u_0'$ by
	\begin{equation}
		u_0'' = u_0' \frac{d}{du_0} u_0', \label{eqn:u0'tou0''}
	\end{equation}
via a straightforward application of the chain rule. At $g = 1$ one finds
	\begin{equation}
		u_1 = -\frac{u_0'}{12} \frac{d^2}{du_0^2} u_0'. \label{eqn:u0'tou1}
	\end{equation}     
It is likely that there are similar relationships expressing each contribution $u_g$ in terms of derivatives of lower order contributions to $u$. A proof of this would probably involve the recursion relation of the GD polynomials and their $\hbar$-expansion. If such a total derivative relation exists at each $g$, it would simplify certain correlation functions in the closed string sector. We will comment on this again later.
    
At the next order in $\hbar$, the string equation is
	\begin{equation}
	\begin{aligned}
		u_2\dot{f} &+ \half\left(u_1^2 - \third u_1''\right)\ddot{f} + \frac{1}{6}\left(\frac{1}{10}u_0^{(4)} - u_0''u_1 - u_0' u_1'\right)\dddot{f} \\
		&+ \half\left(\frac{1}{20}(u_0'')^2 + \frac{1}{15}u_0'u_0''' - \frac{1}{6}(u_0')^2u_2\right)f^{(4)} + \frac{11}{366}(u_0')^2u_0''f^{(5)} + \frac{1}{288}(u_0')^4f^{(6)} = 0.
	\end{aligned}
	\end{equation}
The solution, expressed purely in terms of $f$, is
	\begin{equation}
	\begin{aligned}
		u_2 &= \frac{1}{1440 \dot{f}^9}\Bigg[-980 \ddot{f}^5-420 f^{(4)} \dot{f}^2 \ddot{f}^2+1760 \dddot{f} \dot{f} \ddot{f}^3\\
		&+\dot{f}^3 \left(102 \dddot{f} f^{(4)}-5 f^{(6)} \dot{f}\right)+\dot{f}^2 \left(64 f^{(5)} \dot{f}-545 \dddot{f}^2\right) \ddot{f}\Bigg].
	\end{aligned}
	\end{equation}
Evaluated at $x = \mu$ it is
	\begin{equation}
        \begin{aligned}
		u_2(\mu) &= \frac{-3920 t_2^5+10560 t_1 t_3 t_2^3-5040 t_1^2 t_4 t_2^2+15 t_1^2 \left(128 t_1 t_5-327 t_3^2\right) t_2+18 t_1^3}{180 t_1^9} \\
  &+ \frac{\left(102 t_3 t_4-25 t_1 t_6\right)}{180 t_1^9}.
        \end{aligned}
	\end{equation}	
	
It is possible to determine which coupling constants appear when the function $u_{g}^{(n)}$ is evaluated at $x = \mu$. Let $\overline{k}'_{g,n}$ denote the largest index of the $t_k$ that shows up in $u_{g}^{(n)}(\mu)$. It can be shown by induction that $\overline{k}'_{g,n} = 3g + n$. For $g = 0$, the function $u_0^{(n)}$ will contain one appearance of $f^{(n)}[u_0]$, and $f^{(n)}[0] \propto t_{n}$. For $g \geq 1$, the result follows from analysis of the $\hbar$-expansion of the Gelfand-Dikii polynomials. The base case $g = 1, n = 0$ has an appearance of $f^{(3)}$. Inducting on $g$ follows, and any nonzero value of $n$ is included trivially.

\subsubsection{Examples}

Here we provide a collection of the functions $u_g$ and their derivatives evaluated for interesting examples: JT gravity, the minimal string, and the Virasoro minimal string. These examples will be reconsidered when we look at the perturbative expansion of correlation functions later, but nevertheless this subsection is particularly result-heavy. The ready is welcome to skip to the section \ref{scn:DJM} and refer back here when necessary.

The coupling constants defining JT gravity are \cite{Okuyama:2019xbv, Johnson:2019eik}
    \begin{equation}
        t_k = \half \frac{\pi^{2k-2}}{k!(k-1)!}.
    \end{equation}
They are determined by first computing the spectral density using the matrix model formalism and demanding that it matches the path integral calculation. The leading order contribution $u_0$ satisfies
    \begin{equation}
       \frac{\sqrt{u_0}}{2\pi} I_1(2\pi\sqrt{u_0}) + x = 0.
    \end{equation}
The first derivative $u_0'$ is given in terms of $u_0$ by
    \begin{equation}
        u_0'[u_0] = -\frac{1}{{_0}\tilde{F}_1(1,\pi^2 u_0)},
    \end{equation}
where the tilde denotes that the hypergeometric function is regularized and we have used
    \begin{equation}
        {_0}\tilde{F}_1\left(a+\half,\frac{x^2}{16}\right) = \left(\frac{x}{4}\right)^{\half - a} I_{a-\half}\left(\frac{x}{2}\right).
    \end{equation}
Further, the second and third derivatives are
    \begin{equation}
    \begin{aligned}
        u_0''[u_0] &= -\frac{\pi^2\,{_0}\tilde{F}_1(2,\pi^2 u_0)}{{_0}\tilde{F}_1(1,\pi^2 u_0)^3}, \\
        u_0'''[u_0] &= \pi^4 \frac{{_0}\tilde{F}_1(1,\pi^2 u_0)\,{_0}\tilde{F}_1(3,\pi^2 u_0) - 3\,{_0}\tilde{F}_1(2,\pi^2 u_0)^2}{{_0}\tilde{F}_1(1,\pi^2 u_0)^4}.
    \end{aligned}
    \end{equation}
Hence the first correction to the potential is given by
    \begin{equation}
        u_1[u_0] = \frac{\pi^4}{12} \frac{{_0}\tilde{F}_1(1,\pi^2 u_0)\,{_0}\tilde{F}_1(3,\pi^2 u_0) - 2\,{_0}\tilde{F}_1(2,\pi^2 u_0)^2}{{_0}\tilde{F}_1(1,\pi^2 u_0)^4}.
    \end{equation}
Its first derivative is given by
    \begin{equation}
    \begin{aligned}
        u_1'[u_0] &= \frac{\pi^6}{12}\frac{ 7 {_0}\tilde{F}_1(1,\pi^2 u_0) {_0}\tilde{F}_1(2,\pi^2 u_0) {_0}\tilde{F}_1(3,\pi^2 u_0) - 8{_0}\tilde{F}_1(2,\pi^2 u_0)^3}{ {_0}\tilde{F}_1(1,\pi^2 u_0)^6 } \\
        & - \frac{\pi^6}{12}\frac{{_0}\tilde{F}_1(1,\pi^2 u_0)^2{_0}\tilde{F}_1(4,\pi^2 u_0)}{ {_0}\tilde{F}_1(1,\pi^2 u_0)^6 }.
    \end{aligned}
    \end{equation}
    
The $(2,2p-1)$ minimal string is defined by a particular massive interpolation between the first $p$ minimal models, with the $p$-dependent coupling constants \cite{Mertens:2020hbs}
	\begin{equation}
        t_k = \half \frac{\pi^{2k-2}}{k!(k-1)!}\frac{4^{k-1}(p + k - 2)!}{(p-k)!(2p-1)^{2k-2}}. \label{eqn:minstringcoupling}
    \end{equation}
The leading order contribution $u_0$ satisfies
	\begin{equation}
		u_0\, {_{2}}F_1\left(1-p,p,2,-\frac{4\pi^2u_0}{(2p-1)^2}\right) + x = 0.
    \end{equation}
The first and second derivatives are
	\begin{equation}
	\begin{aligned}
		u_0'[u_0] &= - \frac{1}{{_{2}}F_1\left(1-p,p,1,-\frac{4\pi^2u_0}{(2p-1)^2}\right)},\\
		u_0''[u_0] &= - \frac{4p(p-1)\pi^2 {_{2}}F_1\left(2-p,p+1,2,-\frac{4\pi^2u_0}{(2p-1)^2}\right)}{(2p-1)^2 {_{2}}F_1\left(1-p,p,1,-\frac{4\pi^2u_0}{(2p-1)^2}\right)}.
	\end{aligned}
	\end{equation}
The first correction to $u$ is given by
	\begin{equation}
	\begin{aligned}
		u_1[u_0]&= -\frac{8p^2(p-1)^2\pi ^4  \, _2F_1\left(2-p,p+1,2,-\frac{4 \pi ^2 u_0}{(2p-1)^2}\right)^2}{3 (2p-1)^4 \, _2F_1\left(1-p,p;1;-\frac{4 \pi ^2 u_0}{(2p-1)^2}\right)^4} \\
		&+ \frac{2p(p-1)(p^2-p-2) \, _2F_1\left(1-p,p,1,-\frac{4 \pi ^2 u_0}{(2p-1)^2}\right) \, _2F_1\left(3-p,p+2,3,-\frac{4 \pi ^2 u_0}{(2p-1)^2}\right)}{3 (2p-1)^4 \, _2F_1\left(1-p,p;1;-\frac{4 \pi ^2 u_0}{(2p-1)^2}\right)^4}.
	\end{aligned}
	\end{equation}
	
The matrix model dual to the Virasoro minimal string has coupling constants \cite{Collier:2023cyw, Johnson:2024bue} 
	\begin{equation}
		t_k = 2\sqrt{2}\frac{\pi^{2k+1}}{(k!)^2}\left(Q^{2k} - \hat{Q}^{2k}\right),
	\end{equation}
where
	\begin{equation}
		Q = b + b^{-1},\quad \hat{Q} = b^{-1} - b. \label{eqn:Qdef}
	\end{equation}
Here $b$ is a parameter in Liouvillie theory. The leading order contribution $u_0$ satisfies
	\begin{equation}
		2\sqrt{2}\Bigg[I_0(2\pi Q \sqrt{u_0}) - I_0(2\pi \hat{Q}\sqrt{u_0})\Bigg] + x = 0.
	\end{equation}
The first and second derivatives are
	\begin{equation}
	\begin{aligned}
		u_0'[u_0] &= -\frac{b^2}{2 \sqrt{2} \pi ^3} \frac{1}{\left(b^2+1\right)^2 \, _0\tilde{F}_1\left(2,Q^2 \pi^2 u_0\right)-\left(b^2-1\right)^2 \, _0\tilde{F}_1\left(2,\hat{Q}^2 \pi ^2 u_0\right)}, \\
		u_0''[u_0] &= -\frac{b^2}{8\pi^4}\frac{\left(b^2+1\right)^4 \, _0\tilde{F}_1\left(3,Q^2\pi ^2 u\right)-\left(b^2-1\right)^4 \, _0\tilde{F}_1\left(3,\hat{Q}^2 \pi ^2 u\right)}{ \Big(\left(b^2+1\right)^2 \, _0\tilde{F}_1\left(2,Q^2 \pi ^2 u_0\right)-\left(b^2-1\right)^2 \, _0\tilde{F}_1\left(2,\hat{Q}^2 \pi ^2 u_0\right)\Big)^3}.
	\end{aligned}
	\end{equation}

	
\subsection{DJM equation} \label{scn:DJM}

For the remainder this work, we will redefine the constant $s$ in the DJM equation as $z^2 = -s$. This is the standard uniformizing coordinate transformation familiar from studying the spectral curve in matrix models, and is not a coincidence since standard interpretations of FZZT branes in matrix models link the parameter $z$ to the matrix eigenvalues. 

Unlike the string equation arising from double scaled Hermitian matrix models, the DJM equation admits real solutions on all of $\mathbb{R}$. As discussed above, open string perturbation theory is always recovered in the limit $x \to -\infty$. An interesting feature of supersymmetric theories described in this fashion is that they naturally incorporate D-branes by having this perturbative regime built into their equations of motion, which is in stark contrast to the non-supersymmetric theories where extra work had to be done to add branes. Nevertheless, the $x \to -\infty$ solutions to the DJM equation provide a sort of universal description of the open string sectors of the class of matrix models describing $(2,\#)$ (super) minimal models coupled to (super) Liouville theory.


\subsubsection{Open string sector}

The GD resolvent $\hat{R}$ has the expansion
	\begin{equation}
		\hat{R}(x,z) = \frac{1}{\hbar} \sum_{k = 0}^\infty \frac{(-1)^k}{z^{2k + 1}}\frac{(2k)!}{2^{2k+1}(k!)^2}R_k,
	\end{equation}
in terms of the differently normalized GD polynomials. Notice that we have considered the resolvent with an explicit minus sign in the parameter. To avoid confusion we will change variables $- \rho = z^2$. To obtain the Kostov differential equation we define the shifted variables
	\begin{equation}
	\begin{aligned}
		t_k &\to t_k + 2\hbar\Gamma \frac{(-1)^k(2k)!}{2^{2k+1}(k!)^2}z^{-(2k + 1)}\\
		x&\to x + 2\hbar\Gamma (2z)^{-1}. \label{eqn:shiftedcouplings}
	\end{aligned}
	\end{equation}
Define the shift coefficients
	\begin{equation}
		\zeta_k = \frac{(-1)^k(2k)!}{2^{2k+1}(k!)^2},
	\end{equation}
for $k \geq 0$.

The full string equation of the open-closed model can be written
	\begin{equation}
		\sum_{k = 1}^\infty \Big(t_k + 2\hbar\Gamma\zeta_k z^{-(2k+1)}\Big)R_k + x + \hbar\Gamma z^{-1} = 0.
	\end{equation}
Denote the expansion of $u$ by
	\begin{equation}
		u = \sum_{g,h = 0}^\infty \hbar^{2g + h} \Gamma^h u_{g,h},
	\end{equation}
with $u_{0,0}\equiv u_0$. At leading order we still have the disk level string equation $f(u_0) + x = 0$. The equation for $u_{0,1}$ can be written
	\begin{equation}
		\dot{f}u_{0,1} + 2\varphi + z^{-1} = 0
	\end{equation}
where $\varphi = \sum_{k = 1}^\infty \zeta_k z^{-(2k+1)}u_0^k$. The function $\varphi$ can be written in closed form
	\begin{equation}
		2\varphi + \frac{1}{z} = \frac{1}{\sqrt{u_0 + z^2}},
	\end{equation}
and thus
	\begin{equation}
		u_{0,1} = - \frac{1}{\dot{f}\sqrt{u_0 + z^2}}. \label{eqn:u01}
	\end{equation} 
The function $\varphi + (2z)^{-1}$ is merely the leading solution to the Gelfand-Dikii equation for the resolvent $\hat{R}$. The functions $f$ and $\varphi$ are ubiquitous in perturbation theory of the open-closed string equation, much like $f$ is in solving the undeformed closed string equation.

\begin{figure}[ht]
	\begin{subfigure}[t]{0.49\textwidth}
	\includegraphics[clip,trim=.6in 2.75in .5in 2.7in, width=\textwidth]{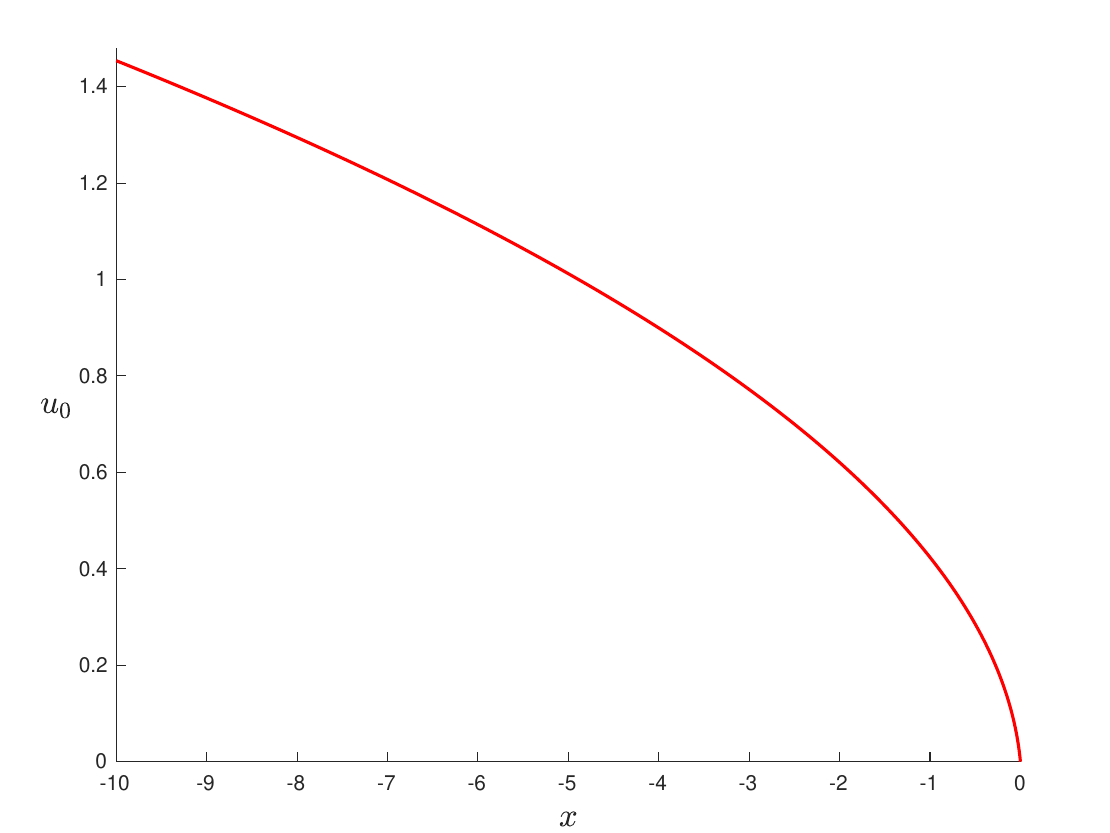}
	\end{subfigure}
	\begin{subfigure}[t]{0.49\textwidth}
	\includegraphics[clip,trim=.6in 2.75in .5in 2.7in, width=\textwidth]{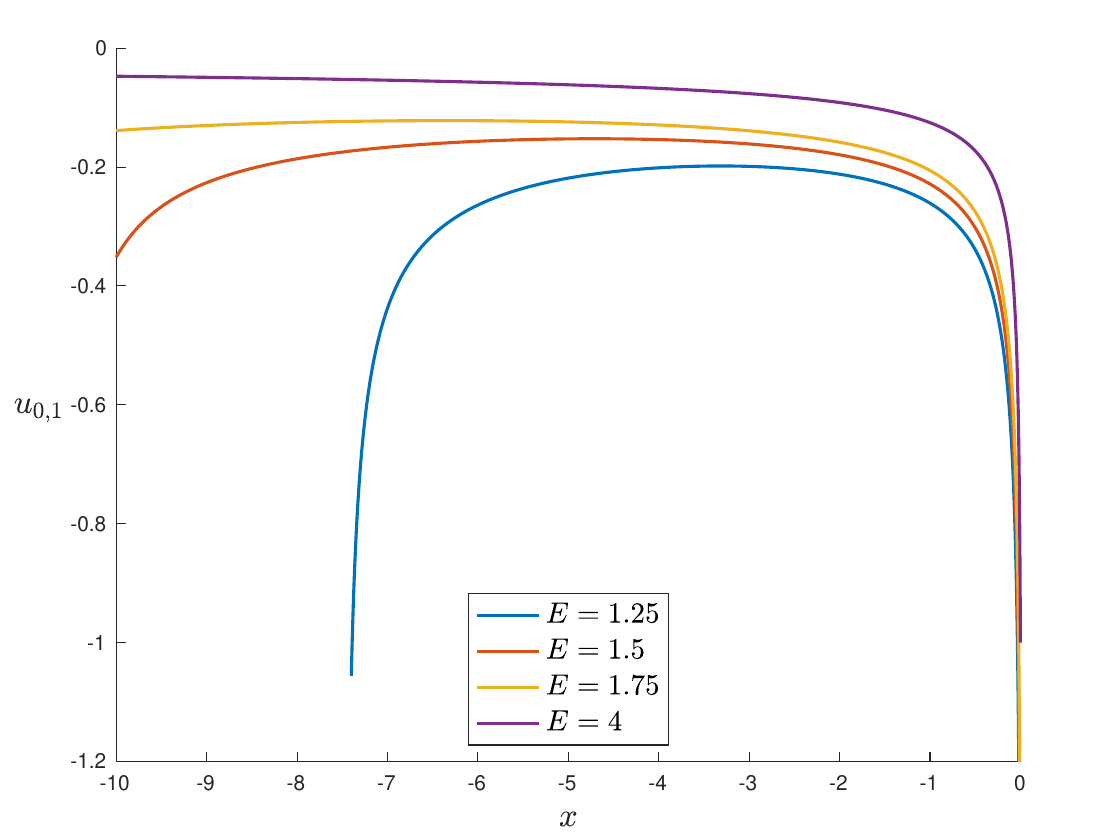}
	\end{subfigure}
\caption{The leading order (left) and first open string sector correction (right) to the potential $u$ for the $(2,3)$ minimal string. As the energy grows, the nonzero vertical asymptote in $u_{0,1}$ gets pushed toward $x \to -\infty$.}
\label{fig:minstringpotential}
\end{figure}

The general structure of the string equation for higher order corrections to $u_0$ is
	\begin{equation}
		\sum_{k = 1}^\infty \Big(t_k r^{(g,h)}_k + 2\zeta_k z^{-(2k+1)}r^{(g,h-1)}_k\Big) = 0,
	\end{equation}
where $r^{(g,h)}_k$ is the full contribution to $R_k$ at order $\hbar^{2g + h}\Gamma^h$. For $g = 0,h = 2$ we have
	\begin{equation}
		\dot{f}u_{0,2} + \half \ddot{f}u_{0,1}^2 + \dot{\varphi}u_{0,1} = 0.
	\end{equation}
The solution $u_{0,2}$ is
	\begin{equation}
		u_{0,2} = \frac{1}{2\dot{f}}\frac{d}{du_0} \Big[\dot{f} u_{0,1}^2 \Big] = -\frac{\dot{f} + 2\ddot{f}(u_0 + z^2)}{4\dot{f}^2(u_0 + z^2)^2}. \label{eqn:u01tou02}
	\end{equation}

For $g = 0,h = 3$ the string equation is
	\begin{equation}
		\dot{f} u_{0,3} + \ddot{f}u_{0,1}u_{0,2} + \frac{1}{6} \dddot{f}u_{0,1}^3 + \dot{\varphi}u_{0,2} + \half \ddot{\varphi}u_{0,1}^2 = 0,
	\end{equation}
with the solution
	\begin{equation}
		u_{0,3} = - \frac{3\dot{f}^2 + 9\dot{f} \big( 1 + 2(u_0 + z^2)\ddot{f} \big) + 8 (u_0 + z^2)^2(3\ddot{f}^2 - \dddot{f})}{48 \dot{f}^4 (u_0 + z^2)^{7/2}}
	\end{equation}
For $g = 1, h = 1$
	\begin{equation}
		\dot{f}u_{1,1} - \frac{1}{12}\Bigg[f^{(4)}u_{0,1}(u_0')^2 + 2\dddot{f}(u_{0,1}u_0')' - 2\ddot{f}(6u_{0,1}u_{1,0} - u_{0,1}'') + \ddot{\varphi} (u_{0}')^2 + 2\dot{\varphi}u_0''\Bigg] + \dot{\varphi}u_{1,0} = 0.
	\end{equation}
Notice that the $g = 1$ equation involves both $h = 0 $ and $h = 1$ functions.

\begin{figure}[ht]
	\begin{subfigure}[m]{0.49\textwidth}
	\includegraphics[width=\textwidth]{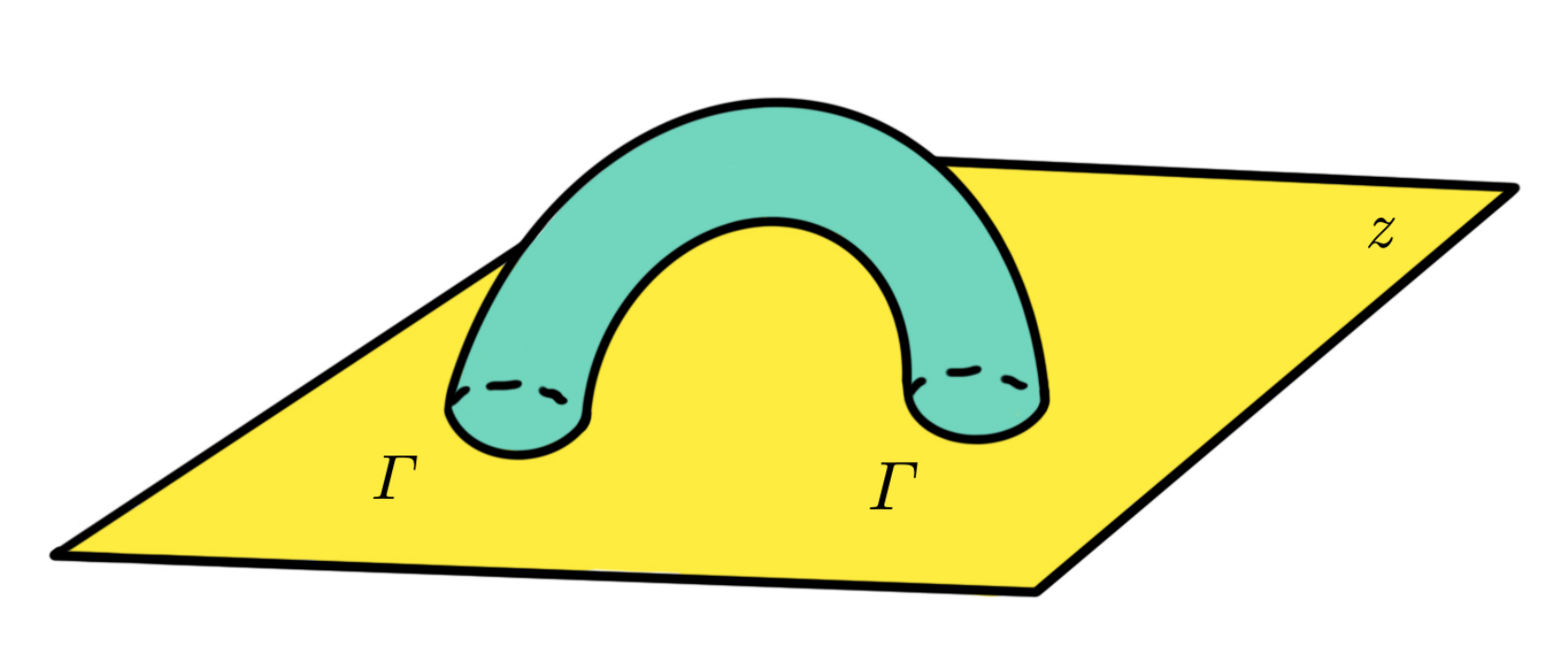}
	\end{subfigure}
	\begin{subfigure}[m]{0.49\textwidth}
	\includegraphics[width=\textwidth]{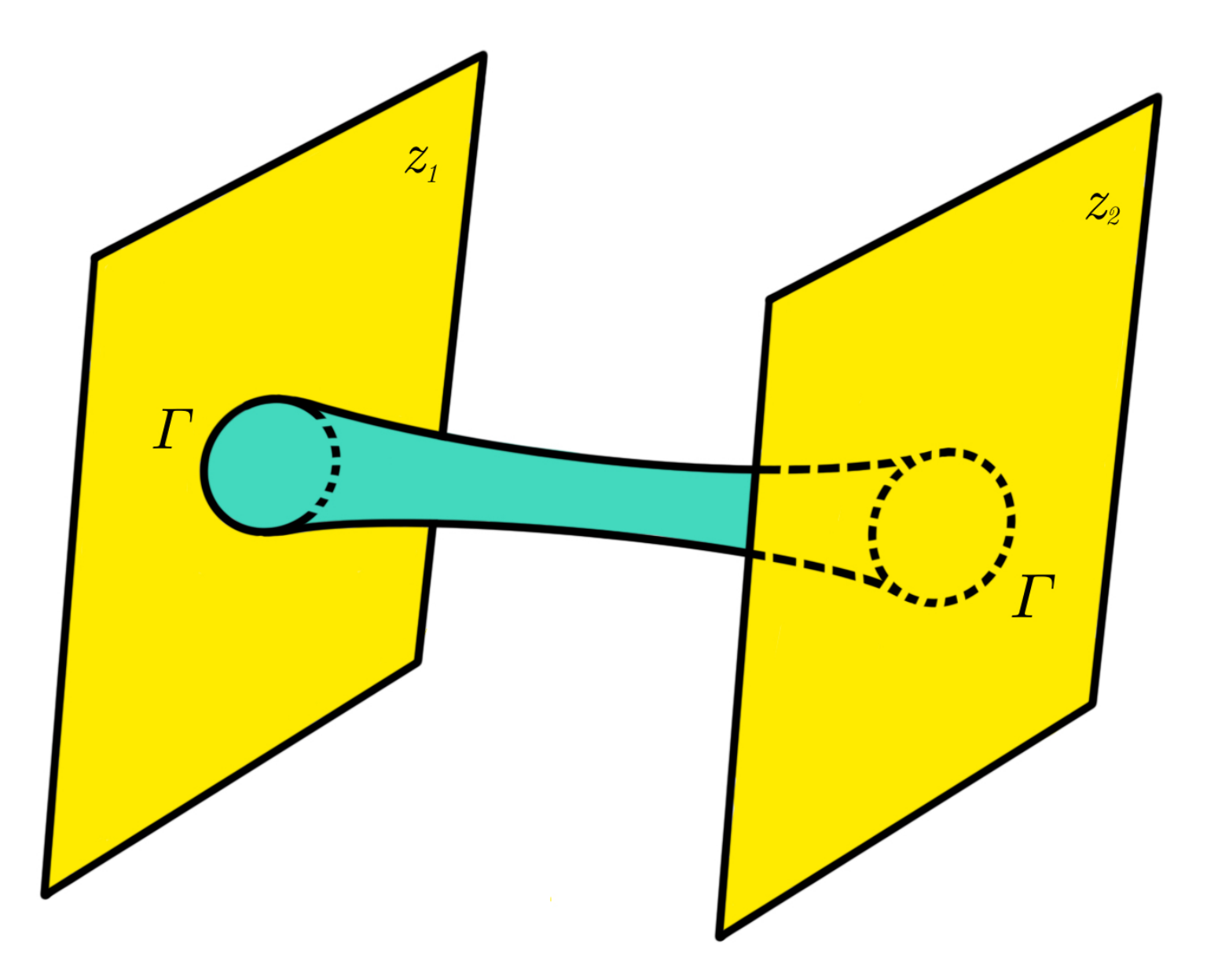}
	\end{subfigure}
\caption{The left panel shows a contribution at $\mathcal{O}(\Gamma^2)$ that involves only one brane.  The right panel shows a contribution at $\mathcal{O}(\Gamma^2)$ that involves two different branes.}
\label{fig:cylindercomparison}
\end{figure}

The formalism so far is adequate to describe any number of open strings boundaries as they all have the same endpoint mass, and in order for the masses to be different we need a slight generalization of string equation. The difference in the two scenarios (i.e. multiple boundaries on the same brane versus multiple boundaries spread out across several branes) is depicted in fig. \ref{fig:cylindercomparison}. The more general framework developed below will be able to simultaneously handle both of these types of interactions.

At the level of the Hermitian matrix model the appropriate change to make is to introduce a potential for each mass
	\begin{equation}
		V_{\text{open}}(M) = \frac{\gamma}{N} \sum_{i} \tr \log \left(1 - m_i^2M^2 \right).
	\end{equation}
After double scaling the string equation is
	\begin{equation}
		\mathcal{R} + 2\hbar \Gamma \sum_i \hat{R}(x;z_i) = 0, \label{eqn:multibraneopenstringeq}
	\end{equation}
and the closed string dual is obtained by the shift in coupling constants
	\begin{equation}
		t_k \to t_k + 2\hbar\Gamma \zeta_k \sum_i z_i^{-2k-1}.
	\end{equation}
The solution $u$ will now have the expansion
	\begin{equation}
		u(x;z_1,\dots) = \sum_{g = 0}^\infty u_{g,0}(x) + \hbar \Gamma \sum_i u_{0,1}(x;z_i) + \hbar^2\Gamma^2 \sum_{i,j}u_{0,2}(x;z_i,z_j) + \cdots,
	\end{equation}
where generally the contribution at $\mathcal{O}(\Gamma^h)$ will depend on $h$ of the $z$ variables at a time. The functions $u_{g,h}$ are not exactly the same as before, but will be related. We will suppress the $x$-dependence for convenience. Observe that at $\mathcal{O}(\Gamma)$ the string equation is
	\begin{equation}
		\dot{f}\sum_i u_{0,1}(z_i) + 2\sum_i \varphi(u_0,z_i) = 0,
	\end{equation}
which is solved by setting $u_{0,1}(z_i)$ to be the same as $\eqref{eqn:u01}$. Define the total contribution at this order as $u_{0,I} \equiv \sum_i u_{0,1}(z_i)$. At the next order
	\begin{equation}
		\sum_{i,j}\bigg[\dot{f}u_{0,2}(z_i,z_j) + \half \ddot{f}u_{0,1}(z_i)u_{0,1}(z_j) + 2\dot{\varphi}(z_i)u_{0,1}(z_j) \bigg] = 0,
	\end{equation}
which has the solution
	\begin{equation}
		u_{0,2}(z_i,z_j) = - \frac{1}{\dot{f}}\left[ \half \ddot{f}u_{0,1}(z_i)u_{0,1}(z_j) + 2\dot{\varphi}(z_i)u_{0,1}(z_j)\right].
	\end{equation}
This is the natural generalization of the result for one string mass, and although each of these is not symmetric under $i \leftrightarrow j$, the total contribution to $u$ is. A succinct way to express the whole contribution at $\Gamma^2$, which is more reminiscent of the one-mass solution, is 
	\begin{equation}
		u_{0,II} = -  \frac{1}{\dot{f}} \left[\half \ddot{f}u_{0,I}^2 + 2 \dot{\Phi} u_{0,I}\right],
	\end{equation}
where $\Phi = \sum_i \varphi(z_i)$. We still have the identity
	\begin{equation}
		u_{0,II} = \frac{1}{2\dot{f}} \frac{d}{du_0} \Big(\dot{f} u_{0,I}^2 \Big).
	\end{equation}

It is unclear a priori how to do the same analysis for the open string sector of supersymmetric theories starting from the DJM equation. However we will be able to make use of open-closed duality to compute brane quantities in both the $0A$ and $\mathcal{N} = 2$ models.


\subsubsection{0A closed string sector}

The results presented above for the bosonic open-closed string matrix model rely entirely upon perturbation theory in the $x \to -\infty$ regime, where the function $u$ is determined via the initial condition $\mathcal{R} = 0$. That this describes open strings is consistent with the type 0 interpretation of the DJM equation. While at face value it may seem inconsistent in the context of considering open strings to consider the different regimes of $x$ values, it is the interpretation of the solution $u$ in those regimes that is important. Although a purely closed bosonic string matrix model is described non-perturbatively by \eqref{eqn:clopenstringeq}, it's perturbation theory is still captured by the $x \to -\infty$ part of the solution (with $\Gamma = 0$). Whereas in \cite{Johnson:2021owr} it was shown that the closed string perturbation theory of the type 0 models is captured by the $x \to +\infty$ part of the solution. In both systems the open string sector is described in the $x \to -\infty$ region, but the closed string contribution to the function $u$ is determined differently. Put more succinctly, we expect that the open string sector is described by the formulae derived from the $\mathcal{R} = 0$ formalism for the bosonic theories as well as the type 0 theories, but the value of $u$ at the Fermi surface $x = \mu$ is dictated by where the closed string contribution comes form.

Before, in the non-supersymmetric theories the initial condition was $f(u_0) + x = 0$, and now it is simply $u_0 = -z^2$ \cite{Johnson:2021tnl}. Typically solutions are found for $z  = 0$, especially in the closed string sector. The results collected below all have smooth transitions from $z \neq 0 $ to $z = 0$, and so we take the point of view that it is worthwhile to present the more general family of solutions, especially because it is useful for doing open string perturbation theory. When we consider perturbation theory in the closed string sector in the next section we will set $ z= 0 $ when considering supersymmetric theories.

The nontrivial solutions for the first couple values of $h$ with $g = 0$ are
	\begin{equation}
	\begin{aligned}
		 u_{0,2} &= \frac{1}{\Big(f(-z^2) + x\Big)\Big(f(-z^2) + 2z^2 \dot{f}(-z^2)+ x\Big)} \\
		 u_{0,4} &= -\frac{\dot{f}(-z^2) \Big[32 x z^2 \Big(f(-z^2)+x\Big) \dot{f}(-z^2) +16 x \Big(f(-z^2)+x\Big)^2\Big]}{8 x^2 \Big(f(-z^2)+x\Big)^3 \Big(2 z \dot{f}(-z^2)+f(-z^2)+x\Big)^3}
	\end{aligned}
	\end{equation}
The contributions at odd powers of $h$ vanish, which is not a feature unique to $g = 0$. At $g =1$ we have
	\begin{equation}
	\begin{aligned}
		u_{1,0} &= -\frac{1}{4\Big(f(-z^2) + x\Big)\Big(f(-z^2) + 2z^2 \dot{f}(-z^2)+ x\Big)}, \\
		u_{1,2} &= \frac{\dot{f}(-z^2) \Big[z^2 \dot{f}(-z^2)+f(-z^2)+x\Big]}{x^2 \Big(f(-z^2)+x \Big)^3 \Big(2 z^2 \dot{f}(-z^2)+f(-z^2)+x\Big)^3} \\
        & \times \frac{\Big[2 z^2 \Big(f(-z^2)+3 x\Big) \dot{f}(-z^2)+\Big( f(-z^2) + 6x \Big)f(-z^2)+5 x^2\Big]}{x^2 \Big(f(-z^2)+x \Big)^3 \Big(2 z^2 \dot{f}(-z^2)+f(-z^2)+x\Big)^3},
	\end{aligned}
	\end{equation}
and at $g = 2$
	\begin{equation}
	\begin{aligned}
		u_{2,0} &= \frac{\dot{f}(-z^2) \Big[-4 z^4 \Big(f(-z^2)+3 x\Big) \dot{f}(-z^2)^2-2 z^2 \Big(3 f(-z^2)+10 x\Big) \Big(f(-z^2)+x \Big) \dot{f}(-z^2)\Big]}{8 x^2 \Big(f(-z^2)+x \Big)^3 \Big(2 z^2 \dot{f}(-z^2)+f(-z^2)+x\Big)^3}\\
		&\quad-\frac{\dot{f}(-z^2)\Big[\Big(f(-z^2)+x\Big)^2 \Big(2 f(-z^2)+9 x \Big)\Big]}{8 x^2 \Big(f(-z^2)+x\Big)^3 \Big(2 z^2 \dot{f}(-z^2)+f(-z^2)+x\Big)^3}.
	\end{aligned}
	\end{equation}
These perturbative contributions are typically summed to be displayed based on overall factors of $\hbar$ as $u = u_0 + \hbar^2 u_1 + \hbar^4 u_2 + \cdots$, where $u_1 = u_{1,0} + \Gamma^2 u_{0,2}$ and $u_2 = u_{2,0} + \Gamma^2 u_{1,2} + \Gamma^4 u_{0,4}$.

There are three features of note in the perturbative solution $u$. First, the result depends only on even powers of $\Gamma$. This has the interpretation that $\Gamma$ counts R-R flux and is consistent with the dependence of the R-R field in the 0A string theory on the Liouville field  \cite{Klebanov:2003wg, Carlisle:2005wa}. Second, notice that when $z = 0$, the function $f(0) = 0$, so the $\mathcal{O}(\hbar^2)$ result becomes independent of the coupling constants and the $\mathcal{O}(\hbar^4)$ contribution depends only on $t_1$. Clearly the dependence of the highest rank coupling constant on the order of perturbation theory of $u(\mu)$ is different as compared to the non-supersymmetric case. Finally, the presence of the function $f$ in each perturbative contribution to $u$ means that there is no longer a universal term at $\mathcal{O}(\hbar)$ for $z \neq 0$. With the parameter $z$ turned off, each 0A model is characterized at leading order by the universal Bessel model. For a generic value of $z$, the FZZT brane associated to it probes theory-dependent information at each order in perturbation theory past leading order.

\begin{figure}[ht]
	\begin{subfigure}[t]{0.49\textwidth}
	\includegraphics[clip,trim=.6in 2.75in .5in 2.7in, width=\textwidth]{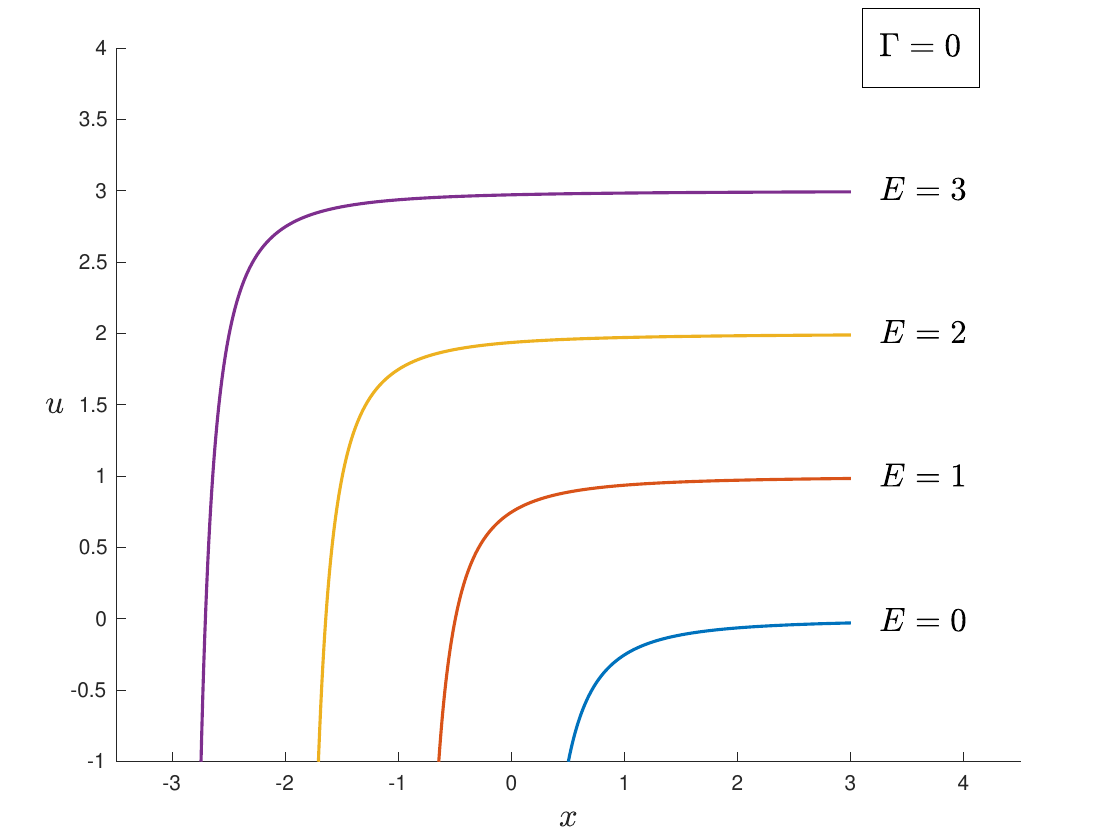}
	\end{subfigure}
	\begin{subfigure}[t]{0.49\textwidth}
	\includegraphics[clip,trim=.6in 2.75in .5in 2.7in, width=\textwidth]{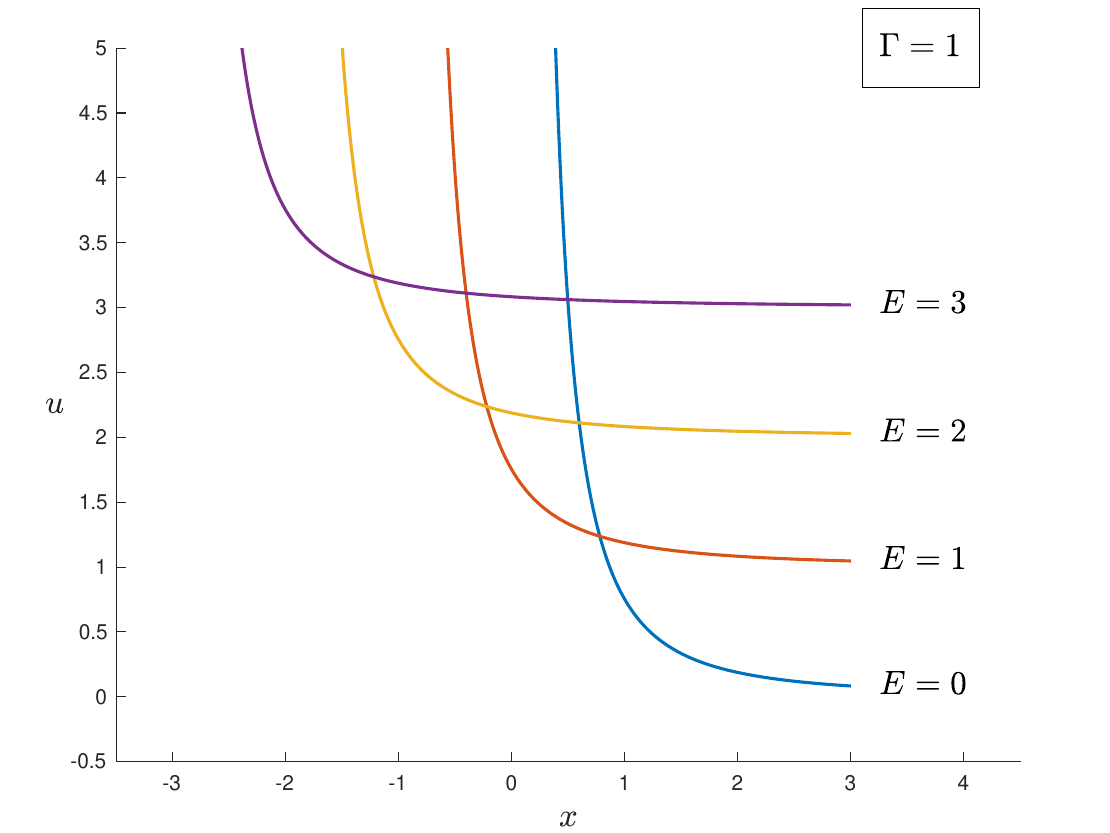}
	\end{subfigure}
\caption{The combined leading and first subleading contributions to $u$ in the $k = 1$ $0A$ multicritical model for different values of $E$, for $\Gamma = 0$ (left) and $\Gamma = 1$ (right), all with $\hbar = 0.1$. We can see as $E$ is increased the function is translated up and to the right. The potential switches concavity at $\Gamma = \half$. Only the positive branch of each solution is displayed.}
\label{fig:k1openstringplots}
\end{figure}

The simplest multicritical 0A model is the $(2,4)$ super-minimal model coupled super Liouville theory and corresponds to having only $t_1$ nonzero. The leading order string equation in the open string sector is defined by $f(u_0) = u_0$, where we have set $t_1 = 1$. The solution up to $\mathcal{O}(\hbar^4)$ is
	\begin{equation}
		u(x;z) = -z^2 + \hbar^2 \frac{ \left(\Gamma^2 - \quarter \right)}{x^2-z^4} - \hbar^4\frac{\left(4 \Gamma^2-1\right) \left(\left(4 \Gamma^2-9\right) x^2+\left(1-4 \Gamma^2\right) z^4\right)}{8 x \left(x^2-z^4\right)^3} + \cdots.
  \label{eqn:0Ak1solution}
	\end{equation}
To visualize this potential, note that $-z^2 = E$ is meant to be positive.

A particular $\mathcal{N} = 1$ JT supergravity theory is defined by the couplings
    \begin{equation}
        t_k = \frac{\pi^{2k}}{(k!)^2},
    \end{equation}
with disk level string equation in the open string sector defined by
    \begin{equation}
       f(u_0) =  I_0(2\pi\sqrt{u_0}) - 1.
    \end{equation}

\subsubsection{2A closed string sector}

There have been several recent advancements in relating JT supergravity with extended supersymmetry to matrix models. With the increase in the amount of supersymmetry comes the added complexity of dealing with R-charge multiplets and BPS states. It was shown in \cite{turiaci2023mathcaln2} that each multiplet is described by a matrix model in the $(1 + 2\Gamma,2)$ AZ class, where $\Gamma$ is the number of BPS states, which is dependent on the R-charge of the supermultiplet and the R-charge of the supercharge. The authors use this to deduce the matrix model describing $\mathcal{N} = 2$ JT gravity. A decomposition of this theory in terms of multicritical models, akin to the ones done for JT gravity and $0A$ JT supergravity, was constructed in \cite{Johnson:2023ofr}. It is apparent from this minimal model decomposition that, although the $\mathcal{N} = 2$ multicritical models are still described by a variation of the DJM equation, the boundary conditions differ from the $0A$ case. Further, the coupling constants $\tilde{t}_k$ and Fermi surface $\tilde{\mu}$ are necessarily functions of $E_0$, the smallest energy of the non-BPS states in the multiplet. For the duration of the discussion of $\mathcal{N} = 2$ models we will use a tilde to implicitly denote when an object has $E_0$ dependence.

An arbitrary $\mathcal{N} = 2$ model is defined by coupling constants $\tilde{t}_k$, which can be written $\tilde{t}_k = t_k \mathcal{F}(E_0)$, where $\mathcal{F}(0) = 1$. The function $\mathcal{F}$ is model dependent and can be fixed by the BPS density of states. The theory will be controlled by the DJM equation, but with some modification, and the closed string perturbation theory will once again be contained in the $x \to + \infty $ regime. The desired solution for $u$ is seeded by the initial condition
	\begin{equation}
		u_0(x) =\frac{\tilde{\mu}^2 E_0}{\Big(\tilde{f}(-z^2) + x\Big)\Big(\tilde{f}(-z^2) + 2z^2 \dot{\tilde{f}}(-z^2)+ x\Big)} -z^2,	
	\end{equation}
which is obtained conceptually by taking the simultaneous limit $\hbar \to 0$ and $\Gamma \to \infty$, such that $\hbar \Gamma = \tilde{\mu} \sqrt{E_0}$ of the $0A$ solution\footnote{Hence why we sometimes refer to these as 2A theories. A similar model could be constructed by taking the same limit of the 0B solutions. The nomenclature `2A' should not be confused with the standard `IIA' naming in critical superstring theory.}, keeping only the leading contributions\footnote{The combination $\hbar\Gamma$ is necessarily small, which is why higher powers of $\hbar\Gamma$ are not included in the limit to obtain the new $u_0$. \label{ftnt:finestructure}} \cite{Johnson:2023ofr}. 

There are barriers to making this a good initial condition for perturbation theory if we use the DJM equation the same way we did in the 0$A$ case. First, the desired $u_0$ simply does not solve the DJM equation with $z = 0$ and $\hbar^2\Gamma^2 = \tilde{\mu}^2 E_0$. Even if we ignore this, the perturbative expansion does not match the 0A expansion (with the appropriate changes made in the limit). To remedy this, consider the slight changes
	\begin{equation}
		\overline{\mathcal{R}} = \hbar^2 \sum_k \tilde{t}_k R_k + x, \quad \hbar^2 \Gamma^2 \equiv \alpha^2,
	\end{equation}
with the DJM equation otherwise unchanged. This can be obtained from the 0A equation by rescaling the coupling constants by $\hbar^2$ in addition to the $E_0$-dependent factor. If we expand $u = u_0 +\cdots$ as before, then the leading order solution to this modified equation is our desired result, and perturbation theory proceeds as before for higher powers of $\hbar$, with $u = \sum_{n = 0}^\infty \hbar^{2n}u_n$. The first correction to $u$ is simply given by
	\begin{equation}
		u_1 = -\frac{1}{4x^2} - \frac{2\alpha^2 \tilde{f}\left(\frac{\alpha^2}{x^2}\right)}{x^3}. 
	\end{equation}
The next perturbative correction is
	\begin{equation}
	\begin{aligned}
		u_2 &=  \frac{4 \alpha ^6 \dddot{\tilde{f}}\left(\frac{\alpha ^2}{x^2}\right) \dot{\tilde{f}}\left(\frac{\alpha ^2}{x^2}\right)+3 x^2 \left[\alpha ^2 x \left(8 \alpha ^2 \tilde{f}\left(\frac{\alpha ^2}{x^2}\right)+9 x\right) \dot{\tilde{f}}\left(\frac{\alpha ^2}{x^2}\right)\right]}{6 x^9}\\
		 &\quad + \frac{4 \alpha ^4 \left(\dot{\tilde{f}}\left(\frac{\alpha ^2}{x^2}\right)+1\right) \ddot{\tilde{f}}\left(\frac{\alpha ^2}{x^2}\right)+x^3 \tilde{f}\left(\frac{\alpha ^2}{x^2}\right) \left(6 \alpha ^2 \tilde{f}\left(\frac{\alpha ^2}{x^2}\right)+x\right)}{6 x^9}.
	\end{aligned}
	\end{equation}
	
The simultaneous limit that keeps $\alpha^2$ constant leaves terms proportional to $\hbar^{2g}\alpha^{2n}$.  Define the deficit $\Delta = 2g-2n$ between the powers of $\alpha$ and the remaining power of $\hbar$ in the 0A solution. Let $c$ be the total power of the coupling constants in a particular term (e.g. $t_1^2$ and $t_1t_2$ both have $c = 2$). Then a term from the 0A solution should appear in the $\mathcal{N} = 2$ solution at $\mathcal{O}(\hbar^{\Delta + 2c})$. To check, consider again the simplest theory, $k = 1$. Leaving the $E_0$-dependent coupling constant $\tilde{t}_1$ arbitrary, we find
	\begin{equation}
	\begin{aligned}
		u &= \frac{\alpha^2}{x^2} - \hbar^2 \left( \frac{2\tilde{t}_1 \alpha ^4 }{x^5} + \frac{1}{4 x^2} \right)
		+  \hbar^4 \left( \frac{7 \tilde{t}_1^2 \alpha ^6 }{x^8} + \frac{5 \tilde{t}_1\alpha ^2 }{x^5}  \right) \\
  &- \hbar^6 \left( \frac{30 \tilde{t}_1^3 \alpha ^8}{x^{11}} + \frac{217  \tilde{t}_1^2\alpha ^4 }{4 x^8} + \frac{9 \tilde{t}_1 }{8 x^5} \right) + \cdots.
	\end{aligned}
	\end{equation}
Comparing this to \eqref{eqn:0Ak1solution} confirms the assertion.

By rescaling the 0A coupling constants by $\hbar^2$, it is possible that we have interfered with the topological interpretation of $u$. Previously, the power of $\hbar$ indicated the Euler characteristic of the associated surface. Even after wrapping the powers of $\hbar$ coming from RR flux insertions into powers of $\alpha$, the remaining factors of $\hbar$ in the 0A solution still correspond to genus. However, the powers of $\hbar$ in the $\mathcal{N} = 2$ solution are also affected by the rescaled coupling constants. In order to maintain the same topological interpretation we would have to undo the rescaling, which corresponds to summing over terms with the same value of $\Delta$ and dropping the portion related to $c$. On the other hand, the fact that the desired leading contribution is a solution to the modified DJM equation should inspire confidence, especially since so little had to be done to the differential equation to obtain it.  The fate of perturbation theory in $\mathcal{N} = 2$ will be reexamined in an upcoming work.


\section{Macroscopic Loop Perturbation Theory} \label{scn:macroscopicloopperturbationtheory}

A convenient way to organize perturbation theory in these matrix models is through an expansion in $\hbar$. This affords us the ability to interpret perturbation theory in the context of an expansion in topologies, which is familiar from string theory. The parameter $\hbar$ used here is also referred to as the closed string coupling in older literature.

We saw in the preceding section that the $\hbar$-expansion of $u$ arises as a way to solve the non-linear equation of motion efficiently. Using \eqref{eqn:utof} we acquire the genus expansion of the closed string free energy
	\begin{equation}
		F = \sum_{g =0}^\infty \hbar^{2g-2} F_g.
	\end{equation}
Notice that the power of $\hbar$ is $\chi(g)$, the Euler characteristic of a surface with $g$ handles. Since $F = -\log Z$ and $Z$ enumerates the random closed surfaces, we are justified in regarding $g$ as counting genus.

All observables in the theory will have an expansion in powers of $\hbar$, with the powers being linked to the Euler characteristic of a surface. In the purely closed string context we will usually limit ourselves to considering surfaces with one type of boundary, where the Euler characteristic is $\chi(g,n) = 2g-2+n$. One such class of boundary-having observables is the correlation functions of macroscopic loop operators $e^{-\beta \mathcal{H}}$, which are dual to path integrals on surfaces with asymptotic boundaries. For this reason we call
	\begin{equation}
		\mathcal{Z}_{n}(\beta_1,\dots,\beta_n) \equiv \expect{e^{-\beta_1\mathcal{H}} \cdots e^{-\beta_n \mathcal{H}}},
	\end{equation}
a path intergral. The topological expansion of this path integral is denoted
	\begin{equation}
		\mathcal{Z}_n(\beta_1,\dots,\beta_n) = \sum_{g = 0}^\infty \hbar^{2g-2+n} \mathcal{Z}_{g,n}(\beta_1,\dots,\beta_n), 
	\end{equation}

The approach to perturbation theory taken here will involve an expansion of $e^{-\beta \mathcal{H}}$ in terms of the point-like operators $\sigma_k$
	\begin{equation}
        e^{-\beta \mathcal{H}} = \frac{\hbar}{2\sqrt{\pi\beta}}\sum_{k = 0}^\infty \frac{(-1)^k\beta^{k}}{k!}\sigma_{k-1},
    \end{equation}
where $\sigma_{-1} \equiv 0$. We pause here to consider a scaling argument that justifies the apparent shift in labelling of the $\sigma_k$. Recall that previously we defined the mass dimension of $u$ to be $[u] = 1$, with $\hbar$ is dimensionless, and the variable $x$ had dimension $[x] = -\half$. Then the mass dimension of $\beta$ is $[\beta] = -1$ to make $\beta \mathcal{H}$ dimensionless. 
Meanwhile, the $k^\text{th}$ Gelfand-Dikii, polynomial has dimension $[R_k] = k$, so that the coupling constants satisfy $[t_k] + [R_k] = [x],\quad \text{or} \quad [t_k] = -k-\half$, in order to make the string equation dimensionless. For the combination $t_k\sigma_k$ to be dimensionless, we must have $[\sigma_k] = k + \half$. This implies that $\left(k - \half\right)[\beta] + [\sigma_{k-1}] = 0$,
as necessary to make $e^{-\beta \mathcal{H}}$ dimensionless. This justifies the powers of $\beta$ relative to the labelling of the $\sigma_k$ in the expansion of the macroscopic loop operator.

The point-like operators $\sigma_k$ are also referred to as closed string operators, coming from the string theory analysis of double scaled matrix models. The intimate relationship that these operators have with both the KdV organization of the theory and the underlying closed string physics means that all formulae derived in this section will apply equally well to supersymmetric and non-supersymmetric theories. The prescription for performing computations in specific models will be to evaluate the functions $u_g^{(n)}(\mu)$ in the model's closed string sector. 

\subsection{General formulae}

Correlation functions of the macroscopic loop operators are computed, by virtue of the relationship \eqref{eqn:sigmatoxi} between the point-like operators $\sigma_k$ and the vector fields $\xi_k$, by repeated action of the KdV vector fields on the function $u$ \cite{BANKS1990279}
	\begin{equation}
        \begin{aligned}
        \langle e^{-\beta_1 \mathcal{H}} \cdots e^{-\beta_n \mathcal{H}} \rangle &= \frac{2^{n-2}\hbar^{n-2}}{\pi^{n/2}\sqrt{\beta_1 \cdots \beta_n}} \sum_{k_i = 0}^\infty \frac{(-1)^{k_1+\cdots + k_n}\beta_1^{k_1}\cdots \beta_n^{k_n}}{(k_1)!\cdots(k_n)!}\\
        &\times \int^\mu dx \int^x dx' \xi_{k_1}\cdots \xi_{k_n} \cdot u(x').
    \end{aligned}
    \end{equation}
Therefore each perturbative contribution to $\mathcal{Z}_n$ will be
	\begin{equation}
        \begin{aligned}
		\mathcal{Z}_{g,n}(\beta_1,\dots,\beta_n) &= \frac{2^{n-2}}{\pi^{n/2}\sqrt{\beta_1 \cdots \beta_n}} \sum_{k_i = 0}^\infty \frac{(-1)^{k_1+\cdots ,+ k_n}\beta_1^{k_1}\cdots \beta_n^{k_n}}{(k_1)!\cdots(k_n)!} \\
  &\times \int^\mu dx \int^x dx' \Big[\xi_{k_1}\cdots \xi_{k_n} \cdot u(x')\Big]_g.
        \end{aligned}
	\end{equation}
The notation $[\cdot]_g$ denotes the full contribution at order $\hbar^{2g}$, after taking into account the $\hbar$-expansion of the Gelfand-Dikii polynomials and the topological expansion of $u$.

In this language the expectation value of a single macroscopic loop operator is
	\begin{equation}
		\langle e^{-\beta \mathcal{H}} \rangle = \frac{\hbar}{2\sqrt{\pi\beta}}\sum_{k = 0}^\infty \frac{(-1)^k\beta^{k}}{k!} \int_{-\infty}^\mu dx\,R_k.
	\end{equation}
To expand this perturbatively, there are two sources of $\hbar$ that must be taken into account. The Gelfand-Dikii polynomials $R_k$ have an arrangement in powers of $\hbar$ (see \eqref{eqn:GDhbarexpansion} above), and so does the function $u$. The full leading order contribution to $R_k$ is $u_0^k$, and hence
	\begin{equation*}
		\mathcal{Z}_{0,1}(\beta) = \frac{\hbar}{2\sqrt{\pi\beta}}\sum_{k = 0}^\infty \frac{(-1)^k\beta^{k}}{k!} \int_{-\infty}^\mu dx\,u_0(x)^k.
	\end{equation*}
This integral can be computed in two steps. First, notice that the bounds of the integral straddle $x = 0$, where there is a change in how $u_0$ is determined. After splitting the integral, the portion from $-\infty$ to 0 can be computed in terms of the coupling constants by changing the integration variable to $u_0$, using the disk level string equation to calculate the Jacobian. The result in this region is
	\begin{equation}
        \mathcal{Z}^{(-)}_{0,1}(\beta) = \frac{\hbar}{2\sqrt{\pi}\beta^{3/2}} \sum_{k = 1}^\infty k!t_k \beta^{-k+1}. \label{eqn:diskpathintegral}
    \end{equation}
This contribution is the same in every model we study here. However, the portion from 0 to $\mu$ depends on which model we are studying. In the non-supersymmetric case where $\mu = 0$ this portion actually doesn't exist. In the 0A models where $u_0 = 0$, the only non-vanishing term in the sum is at $k = 0$, giving $\mathcal{Z}^{(+)}_{0,1} = \hbar \mu/2\sqrt{\pi\beta}$. Actually, since the KdV time $t_0$ is identified by our variable $x$, it is natural to also make the connection $t_0 \leftrightarrow \mu$. Doing so, the full expression fo $\mathcal{Z}_{0,1}$ in 0A theories can be obtained by changing the lower bound of summation to $k = 0$ in the expression for $\mathcal{Z}_{0,1}^{(-)}$ \eqref{eqn:diskpathintegral}. In the $\mathcal{N} = 2$ models, where $u_0$ is nonzero in this region, it is convenient to once again pull the infinite sum inside the integral to get
	\begin{equation}
		\mathcal{Z}^{(+)}_{0,1}(\beta) = \frac{\hbar}{2} \left(\frac{\tilde{\mu}  e^{- E_0 \beta}}{\sqrt{\pi\beta }}-\tilde{\mu}^2E_0 \text{Erfc}\left(\sqrt{E_0\beta}\right)\right).
	\end{equation}
In totality, we have 
	\begin{equation}
		\mathcal{Z}_{0,1}(\beta) = 
			\begin{cases}
				\mathcal{Z}^{(-)}(\beta), & \text{Non-supersymmetric} \\
				\mathcal{Z}^{(-)}(\beta) + \frac{\hbar \mu}{2\sqrt{\pi\beta}} & \text{0A} \\
				\mathcal{Z}^{(-)}(\beta) + \frac{\hbar}{2} \left(\frac{\tilde{\mu}  e^{- E_0 \beta}}{\sqrt{\pi\beta }}-\tilde{\mu}^2E_0 \text{Erfc}\left(\sqrt{E_0 \beta }\right)\right) & \mathcal{N} = 2
			\end{cases}
	\end{equation}
	\begin{figure}[ht]
	\begin{center}
	\includegraphics[width = 0.5\textwidth]{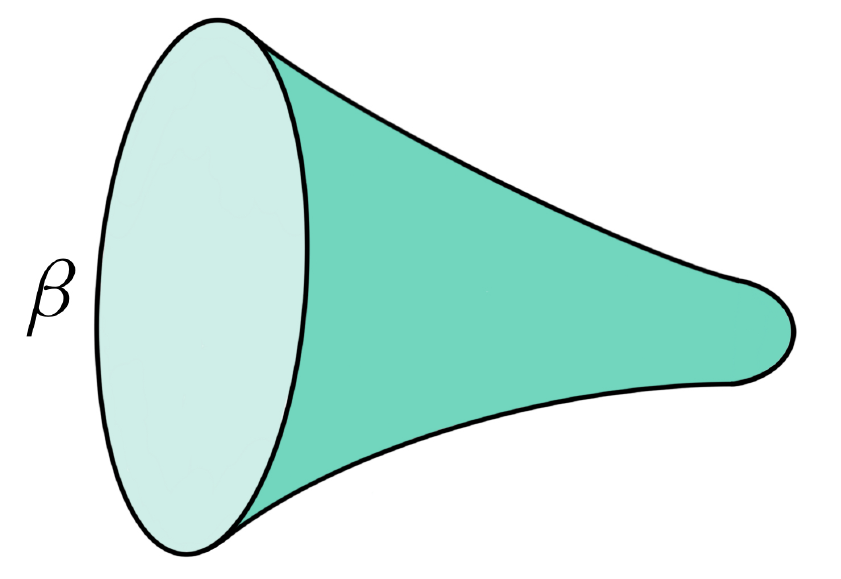}
	\caption{The disk geometry as it is usually presented in nAdS$_2$ with an asymptotic boundary of 	renormalized length $b$. Even though a matrix model with arbitrary $t_k$ may not have an explicit geometric interpretation, it is useful to keep this picture in mind in the topological expansion.}
	\label{fig:disk}
	\end{center}
	\end{figure}
It is often said that the supersymmetric models incorporate a second topological point at $k = 0$, which we can see borne out by the additional term here in the macroscopic loop expectation value. 
    
The first perturbative correction $\mathcal{Z}_{1,1}(\beta)$ can also be computed exactly in terms of the coupling constants. The full contribution to $R_k$ at order $\hbar^2$ is 
	\begin{equation}
		R_k = \cdots + \hbar^2 \left(ku_0^{k-1}u_1 - \frac{k(k-1)}{12}u_0^{k-3}\Big[(k-2)(u_0')^2 + 2 u_0 u_0''\Big]\right) + \cdots.
	\end{equation}
The correction is then
	\begin{equation*}
        \mathcal{Z}_{1,1}(\beta) = - \frac{1}{2\sqrt{\pi \beta}} \int_{-\infty}^\mu dx\,e^{-\beta u_0(x)} \left[ \beta u_1(x) + \frac{\beta^2}{12}\Big( 2u_0''(x) - \beta (u_0'(x))^2 \Big)  \right]. \label{eqn:g1n1}
    \end{equation*}
Since this expression is meant to be perturbative, the integral should only be done in the region where closed string perturbation theory is defined for the model. In the non-supersymmetric models, by invoking the relationships \eqref{eqn:u0'tou0''} and \eqref{eqn:u0'tou1} and integrating by parts, one is left with purely surface terms
	\begin{equation*}
		\mathcal{Z}_{1,1}(\beta) = -\frac{\sqrt{\beta}}{24\sqrt{\pi}} \left( u_0'(\mu) \beta + \frac{u_0''(0)}{u_0'(0)}\right),
	\end{equation*}
or written in terms of the coupling constants
	\begin{equation}
        \mathcal{Z}_{1,1}(\beta) = \frac{\sqrt{\beta}}{24t_1\sqrt{\pi}}\left(\beta + \frac{2t_2}{t_1}\right). \label{eqn:Z11}
    \end{equation}
The result is also obtainable by exploiting an interesting feature of Laplace transforms, see appendix \ref{apdx:laplacetransformequivalence} for more. In both types of supersymmteric model the integration is restricted to be from 0 to $\mu$.

The leading order contribution to the two-point function, $\mathcal{Z}_{0,2}$, relies on the fact that
	\begin{equation}
		 \int^x dx' \Big[\xi_{k_1} \cdot \xi_{k_2} \cdot u(x')\Big]_0 = k_1k_2 u_0'(x)u_0(x)^{k_1 + k_2 - 2}. \label{eqn:sigmatwopoint}
	\end{equation}
After performing the sum and computing the remaining integral the result is
	\begin{equation}
		\mathcal{Z}_{0,2}(\beta_1,\beta_2) = \frac{\sqrt{\beta_1\beta_2}}{\pi(\beta_1 + \beta_2)} e^{-(\beta_1 + \beta_2)u_0(\mu)}.
	\end{equation}
As mentioned previously the bosonic theories usually have $u_0(\mu) = 0$, and the so the exponential factor is trivial. The matrix model calculation confirms the well-known fact from topological recursion that the path integral `double trumpet' geometry is universal (independent of the coupling constants). The $g = 1$ correction to the two-point function is given by
	\begin{equation}
		\mathcal{Z}_{1,2}(\beta_1,\beta_2) = \frac{\sqrt{\beta_1\beta_2}}{\pi} \left( u_1 - \left( \frac{\beta_1^2 + \beta_1\beta_2 + \beta_2^2}{12}\right) (u_0')^2 + \left(\frac{\beta_T}{6}\right)u_0''\right) e^{-\beta_Tu_0}\Bigg|_{x = \mu},
	\end{equation}
where $\beta_T = \beta_1 + \beta_2$. The details of this calculation are in appendix \ref{apdx:macroscopicloopcalculations}.

The $g = 0$ contribution to the three-point function uses
	\begin{equation}
		\Big[\xi_{k_1}\cdot \xi_{k_2} \cdot \xi_{k_3} \cdot u \Big]_0= k_1k_2k_3 \frac{d^2}{dx^2} u_0'u_0^{k_1 + k_2 + k_3 - 3}. \label{eqn:3ptidentity}
	\end{equation}
Performing the three sums yields
	\begin{equation}
		\mathcal{Z}_{0,3}(\beta_1,\beta_2,\beta_3) = -\frac{2\sqrt{\beta_1\beta_2\beta_3}}{\pi^{3/2}}u_0'(\mu) e^{-\beta_T u_0(\mu)},
	\end{equation}
where here $\beta_T = \beta_1 + \beta_2 + \beta_3$. The $g = 1$ correction to the three-point function is
	\begin{equation}
    \begin{aligned}
        \mathcal{Z}_{1,3}(\beta_1,\beta_2,\beta_3) &= \frac{2\sqrt{\beta_1\beta_2\beta_3}}{\pi^{3/2}}e^{-\beta_Tu_0}\Bigg[ -u_1' + \beta_T u_1u_0' - \Bigg( \frac{\beta_1^3 + \beta_2^3 + \beta_3^3}{12}\\
        &\quad + \frac{\beta_1\beta_2\beta_3 + \beta_1\beta_2(\beta_1 + \beta_2) + \beta_1\beta_3(\beta_1 + \beta_3) + \beta_2\beta_3(\beta_2 + \beta_3)}{6}\Bigg)(u_0')^3\\
        &\quad + \Bigg( \frac{2(\beta_1^2 + \beta_2^2 + \beta_3^2) + 3(\beta_1\beta_2 + \beta_1\beta_3 + \beta_2\beta_3)}{6}\Bigg)u_0'u_0'' - \frac{\beta_T}{6}u_0''' \Bigg] \Bigg|_{x = \mu}.
    \end{aligned}
    \end{equation}
The details of this calculation can also be found in appendix \ref{apdx:macroscopicloopcalculations}.

A general formula exists for the $g = 0$ part of the higher order correlation functions starting with $n = 3$ \cite{ginsparg1993lectures,AMBJORN1990517, MOORE1991665}
	\begin{equation}
        \mathcal{Z}_{0,n}(\beta_1,\cdots,\beta_n) = -\frac{\sqrt{\beta_1\cdots\beta_n}}{2\pi^{\frac{n}{2}}}\left[ \left( \frac{\partial}{\partial x} \right)^{n-3} \left( u_0'(x) e^{-\beta_T u_0(x)} \right)  \right]_{x \to \mu}, \label{eqn:g0nptfunction}
    \end{equation}
which is easily confirmed using the matrix model technology here. For $n \geq 3$ one finds that
	\begin{equation}
		\Big[\xi_{k_1} \cdots \xi_{k_n}\cdot u\Big]_0 = k_1 \cdots k_n \frac{d^{n-1}}{dx^{n-1}} u_0' u_0^{k_1 + \cdots k_n - n},
	\end{equation}
which can be shown by repeated application of $u^k(\partial/\partial u)$ and taking the $\hbar^0$-part. Integrating this twice and performing the sum reproduces \eqref{eqn:g0nptfunction}.


\subsection{Examples}

\subsubsection{Non-supersymmetric theories}

The individual multicritical models display an unusual aversion to macroscopic loop perturbation theory for $k \geq 2$. In fact the only finite results are for $\mathcal{Z}_{0,1}$ and $\mathcal{Z}_{0,2}$. For the $p^\text{th}$ model, the former is evaluated by setting $t_k = \delta_{kp}$ in \eqref{eqn:diskpathintegral}, while the latter is universal. Most other quantities diverge due to the combination of having $u_0 = (-x)^{1/p}$ and the Fermi surface at $\mu = 0$.

The failure of the multicritical models to yield finite results can be traced to the fact that the string equation is defined by the monomial $f(u_0) = u_0^p$, which for $p \geq 2$ satisfies $\dot{f}(0) = 0$. The only model in this family lacking this behavior is the Gaussian model $p = 1$, and in order for the matrix model to provide nontrivial finite results we have to consider an interpolation including the topological Gaussian point. An interesting set of models defined by such an interpolation are the minimal strings. The $(2,2p-1)$ minimal string is a particular massive interpolation between the models $k = 1$ through $k = p$, with coupling constants $t_k$ that depend on $p$ \cite{Mertens_2021}
    \begin{equation}
        t_k = \half \frac{\pi^{2k-2}}{k!(k-1)!}\frac{4^{k-1}(p + k - 2)!}{(p-k)!(2p-1)^{2k-2}}.
    \end{equation}
A nontrivial check of the formalism above is that the $g = 1$ correction to $\mathcal{Z}_1$ is computed to be
	\begin{equation}
        \mathcal{Z}_{1,1}(\beta) = \frac{\sqrt{\beta}}{24\sqrt{\pi}} \left[ \beta + \left(1 - \frac{1}{(2p-1)^2}\right)\pi^2\right],
    \end{equation}
which matches the result obtained in \cite{Mertens_2021}. The first perturbative correction to the macroscopic loop two-point function is readily computed to be
	\begin{equation}
        \begin{aligned}
		\mathcal{Z}_{1,2}(\beta_1,\beta_2) &= \frac{\sqrt{\beta_1\beta_2}}{\pi} \Bigg[ \frac{2 \pi ^4 p (p-1)  \Big(3 p(p-1) +2\Big)}{3 (2 p - 1)^4} + \left( \frac{\beta_1^2 + \beta_1\beta_2 + \beta_2^2}{12}\right) \\
  &+ \frac{4 \pi ^2 p(p-1) }{(2 p - 1)^2}\left(\frac{\beta_T}{6}\right)\Bigg].
        \end{aligned}
	\end{equation}

JT gravity can be thought of as the $p \to \infty$ limit of the minimal string, or equivalently as the infinite interpolation between multicritical models defined by the coupling constants
	\begin{equation}
		t_k = \half \frac{\pi^{2k-2}}{k!(k-1)!}.
	\end{equation}
The disk path integral $\mathcal{Z}_{0,1}$ is confirmed to be \cite{Saad:2019lba}
	\begin{equation}
		\mathcal{Z}_{0,1}(\beta) = \frac{1}{2\sqrt{\pi}\beta^{3/2}}e^{\frac{\pi^2}{\beta}}.
	\end{equation}
The first perturbative correction to this is also confirmed to be
	\begin{equation}
		\mathcal{Z}_{1,1}(\beta) = \frac{\sqrt{\beta}}{24\sqrt{\pi}} \Big[\beta + \pi^2 \Big].
	\end{equation}
In \cite{Saad:2019lba}, the authors do not directly compute $\mathcal{Z}_{1,2}$ in the matrix model, but instead derive the general procedure to compute it and other perturbative corrections using Weil-Petersson volumes. The macroscopic loop formalism correctly gives
	\begin{equation}
		\mathcal{Z}_{1,2}(\beta_1,\beta_2) = \frac{\hbar \sqrt{\beta_1\beta_2}}{24\pi} \Big[ 3\pi^4 + 2\left(\beta_1^2 + \beta_1\beta_2 + \beta_2^2 \right) + 4\pi^2 \beta_T \Big].
	\end{equation}

Another interesting model defined by an infinite interpolation is the Virasoro Minimal String   \cite{Collier:2023cyw, Johnson:2024bue}. The couplings constants of this model are
	\begin{equation}
		t_k = 2\sqrt{2}\frac{\pi^{2k+1}}{(k!)^2}\left(Q^{2k} - \hat{Q}^{2k}\right),
	\end{equation}
where
	\begin{equation}
		Q = b + b^{-1},\quad \hat{Q} = b^{-1} - b.
	\end{equation}
The parameter $b$ and the two combinations $Q,\hat{Q}$ are familiar from Liouville theory, which similarly to the non-supersymmetric multicritical matrix models has a direct connection to the theory. The disk path integral is given by
	\begin{equation}
		\mathcal{Z}_{0,1}(\beta) = \sqrt{\frac{2\pi}{\beta}} \left[e^{\frac{c-1}{6\beta}} - e^{-\frac{\hat{c}-1}{6\beta}}\right],
	\end{equation}
where $c = 1 + 6Q^2$ and $\hat{c} = 1 - 6\hat{Q}^2$ are central charges. This is the result one would get from Laplace transforming the universal Cardy density of states for two-dimensional CFTs\footnote{Indeed this is in a way by construction, since this density of states is used to determine the coupling constants \cite{Johnson:2024bue}.}. The first correction to the macroscopic loop one-point function, typically called the torus path integral in the CFT context, is given by
	\begin{equation}
		\mathcal{Z}_{1,1}(\beta) = \frac{\sqrt{\tilde{\beta}}}{96\sqrt{\pi}}\left[\tilde{\beta} + \frac{c-13}{24}\right],
	\end{equation}
where we have introduced a rescaled length $\tilde{\beta} = \beta/2\pi^2$. This includes the familiar $(c-13)/24$ from torus calculations.

\subsubsection{Supersymmetric theories}

One 0A theory of note is $\mathcal{N} = 1$ JT supergravity, defined by the coupling constants \cite{Johnson:2020heh}
	\begin{equation}
		t_k = \frac{\pi^{2k}}{(k!)^2}.
	\end{equation}
Using the appropriately modified version of \eqref{eqn:diskpathintegral} we get the disk path integral
	\begin{equation}
		\mathcal{Z}_{0,1} = \frac{1}{2\sqrt{\pi\beta}} e^{\frac{\pi^2}{\beta}}.
	\end{equation}
A more recent interesting 0A theory is the $\mathcal{N} = 1$ generalization of the Virasoro Minimal String considered above. It's coupling constants are \cite{Johnson:2024fkm}
	\begin{equation}
		t_k = 2\sqrt{2}\frac{\pi^{2k+1}}{(k!)^2} \left(Q^{2k} + \hat{Q}^{2k}\right),
	\end{equation}
giving the disk path integral
	\begin{equation}
		\mathcal{Z}_{0,1}(\beta) = \sqrt{\frac{2\pi}{\beta}} \left[e^{\frac{c-1}{6\beta}} + e^{-\frac{\hat{c}-1}{6\beta}}\right].
	\end{equation}
Note that this differs from the ordinary Virasoro Minimal String only in the relative sign between the exponentials.

An interesting feature of the closed string sector of the DJM equation is that, while we typically set $z = 0$, is is entirely permissible to have $\Gamma \neq 0$ and to interpret it in terms of RR flux. As we saw at the level of the function $u$, there is a dramatic difference between $\Gamma = 0$ and $\Gamma \geq 1$. This difference continues to manifest itself in the perturbative corrections to macroscopic loop correlators. For example, in $0A$ theories with the initial condition $u_0 = 0$, the term $\mathcal{Z}_{1,1}$ is given by
	\begin{equation}
		\mathcal{Z}_{1,1}(\beta)  = -\frac{\sqrt{\beta}}{2\sqrt{\pi}} \int_0^\mu dx\frac{\Gamma^2 - \quarter}{x^2} = \frac{1}{2\mu\sqrt{\pi}}\left(\Gamma^2 - \quarter\right)\sqrt{\beta},
	\end{equation}
where we've dropped a divergence at $x = 0$ and left the Fermi surface unfixed, but nonzero. The bounds of integration are different, starting at $x = 0$ and ending at $x = \mu$, because the closed string sector of the theory is located in the $x > 0$ region and the matrix model naturally depends only on sub-Fermi level information \cite{Johnson:2021owr}. There are two features of $\mathcal{Z}_{1,1}$ worth mentioning. First, since the total contribution to $u$ at $\mathcal{O}(\hbar^2)$ is universal in $0A$ models with $z = 0$, the $g = 1$ correction to the one-point function is also universal. Second, $\mathcal{Z}_{1,1}$ changes sign as $\Gamma$ is increased from $0$ to $1$, being identically $0$ for $\Gamma = \half$. The special value $\Gamma = \half$ is the appropriate choice for $0A$ JT supergravity \cite{Stanford:2019vob, Johnson:2020heh}. For $n > 2$, the leading order contributions $\mathcal{Z}_{0,n} = 0$ irrespective of the value of $\Gamma$ because $u_0 = \text{const}$., which is true even if $z$ is turned on. For $z = 0$, the $g = 1$ corrections $\mathcal{Z}_{1,n}$ all have the form $\sqrt{\beta_1\cdots\beta_n} \left(\Gamma^2 - \quarter\right)$, which can easily be predicted using the fact that $u_1$ is the highest order correction to $u$ that can appear.

To conclude the supersymmetric examples, we consider the $\mathcal{N} = 2$ theory. As discussed in the previous section, it is not yet clear how to do perturbation theory past leading order. So, we will only examine $g = 0$ quantities. Due to the unique nature of $u_0$, the $g = 0$ contributions to both the macroscopic loop two- and three-point functions have nontrivial exponential factors
	\begin{equation}
		\mathcal{Z}_{0,2}(\beta_1,\beta_2) = \frac{\sqrt{\beta_1\beta_2}}{\pi\beta_T}e^{-\beta_T E_0},\quad \mathcal{Z}_{0,3}(\beta_1,\beta_2,\beta_3) = \frac{4E_0\sqrt{\beta_1\beta_2\beta_3}}{\tilde{\mu}\pi^{3/2}}e^{-\beta_TE_0}.
	\end{equation}
In $\mathcal{N} = 2$ JT supergravity the Fermi surface is given by $\tilde{\mu}(E_0) = \frac{\sin(2\pi\sqrt{E_0})}{4\pi^2\sqrt{E_0}}$, and at leading order in small $E_0$ is $\tilde{\mu} \approx 1/2\pi$. Inserting this into the expression for $\mathcal{Z}_{0,3}$ gives precise agreement with \cite{turiaci2023mathcaln2}, up to an overall factor of 4. Most importantly, the factor of $2\pi$ that comes from the $U(1)$ R-symmetry and the exponential dependence on $E_0$ and $\beta$ are accounted for. Evidently the matrix model agrees with the path integral calculation only in the small $E_0$ limit. Since $\tilde{\mu}$ and $E_0$ are related to the density of BPS states, $E_0$ is constrained to have a relatively small value. It is possible that the matrix model definition captures physics away from the strictly small $E_0$ limit.

\subsection{KdV flows}

An important fact about the nonsupersymmetric examples considered here has to do with which coupling constants the perturbative corrections to $\mathcal{Z}_n$ depend on. What we have seen is that the quantity $\mathcal{Z}_{g,n}$ generically depends on the coupling constants $t_1$ through $t_{3g-2+n}$. Notice that the maximum possible index is very nearly the virtual dimension of the moduli space $\mathcal{M}_{g,n}$ of genus-$g$ Riemann surfaces with $n$ punctures. In practice, the maximum value of the index is determined by the minimum of $3g-2+n$ and the largest index of the nonzero coupling constants. The models that always have $3g-2+n$ as the maximum index in $\mathcal{Z}_{g,n}$ must have all coupling constants turned on.

This feature can be proven generally. Due to the origin of the contributions $\mathcal{Z}_{g,n}$ in the topological expansion, each one contributes at order $\hbar^{\chi(g,n)}$, where $\chi(g,n) = 2g - 2 + n$ is the Euler characteristic. On the other hand, as we have seen these partition functions depend on the $\hbar$-expansion of $u$ and its derivatives evaluated at $x = \mu$. Since the function $u_{2g'}^{(n')}$ contributes at $\hbar^{2g'+n'}$, the partition function $\mathcal{Z}_{g,n}$ can depend on combinations of the functions $u_{2g'}^{(n')}$ as long as $2g'+n' \leq \chi(g,n)$. Let $\overline{k}_{g,n}$ denote the largest index of the coupling constants appearing in $\mathcal{Z}_{g,n}$, and let $\overline{k}'_{g',n'}$ denote the same quantity for $u_{2g'}^{(n')}(\mu)$. The index $\overline{k}_{g,n}$ will be the maximum value of  $\overline{k}'_{g',n'}$ subject to the constraint $2g' + n' \leq 2g - 2 + n$. The maximum occurs when $g' = g$ and $n' = n-2$ and is
    \begin{equation}
        \overline{k}_{g,n} = \underset{g',n'}{\text{max}}\, \overline{k}'_{g',n'} = 3g -2 + n,
    \end{equation}
as desired. A possible interpretation for why $ \overline{k}_{g,n}$ is one more than the dimension of the corresponding moduli space is the fact that $t_1$ is a special coupling constant. It is the variable dual to the topological $k = 1$ model and therefore can be discarded as a ``degree of freedom'' when compared to the models with higher values of $k$. Evidently, since perturbation theory forces most quantities to depend on $t_1^{-1}$, the topological model must be included to provide a sort of stability, because otherwise setting it to 0 introduces divergences. Excluding $t_1$ from the count decreases $ \overline{k}_{g,n}$ by one, matching the moduli space dimension.

Macroscopic loop perturbation theory is able to access information about an underlying moduli space because of the deep and rich connection between the KdV hierarchy and topological gravity. The point-like KdV operators $\sigma_k$ can be thought of as being dual, in a loose sense, to the tautological classes $\psi_k$ on $\mathcal{M}_{g,n}$, whose intersection numbers are generated by the matrix model free energy $F$. There is a strict limit on how many of the $\psi_k$ can be considered in an intersection number determined precisely by the dimension $d_{g,n} = 3g-3+n$.


\section{Geodesic Loop Perturbation Theory} \label{scn:geodesicloopperturbationtheory}

In the preceding section we studied the perturbative expansion of certain correlation functions by computing them entirely in the closed string sector, since the function $u$ was evaluated in the closed string regime of its string equation. In this section we incorporate the open string solutions obtained in section 3. This will allow us to accomplish two things: we will study surfaces with both asymptotic and geodesic boundaries, and we will establish a duality between operators in the closed string sector and operations done on the free energy in the open string sector.

Since the function $u$ can be determined in two different regimes, corresponding to open and closed strings, we can define two different free energies, each determined using \eqref{eqn:utof}. In the context of the DJM equation, open-closed duality has historically consisted either of the statement that there is a transformation which maps open string perturbation theory to closed string perturbation theory in nonsupersymmetric models \cite{Johnson:1993vk}, or that the $x \to \pm \infty$ regimes of perturbation theory of the DJM equation describe the different sectors \cite{Johnson:2003hy}. We will exhibit a different kind of duality here that applies for nonsupersymmetric and supersymmetric theories alike, which relates the correlation functions of a certain operator in the closed string sector to derivatives of the open string free energy. 

\begin{figure}[htbp]
\begin{center}
\includegraphics[width = 0.5\textwidth]{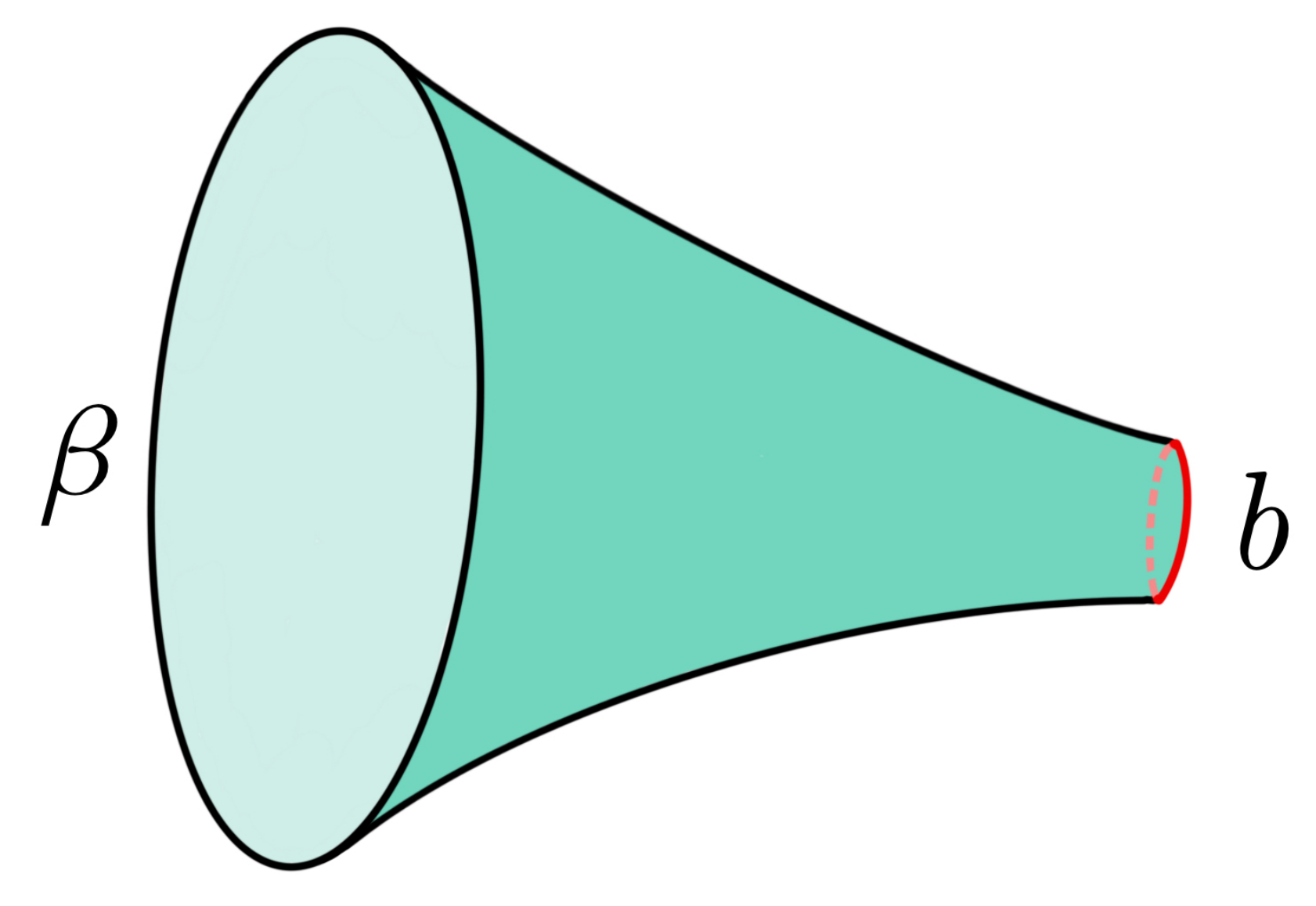}
\caption{The trumpet geometry with one asymptotic boundary of renormalized length $\beta$, and one geodesic boundary (shown in red) of length $b$.}
\label{fig:trumpet}
\end{center}
\end{figure}

The general topic of boundaries in two-dimensional quantum gravity, string theory, and matrix models has been studied for a long time, often with different terminology used to discuss the same things. For example, there is a strong connection between FZZT branes and macroscopic loops, related to one another using identities for the determinant and taylor series. For this reason the two terms have historically been linked in a way that could be confusing, especially in the context of the results presented here. As will become evident in this section, there are two types of boundaries that one can naturally consider: ones created by the insertion of a macroscopic loop operator into a closed string model, and ones created by the insertion of an FZZT brane. The holographic dictionary established for dilaton gravity in \cite{Saad:2019lba} relates macroscopic loop correlators in the matrix model to gravity path integrals on hyperbolic surfaces with asymptotic boundaries. For this reason, it is convenient to picture the macroscopic loops of the previous section in this fashion, as depicted in fig. \ref{fig:disk}. The new type of boundary that is introduced by the brane has a natural geometric interpretation as well. To continue the comparison to \cite{Saad:2019lba}, the authors utilize a gluing procedure that involves hyperbolic surfaces with geodesic boundaries. The worldline of an open string with a nonzero endpoint mass inside the worldvolume of a brane will naturally be a geodesic with a finite length. For this reason, the boundaries introduced in the matrix model by the brane will be considered to be geodesic boundaries. This identification will be justified by computing the trumpet partition function, which in JT gravity is the path integral on the surface shown in fig. \ref{fig:trumpet}, as well as the matrix model resolvents $W_{g,n}$. The trumpet partition function is a feature that is universal to the types of models considered here, including the supersymmetric ones. 

We will continue to use the letter $h$ to count powers of $\Gamma$, which as we will see corresponds to the number of geodesic boundaries. The free energy in the open string sector has the topological expansion
	\begin{equation}
		F = \sum_{g,h = 0}\hbar^{2g-2+h}\Gamma^h F_{g,h},
	\end{equation}
where each $F_{g,h}$ is found by integrating $u_{g,h}$. It is once again possible to consider correlation functions of the point-like operators. When it is not clear from context we will write $\langle \cdot\rangle_{g,h}^{\text{open}}$ to indicate that the open string free energy is involved, with a similar notation for the closed string sector.  The statement of the open-closed duality involves a special operator which we will call $\omega_z$, defined by
	\begin{equation}
		\omega_z =4\hbar \Gamma \sum_{k = 1}^\infty \zeta_k z^{-2k}\sigma_{k-1}. \label{eqn:omegadef}
	\end{equation}
The duality is expressed by
	\begin{equation}
		\expect{\omega_{z_1}\cdots \omega_{z_h} \mathcal{O}}_g^{\text{closed}} = \frac{2}{h!} \frac{\partial}{\partial z_1}\cdots \frac{\partial}{\partial z_h} \langle \mathcal{O} \rangle_{g,h}^{\text{open}}\Big|_{\text{conn.}}, 
		\label{eqn:duality}
	\end{equation}
where $\mathcal{O}$ is some operator consisting of the $\sigma_k$'s. The interpretation of this is that the operator $\omega_z$ represents the insertion of a geodesic boundary with open string endpoint mass $-z^2$. The right hand side of \eqref{eqn:duality} is to be evaluated using the solution to the multi-brane string equation, but because that solution takes into account the possibility of the worldsheet ending on the same brane multiple times, we only want to look at the case where it lands on each brane once. Even though a multi-brane generalization of the DJM equation to describe multiple open strings in supersymmetric theories does not currently exist, the desired open string quantities will be computed using the left hand side of \eqref{eqn:duality}, in the closed string sector. Actually, a way to generate the solutions that should exist -- although they aren't obtained from a differential equation -- is to apply the duality relation to the identity operator to calculate the open string free energy, and then invert the relationship between $F$ and $u$.

\subsection{Boundary operator}

Thinking of the coupling constants $t_k$ and the open string endpoint mass $-z^2$ as parameters of the theory, since derivatives with respect to $t_k$ correspond to inserting the operators $\sigma_k$ into correlation functions, it is natural to expect that a derivative with respect to $z$ should be represented by an operator insertion as well. This was shown to be true in the open string matrix model for the case of an individual multicritical model in \cite{Johnson:1993vk}. One can show by direct matrix model computation that \cite{Dalley:1992br}
	\begin{equation}
		\frac{\partial u}{\partial z} = 4z \hbar \Gamma \hat{R}'. \label{eqn:openstringzflow}
	\end{equation}
By expanding $\hat{R}'$ in terms of the Gelfand-Dikii polynomials and using the KdV flow equations it is clear that the definition of $\omega_z$ in \eqref{eqn:omegadef} is the appropriate operator dual to the $z$-derivative.

By construction, the expectation value of $\omega$ (in the closed string sector) will be intimately related to the Gelfand-Dikii resolvent. In particular
	\begin{equation}
		\langle \omega_z \rangle = \frac{4\Gamma z}{\hbar} \int_{-\infty}^\mu \hat{R}\,dx.
	\end{equation}
Therefore knowing the Gelfand-Dikii resolvent, which is the solution to the Gelfand-Dikii differential equation \eqref{eqn:GDeqn}, is enough information to determine the expectation value of $\omega_z$ at any order in perturbation theory. But, this integral of the Gelfand-Dikii resolvent is nothing by the complexified spectral density of the closed string theory, the leading order contribution to which is often referred to as the spectral curve. The spectral density contains important statistical information about the model, making its connection to a brane insertions especially interesting.

It is common to work in terms of the complexified variable $z$ when computing the correlation functions of the matrix resolvent $(M - \lambda)^{-1}$ in the double scaling limit. Recall that it is natural to redefine the components of the topological expansion of one of these correlation functions in terms of the objects $W_{g,n}(z_1,\dots,z_n)$. Given the strong connection between the matrix resolvent and the determinant, it is no surprise that there is still a close connection between FZZT brane quantities and the functions $W$. In fact, using open-closed duality the connection can be written as
	\begin{equation}
		W_{g,n}(z_1,\dots,z_n) = 2\langle \omega_{z_1}\cdots \omega_{z_n}\rangle_g = \frac{2}{n!} \frac{\partial^n F_{g,n}}{\partial z_1\cdots \partial z_n}
	\end{equation}
This claim will be explored more thoroughly in the next section. It is also well-known that the (real) spectral density is given by the discontinuity of the resolvent $R$ across the real axis. A web of connections is beginning to unfold. The story continues with the fact that the resolvent functions $W_{g,n}$ are the Laplace transforms of the generalized Weil-Petersson volumes \cite{Eynard:2007fi}. Hence the functions $W_{g,n}$ contain information about the intersection numbers on a generalized moduli space, the generating function for which -- at least in the case where the matrix model is dual to JT gravity -- is the open string free energy $F$ \cite{mulase2006mirzakhani}. The connections are summarized below in fig. \ref{fig:web}.

\begin{figure}[htbp]
\begin{center}
	\includegraphics[clip,trim=.6in 2.75in .5in 2.7in, width = 0.75\textwidth]{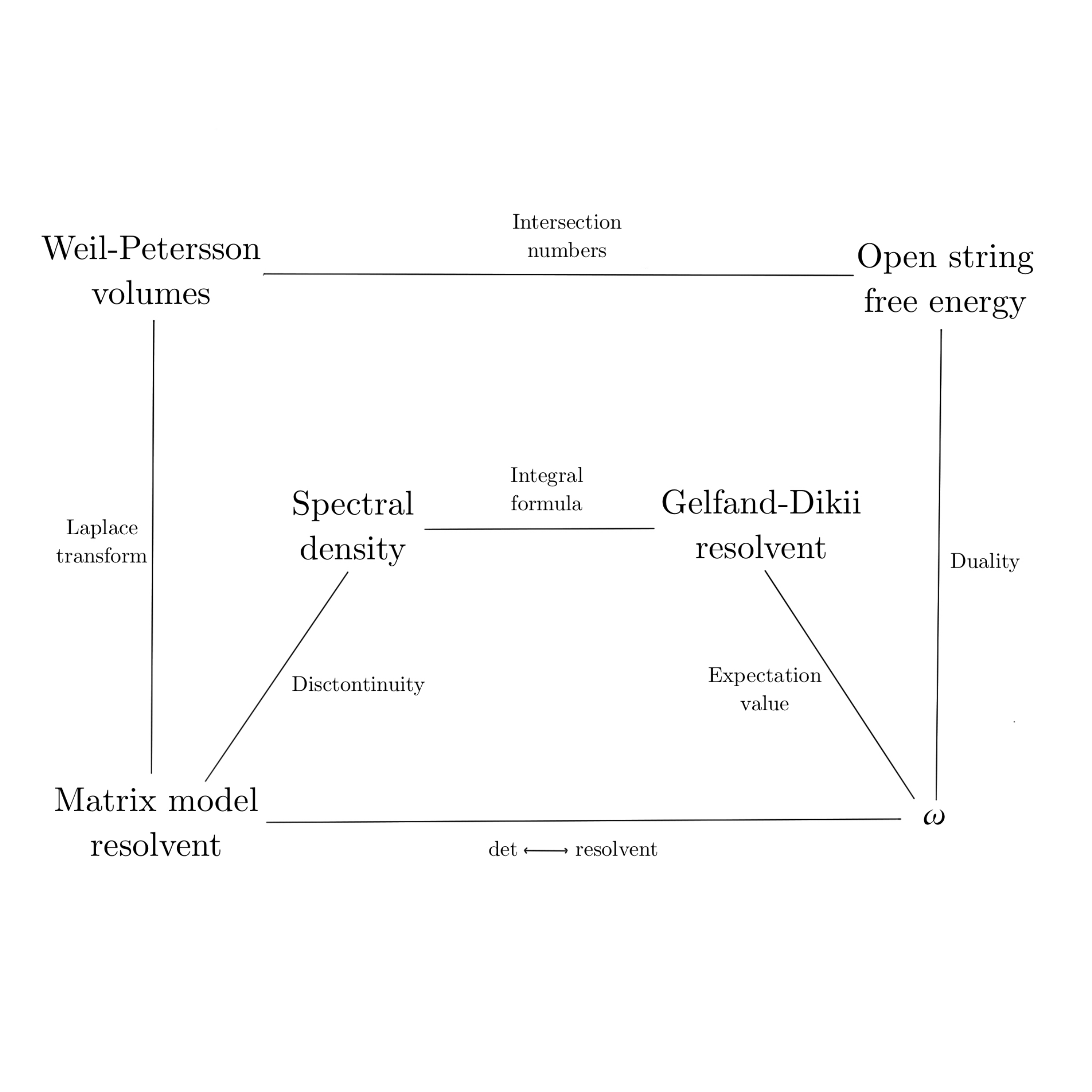}
\caption{The web of connections forming in the string equation formalism, which includes important matrix model objects and the Weil-Petersson volumes.}
\label{fig:web}
\end{center}
\end{figure}

\subsection{Trumpet partition function}

Studying macroscopic loops is still a fruitful endeavor in the open-closed theory. A modern motivation for studying them is the trumpet partition function defined in the JT gravity context in \cite{Saad:2019lba}
	\begin{equation}
		Z_{\text{tr}}(\beta,b) = \frac{1}{\sqrt{4\pi \beta}}e^{-\frac{b^2}{4\beta}}, \label{eqn:trumpet}
	\end{equation}
which is the Schwarzian path integral on a (hyperbolic) surface with a geodesic boundary of length $b$ and an asymptotic boundary of length $\beta$. The same quantity was computed in \cite{Stanford:2019vob}, where they found it has the same functional dependence on $b$ and $\beta$. In both instances $Z_{\text{tr}}$ was computed specifically using the (super) Schwarzian action, and so given the specificity one might expect that the corresponding object computed in the matrix model would at least be dependent on the coupling constants $t_k$. In \cite{Okuyama:2021eju} it was argued that FZZT branes have a natural place in the landscape of general topological gravity. Moreover they argued that the trumpet partition function was related to an FZZT brane insertion, and that it should be independent of the coupling constants $t_k$. Their argument relates the trumpet partition function to the Liouville wavefunction, which has no explicit knowledge of the coupling constants used to define the matrix model. In the context of the present work, this independence will arise from a specific dependence on the function $f$ that defines the leading order open string equation. In order to begin justifying the duallity, we will compute the trumpet partition function three ways.

The trumpet is an asymptotic disk with a geodesic boundary insertion in the bulk. Hence it should show up at order $\Gamma$ in the macroscopic loop expectation value $\langle e^{-\beta \mathcal{H}}\rangle^{\text{open}}_{0,1}$ in the open string theory.
The expectation value of each $\sigma_k$ is computed as before in terms of the KdV vector fields, but now we expand the Gelfand-Dikii polynomial $R_k$ to order $\Gamma$. Therefore the $g = 0,h = 1$ contribution to $\mathcal{Z}_1$ is given by
	\begin{equation}
		\langle e^{-\beta \mathcal{H}} \rangle _{0,1} = -\half \sqrt{\frac{\beta}{\pi}} \int_{-\infty}^\mu dx\,e^{-\beta u_0} \, u_{0,1}.
	\end{equation}
Plugging in the solution for $u_{0,1}$ in terms of its Taylor series in $u_0$ and changing the integration variable to $u_0$ using $x = -f(u_0)$, the bounds of integration go from $u_0(\mu) = E_0$ to $\infty$. The result is
	\begin{equation}
		\langle e^{-\beta \mathcal{H}} \rangle _{0,1} = \frac{1}{\sqrt{4\pi\beta}}\sum_{k = 0}^\infty \zeta_k \Gamma(k+1,\beta E_0) z^{-2k-1}\beta^{-k},
		\end{equation}
where $\Gamma(a,b)$ is the incomplete gamma function. The mass of the endpoint of the open string serves as a cosmological constant for the open string boundaries. This fact is captured in the FZZT language by the fact that the determinant operator is really related to the Laplace transform of a fixed-length insertion. The open string endpoint mass is therefore Laplace conjugate to geodesic boundary length. In order to retrieve the trumpet partition function we must take the inverse Laplace transform of $\langle e^{-\beta \mathcal{H}} \rangle _{0,1}$. This gives
	\begin{equation}
		\mathcal{L}^{-1}_z\Big[\langle e^{-\beta \mathcal{H}} \rangle _{0,1}  \Big](b) = \frac{1}{\sqrt{4\pi\beta}} \sum_{k = 0}^\infty \frac{\Gamma(k+1,\beta E_0)}{(k!)^2} \left(-\frac{b^2}{4\beta}\right)^k,
	\end{equation}
By expanding the gamma function in a power series and resuming, we get
	\begin{equation}
		\mathcal{L}^{-1}_z\Big[\langle  e^{-\beta \mathcal{H}}\rangle_{0,1} \Big](b) =   Z_{\text{tr}}(b,\beta) - \frac{ \beta E_0}{\sqrt{4\pi\beta}} \sum_{k = 0}^\infty \frac{(-\beta E_0)^k}{k!(k+1)} {_1}F_2 \left(k+1;1,k+2;-\frac{E_0b^2}{4} \right).
	\end{equation}
For nonzero $E_0$, there is an infinite tower of corrections to the trumpet partition function, whereas for $E_0 = 0$ the corrections disappear leaving the desired result. This is a departure from the result obtained in \cite{turiaci2023mathcaln2}, where they found that the $\mathcal{N} = 2$ trumpet was simply $e^{-\beta E_0} Z_\text{tr}$. In the following section we will establish a different gluing procedure that utilizes just $Z_\text{tr}$, with a volume that takes into account the exponential $E_0$-dependence. Although the result for that volume will also differ from \cite{turiaci2023mathcaln2}, the path integral calculations will match up. Crucially, this expression is independent of the coupling constants, making it universal for the class of models considered here.

Next consider the correlator $\langle \omega_z e^{-\beta \mathcal{H}}\rangle$, computed in the closed string sector. We can utilize the previously stated $g = 0$ two-point function of the $\sigma_k$'s from \eqref{eqn:sigmatwopoint}. There will be one remaining integral with respect to $x$, which we can change into an integral over $u_0$ by using the factor of $u_0'$ present in the integrand. The bounds of the integral are once again from $E_0$ to $\infty$. Integrating once with respect to $z$ we get
	\begin{equation}
		\mathcal{L}^{-1}_z\Big[\langle \omega_z e^{-\beta \mathcal{H}}\rangle_0 \Big](b) = \frac{2\Gamma}{\sqrt{4\pi\beta}} \sum_{k = 0}^\infty \frac{\Gamma(k+1,\beta E_0)}{(k!)^2} \left(-\frac{b^2}{4\beta}\right)^2,
	\end{equation}
which is equal to the result derived in the open string sector, up to the desired factor of $2\Gamma$. Apart from computing the trumpet partition function purely using the matrix model, we have also established the duality in this particular case.

Consider the multi-brane correlator $\langle e^{-\beta \mathcal{H}} \rangle_{0,2}$ in the open string sector. Expanding the Gelfand-Dikii polynomial $R_k$ to order $\hbar^2\Gamma^2$ and performing the sum,
	\begin{equation}
		\langle e^{-\beta \mathcal{H}} \rangle_{0,2} = -\frac{\sqrt{\beta}}{4\sqrt{\pi}} \int_{E_0}^\infty du_0 e^{-\beta u_0} \dot{f}(u_0)\left(\beta u_{0,I}^2 - 2u_{0,II}\right).
	\end{equation}
Using the multi-brane generalization of the identity \eqref{eqn:u01tou02}, integrating by parts, and keeping the term with no repeated factors of $z_1$ or $z_2$ yields the surface term
	\begin{equation}
		\langle e^{-\beta \mathcal{H}} \rangle_{0,2} = - \frac{\sqrt{\beta}}{2\sqrt{\pi}} \frac{u_0'(\mu) }{\sqrt{E_0 + z_1^2}\sqrt{E_0 + z_2^2}}e^{-\beta E_0}.
	\end{equation}
In the closed string sector, we get the result
	\begin{equation}
		\langle \omega_{z_1} \omega_{z_2} e^{-\beta \mathcal{H}} \rangle_0 = - \frac{\Gamma^2}{2}\sqrt{\frac{\beta}{\pi}} \frac{z_1z_2 u_0'(\mu)}{(E_0 + z_1^2)^{3/2}(E_0 + z_2^2)^{3/2}}e^{-\beta E_0}
	\end{equation}
which establishes the duality for this case as well.
	

\subsection{FZZT brane partition function}

It is worth pausing briefly to compare the open-closed duality methodology to the more standard way of dealing with FZZT branes in the closed string sector of matrix models . FZZT branes are represented by determinant operators
    \begin{equation}
        \text{FZZT brane} \quad \longleftrightarrow \quad \det(\lambda - M).
    \end{equation}
By using the identity $\det(\lambda + M) = \exp\{\tr \log(\lambda + M)\}$\footnote{This is precisely the reason why the matrix model describing open strings has a logarithmic potential.}, it is possible to rewrite the FZZT brane operator in terms of macroscopic loop insertions
    \begin{equation}
        \tr \log (\lambda + M) \cong -\int_\ve^\infty \frac{d\beta}{\beta} e^{-\lambda\beta} \tr e^{-\beta M},
    \end{equation}
where $\ve \to 0$ and we have dropped a $\log \ve$ divergence. It must also be understood that we will need to make the substitution $\lambda \to -\lambda$ later. Therefore
    \begin{equation}
        \det(\lambda + M) = \sum_{n = 0}^\infty \frac{(-1)^n}{n!}\prod_{i = 1}^n \int_0^\infty \frac{d\beta_i}{\beta_i}e^{-\lambda\beta_i} \tr e^{-\beta_i M}.
    \end{equation}
The operator $\tr e^{-\beta_iM}$ represents a macroscopic loop prior to double scaling. The expectation value of a single brane operator is denoted by $\langle \Psi(E)\rangle = \psi(\mu,E)$ in the double scaling limit:
    \begin{equation}
        \psi(\mu,E) = \sum_{n = 0}^\infty \frac{(-1)^n}{n!}\left\langle\prod_{i = 1}^n \int_0^\infty \frac{d\beta_i}{\beta_i}e^{\beta_i E}  e^{-\beta_i \mathcal{H}} \right\rangle,
    \end{equation}
where now each expectation value includes connected and disconnected parts. The leading contribution comes from the leading parts of the totally disconnected parts of the correlation functions. That is,
    \begin{equation}
        \psi(\mu,E) \approx \sum_{n = 0}^\infty \frac{(-1)^n}{n!}\prod_{i = 1}^n \int_0^\infty \frac{d\beta_i}{\beta_i}e^{\beta_i E}  \left\langle e^{-\beta_i \mathcal{H}} \right\rangle_0.
    \end{equation}
Using the result derived in section \ref{scn:macroscopicloopperturbationtheory}, the brane expectation value is
    \begin{equation}
        \psi(\mu,E) \approx  \sum_{n = 0}^\infty \frac{(-1)^n}{n!}\prod_{i = 1}^n \frac{1}{ 2\hbar\sqrt{\pi}}\int_{-\infty}^\mu dx\int_0^\infty \frac{d\beta_i}{\beta_i^{3/2}}e^{-\beta_i (-E+u_0)}.
    \end{equation}
Noting that
    \begin{equation}
        \int_0^\infty \frac{d\beta_i}{\beta_i^{3/2}}e^{-\beta_i (-E+u_0)} = -2\sqrt{\pi}\sqrt{-E + u_0},
    \end{equation}
we find
    \begin{equation}
        \psi(\mu,E) \approx \exp\left\{\frac{i}{\hbar} \int^\mu \sqrt{E - u_0}\right\}.
    \end{equation}

This is the expected result, since before double scaling the brane expectation value computes the $N^{\text{th}}$\footnote{Recall $N \times N$ is the size of the random matrix.} orthogonal polynomial $p_N(\lambda)$ associated to the matrix potential. As discussed, these orthogonal polynomials double scale to wavefunctions solving the Schrodinger equation with potential $u$. The leading order WKB approximation to the wavefunction is given by
    \begin{equation}
        \psi(x,E) \approx \exp\left\{\pm\frac{i}{\hbar} \int^x \sqrt{E - u_0}\right\},
    \end{equation}
and the index $N$ is located at the Fermi surface $x = \mu$. By incorporating higher order corrections to $R_k$, we can compute higher orders of the WKB expansion of the wavefunction\footnote{Doing the WKB expansion is much easier than doing the calculation with the Gel'fand-Dikii polynomials. It is still interesting that the expressions involving $R_k$ must reproduce the WKB expansion.}.


Although the preceding result was derived via the $\hbar$-expansion, it should be noted that the outcome represents non-perturbative physics. The next term in the perturbative expansion of $\psi(\mu,E)$, which is really the first {\it true} perturbative contribution, comes from breaking the terms with an even number of macroscopic loops into connected two-point functions and keeping only the leading contributions there. Each of those terms contributes a factor of $\hbar^0$. The calculation requires the integral
    \begin{equation}
       \mathcal{I}_1 = \frac{1}{\pi}\int_0^\infty \frac{d\beta_1d\beta_2}{\beta_1\beta_2} \frac{\sqrt{\beta_1\beta_2}}{\beta_1 + \beta_2}e^{-(\beta_1+\beta_2)(-E+u_0(\mu))} \cong -\frac{1}{2}\log(-E+u_0(\mu)),
    \end{equation}
where we've dropped a divergent piece. Each term with an even number $2n$ of macroscopic loops will contribute $\mathcal{I}^n$ with a degeneracy of $\frac{(2n)!}{2^nn!}$. Thus we find
    \begin{equation}
        \sum_{n = 0}^\infty \frac{1}{(2n)!} \frac{(2n)!}{2^nn!}\mathcal{I}^n = e^{-i\pi/4}(E-u_0(\mu))^{-1/4}.
    \end{equation}
At this order the brane wavefunction is
    \begin{equation}
        \psi(\mu,E) \approx \frac{1}{(E-u_0(\mu))^{1/4}}\exp\left\{\frac{i}{\hbar} \int^\mu \sqrt{E - u_0}-\frac{i\pi}{4}\right\},
    \end{equation}
which is consistent with what one calculates in the Airy model, including the phase.

The natural object to describe just an FZZT brane insertion in the open-closed duality language is the free energy of the open string sector. Using the relationship between $F$ and $u$, we find
	\begin{equation}
		F_{0,1} =  \int^\mu \sqrt{u_0 + z^2}.
	\end{equation}
Recalling that $z^2 = -E$, the leading order contribution to the brane partition function would then be $e^{F_{0,1}/\hbar} = \psi(\mu,E)$ at leading order. We have to divide by $\hbar$ to account for the fact that $F_{0,1}$ is defined without any explicit $\hbar$ present. The first perturbative correction to the partition function comes from exponentiating $F_{0,2}/2$, where the free energy is evaluated using the single-brane solution. We only need to consider $F_{0,2}$ at this order because we must have $h \neq 0$ in order to be describing the brane. The factor of 2 is inserted to account for the symmetry to swap the endpoints of the string. Using \eqref{eqn:u01tou02} we have
	\begin{equation}
		F_{0,2} = -\half \log(E_0 + z^2).
	\end{equation}
The exponential and extra factor of 2 convert this into the desired portion of the WKB wavefunction. The calculations done the more traditional way become increasingly tedious at higher orders, and it can be somewhat hard to parse when disconnected versus connected correlators need to be used. In the open-closed duality language, one simply computes the open string free energy and exponentiates. The next contribution would come from $F_{1,1}$.

\subsection{The matrix kernel}

The double scaled matrix kernel $K(E,E')$ primarily exists to calculate statistical properties of the random matrix model, and as such provides a powerful tool for probing the spectrum of the theory. It has been utilized recently to access the discrete nature of the black hole spectrum in JT gravity using a proposed non-perturbative completion (see for example \cite{Johnson:2022wsr}). The kernel itself is typically calculated using the double scaled wavefunctions $\psi$ via
	\begin{equation}
		K(E,E') = \int_{-\infty}^\mu dx\,\psi(x,E)\psi(x,E').
	\end{equation}
As discussed in the previous subsection, the wavefunction $\psi(\mu,E)$ is the partition function of an FZZT brane with cosmological constant $E$. This implies that the kernel is a ``composite D-brane probe that is well-suited to detecting the `location' of individual energies" \cite{Johnson:2022wsr}. 

Since the one-point function of $\omega_z$ computes the full eigenvalue density, we may expect that its correlation functions compute the correlations between multiple eigenvalues. The two-point function of $\omega$ is 
	\begin{equation}
		\langle \omega_{z_1} \omega_{z_2} \rangle = 16\hbar^2 \Gamma^2 \sum_{k_1,k_2 = 0}^\infty \zeta_{k_1} \zeta_{k_2} z_1^{-2k_1}z_2^{-2k_2} \iint^\mu \xi_{k_1} \cdot R_{k_2}' .
	\end{equation}
Performing the sums, we can write
	\begin{equation}
		\langle \omega_{z_1} \omega_{z_2} \rangle = 16\hbar^2\Gamma^2z_1z_2 \iint^\mu \hat{\xi}(z_1) \cdot \hat{R}'(z_2),
	\end{equation}
where we have introduced $\hat{\xi}(z) = \sum_{k = 0}^\infty \hat{R}^{(k+1)}(z) \frac{\partial}{\partial u^{(k)}}$. We note that the partial derivatives with respect to $u^{(k)}$ are understood to only act on functions in the closed string sector. Two interesting observations arise from this. First, using open-closed duality we have the double-flow of the two-brane open string sector solution
	\begin{equation}
		\frac{\partial^2 u(z_1,z_2)}{\partial z_1 \partial z_2} = 16\hbar^2 \Gamma^2 \hat{\xi}(z_1) \cdot \hat{R}'(z_2),
	\end{equation}
which generalizes \eqref{eqn:openstringzflow}. This is a specific example of the comment made at the beginning of this section about constructing the supersymmetric multi-brane solutions without having a string equation. Second, the connection between the matrix resolvent and the kernel can be exploited to arrive at $W_{g,2}(z_1,z_2) \sim 4z_1z_2 \hbar^2 K(-z_1^2,-z_2^2)$ \cite{johnson_email}. Therefore 
	\begin{equation}
		\iint^\mu \frac{\partial^2 u(z_1,z_2)}{\partial z_1 \partial z_2} dx = 4\hbar^2 K(-z_1^2,-z_2^2) = 16 \iint^\mu \hat{\xi}(z_1) \cdot \hat{R}'(z_2).
	\end{equation}
The left equality relates the kernel to the open string sector solution, while the right one relates the kernel to the closed string sector solution. These equalities provide a sharper image of the relationship between the kernel and its statistical importance, and the presence of branes.


\section{Generalized Weil-Petersson Volumes} \label{scn:generalizedweilpeterssonvolumes}

Similar to the trumpet partition function in the preceding section, our aim in this section is to further justify the claim that correlation functions of $\omega$ compute the matrix model resolvent functions $W_{g,n}$. By computing them in the matrix model using the closed string operator formalism, we will show that it is possible to study more complicated supersymmetric theories and calculate their generalized Weil-Petersson volumes. Using the known results as a testing ground, we will be able to confidently make predictions about these quantities in $\mathcal{N} = 2$ theories.

By computing the resolvent functions $W_{g,n}$ in terms of the closed string solution $u$ we will be able to compute generalized Weil-Petersson volumes for arbitrary coupling constants. A similar procedure was performed in \cite{Okuyama:2021eju}. The added benefit of performing the analysis using the string equation is that it is much easier to define the volumes for supersymmetric theories. In both the non-supersymmetric and supersymmetric cases, we will define the volumes
	\begin{equation}
		b_1 \cdots b_nV_{g,n}(\bm{b};t_k) = \hbar^{-2g-n}\Gamma^{-n} \mathcal{L}^{-1}_{\bm{z}} \Big[\langle \omega_{z_1} \cdots \omega_{z_n} \rangle_g \Big](\bm{b}),
	\end{equation}
where the boldface notation indicates taking the multi-variable inverse Laplace transform.

It is clear from these results that the operator $\omega_z$ makes a lovely connection between objects naturally considered the KdV organization of double scaled matrix models and quantities typically computed using algebraic geometry. 
Of course this is related to the content of the Witten-Kontsevich theorem \cite{Witten:1990hr, cmp/1104250524}. The relationship between matrix models and intersection theory on the moduli space of stable curves is by no means a new story. The operator $\omega_z$ and the formalism of open-closed duality provides an explicit construction of this relationship. That the Gelfand-Dikii resolvent, which is naturally constructed using the operator $\omega_z$, should be so closely related to the matrix model resolvent $W_{g,n}$ was anticipated recently in \cite{Johnson:2024bue}.

\subsection{Non-supersymmetric theories}

Let us first explicitly show that $\langle \omega_z \rangle = 4\Gamma z y(z)$ with some examples, starting at leading order. It can be shown by other means that
	\begin{equation}
		y_0(z) = \frac{1}{2\hbar} \int_{-\infty}^\mu \frac{dx}{\sqrt{u_0 + z^2}} =  \frac{1}{2\hbar} \sum_{k = 1}^\infty \frac{\sqrt{\pi}\Gamma(k+1)}{\Gamma\left(k + \half\right)}t_k z^{2k-1}. \label{eqn:spectraldensity}
	\end{equation}
At leading order the Gelfand-Dikii resolvent is $\hat{R} = \half(u_0 + z^2)^{-1/2}$, which we can easily see produces the equivalence between $\langle \omega_z \rangle_0$ and $y_0(z)$. Due to the stated relationship between the matrix model resolvent and the spectral density, the first resolvent function $W_{g,1}(z)$ is equivalent to $y_g(z)$. At $g = 1$, the Gefland-Dikii resolvent is given by
	\begin{equation}
		\hat{R}_1 = \frac{5 (u_0')^2}{64 \left(z ^2+u_0\right)^{7/2}}-\frac{u_0''}{16 \left(z^2+u_0\right)^{5/2}}-\frac{u_1}{4 \left(z^2+u_0\right)^{3/2}}.
	\end{equation}
After changing the integration variable to $u_0$, the $g = 1$ contribution to $\langle \omega_z \rangle$ is
	\begin{equation}
		\langle \omega_z \rangle_1 = 4\hbar\Gamma z \int_{0}^\infty du_0\,\frac{1}{u_0'} \left[\frac{5 (u_0')^2}{64 \left(z ^2+u_0\right)^{7/2}}-\frac{u_0''}{16 \left(z^2+u_0\right)^{5/2}}-\frac{u_1}{4 \left(z^2+u_0\right)^{3/2}}\right].
	\end{equation}
By using the special relationships between $u_0'$, $u_0''$, and $u_1$ and integrating by parts, the resulting surface terms give
	\begin{equation}
		\langle \omega_z \rangle_1 = \frac{
		\hbar^3\Gamma}{24} \frac{2(u_0''/u_0') z^2 - 3u_0'}{z^4} \Bigg|_{x = \mu}.
	\end{equation}
with the rest cancelling. Using the JT gravity solution, the coefficients in $\langle \omega_z \rangle_1$ are $\pi$ and $-1$, respectively, which agrees with the result obtained via topological recursion. Another noteworthy case that can be checked quickly is the Airy model, where $u_0 = -x$. Generally, we can refer to \eqref{eqn:nonsusy_fermieval} to write this in terms of the coupling constants. By taking the inverse Laplace transform we obtain the generalized volume
	\begin{equation}
		V_{1,1}(b;t_k) = \frac{1}{24t_1} \left(b^2 + \frac{4t_2}{t_1^2} \right).
	\end{equation}

Consider the $g = 0$ contribution to the $\omega$ three-point function. Using \eqref{eqn:3ptidentity} one finds that
	\begin{equation}
		\langle \omega_{z_1}\omega_{z_2}\omega_{z_3} \rangle_0 = -\frac{ u_0'(\mu)}{z_1z_2z_3}.
	\end{equation}
When we take the inverse Laplace transform to change this into a fixed length quantity, one finds
	\begin{equation}
		V_{0,3}(b) = -u_0'(\mu) = \frac{1}{t_1}. \label{eqn:3ptfcn}
	\end{equation}
It is no surprise that, given the poor behavior of the macroscopic loop correlators in individual multicritical models, it is not possible to define $V_{0,3}$ for such a model. Once again in order to get finite results the topological point $k = 1$ needs to be included in an interpolation.


\subsection{\texorpdfstring{$\mathcal{N} = 1$}{N1}}

Before proceeding to compute the volumes in supersymmetric theories we stress again here that the overall formalism does not change, only over what ranges certain integrals are performed and where the function $u$ is evaluated. As we will see, minimal extra effort is required to calculate the 0A volumes, and we never have to directly interact with supermanifolds.

The first notable change occurs in the spectral density. Since the Fermi surface is nonzero in 0A theories, the integration extends into the closed string sector of the theory, where we take the solution $u_0 = 0$. Recall that this was important for the macroscopic loop expectation value as well. If we take $t_0 = \mu$, then the leading order spectral density is given by \eqref{eqn:spectraldensity} but with the sum starting at $k = 0$, the second topological point in the KdV hierarchy.

The integral defining $\langle \omega_z \rangle_1$ must be once again modified to only go from 0 to $\mu$. The Gelfand-Dikii resolvent at $g = 1$ is $\hat{R}_{1} = - \frac{u_1}{4z^3}$ and hence
	\begin{equation}
		\langle \omega_z \rangle_{1} = \hbar \Gamma \frac{\Gamma^2 - \quarter}{\mu z^2}.
	\end{equation}
This implies that the moduli space volume is given by
	\begin{equation}
		V_{1,1} = \frac{\Gamma^2 - \quarter}{\mu}.
	\end{equation}
Recall that in the closed string sector the parameter $\Gamma$ counts RR flux insertions. This result is consistent with the discussion in \cite{Stanford:2019vob}, in which the authors point out that the flux insertions are intrinsic to the theory and therefore constitute extra degrees of freedom or moduli. When $\Gamma$ is set to 0 here, it reproduces the negative volume one would expect in a 0A theory, although our result differs from \cite{Stanford:2019vob} by a normalization choice.

The Gelfand-Dikii resolvent at $g = 2$ is
	\begin{equation}
		\hat{R}_{2}(x,z^2) = -\frac{u_{1}''-3 u_{1}^2+4 z^2 u_{2}}{8 z^5}.
	\end{equation}
Performing the $x$ integral and taking the inverse Laplace transform gives the volume
	\begin{equation}
		V_{2,1}(b;t_k) = -\frac{\left(4\Gamma^2 - 1\right)\left(4\Gamma^2 - 9 \right) \left(b^2 \mu +12 t_1\right)}{384 \mu ^4}
	\end{equation}
The coupling constant takes the value $t_1 = \pi^2$ and the Fermi surface is $\mu = 1$ in this particular $\mathcal{N} = 1$ JT supergravity. Plugging those values in and turning the RR flux off, we once again have agreement with the expected result, up to the same normalization factor of $\half$. We note also that any theory with $t_1 = 0$, for example any of the individual multicritical models, will compute the same volume $V_{2,1}^{0A}(b,0)$. This is another example of how the 0A multicritical models are better behaved than their non-supersymmetric counterparts.

In order to calculate the volume $V_{1,2}$, we no longer can rely on the Gelfand-Dikii resolvent. However we can once again adapt the result from \ref{apdx:macroscopicloopcalculations} for the $\sigma$ two-point function. We get
	\begin{equation}
		\langle \omega_{z_1}\omega_{z_2} \rangle_1 = -\frac{\hbar^2 \Gamma^2 \left(4\Gamma^2 - 1 \right)}{4 \mu^2 z_1^2 z_2^2},
	\end{equation} 
which produces the constant volume
	\begin{equation}
		V_{1,2} = -\frac{4\Gamma^2 - 1}{4 \mu^2}.
	\end{equation}
Once again, this agrees with the expected result up to the normalization choice when we set $\Gamma = 0$.

The recent $\mathcal{N} = 1$ generalization of the Virasoro Minimal String \cite{Johnson:2024fkm} is a 0A theory. Hence the results for $V_{1,1}$ and $V_{0,3}$ are valid there, since they are independent of the closed string coupling constants, which in this theory are given by
	\begin{equation}
		t_k = 2\sqrt{2}\frac{\pi^{2k+1}}{(k!)^2} \left(Q^{2k} + \hat{Q}^{2k}\right),
	\end{equation}
with $Q$ given above in \eqref{eqn:Qdef}. The first coupling constant is $t_1 = 4\sqrt{\pi}/b$, which means the $g = 2, n =1$ volume is
	\begin{equation}
		V_{2,1}(L) = -\frac{3}{2^9} \left( L^2 + \frac{16\sqrt{2}\pi}{b}\right).
	\end{equation}

\subsection{\texorpdfstring{$\mathcal{N} = 2$}{N2}}

We continue here to only look at $g = 0$ quantities, starting with the three-point function. During the derivation of the result for the theories with $u_0(\mu) = 0$, that fact was used to simplify the result. For a model where this does not happen, the result is
	\begin{equation}
		\langle \omega_{z_1}\omega_{z_2}\omega_{z_3} \rangle_0 = - \frac{\hbar^3\Gamma^3 z_1z_2z_3}{\Big[(E_0 + z_1^2)(E_0 + z_2^2)(E_0 + z_3^2)\Big]^{3/2}},
	\end{equation}
The inverse Laplace transform gives the volume
	\begin{equation}
		V_{0,3}(b_1,b_2,b_3) = -J_0(b_1\sqrt{E_0})J_0(b_2\sqrt{E_0})J_0(b_3\sqrt{E_0})u_0'(\tilde{\mu}).
	\end{equation}
By gluing the exponential part of the $\mathcal{N} = 2$ trumpet to this volume three times, we recover the matrix model prediction for $\mathcal{Z}_{0,3}$.

	
\section{Topological Recursion} \label{scn:topologicalrecursion}


There is a common thread beneath the surface of the web in fig. \eqref{fig:web}. The Gelfand-Dikii polynomials, which as we have seen are central to the KdV organization of these matrix models, obey the recursion relation \eqref{eqn:GDrecursion}. The solutions to the string equation and the perturbative expansions of correlators are intimately linked to the structure the Gelfand-Dikii polynomials are endowed with by virtue of this recursion relation. Also within the KdV organization of the models, the closed string operators $\sigma_k$, which are related to the vector fields that generate the KdV flows, obey a sort of recursion relation in the form of Ward identities that result from the Virasoro conditions \cite{Dijkgraaf:1990rs, Johnson:1993vk, Johnson:2004ut}. Finally, the Weil-Petersson volumes famously obey Mirzakhani's recursion relation \cite{Mirzakhani:2006fta}. This can be thought of as a special case of the matrix resolvent recursion relation in \cite{Eynard:2007fi}, where the matrix model is taken to be dual to JT gravity. Still though, if we define the volumes $V_{g,n}(\bm{b};t_k)$ via the inverse Laplace transform of $W_{g,n}(\bm{z};t_k)$, the generalized volumes should satisfy a recursion relation as well. Given the fact that on the surface of fig. \eqref{fig:web} each of these topics is related to each other, it would be surprising if one could not somehow map their respective recursion relations onto one another.

In this section we will focus on the connection between the Virasoro conditions and the generalized volumes, although we will make some comments about the connection between the Gelfand-Dikii polynomials and topological recursion. The fact that the Virasoro conditions are tied to the recursion between the Weil-Petersson volumes was noticed immediately after Mirzakhani's discovery in \cite{mulase2006mirzakhani}. In that work, the authors approached the problem more from the point of view of intersection theory, and as a result dealt with matrix models more formally, in the spirit of the Kontsevich model's role in proving the Witten-Kontsevich theorem. In this sense, the results of \cite{mulase2006mirzakhani} are somewhat limited in scope compared to what can be accomplished using the open-closed duality described here, since there is clearly a difference between using the matrix model purely as a generating function for a specific set of intersection numbers and volumes, and defining a general family of such objects which depend on the coupling constants $t_k$ and that reproduce the Weil-Petersson case as an example. Moreover, the matrix model technology displayed above allows for the computation of volumes in both 0A and $\mathcal{N} = 2$ supersymmetric theories. Since the Virasoro conditions are attached to the KdV organization of the matrix model, and not specifically the string equation\footnote{The non-supersymmetric closed string equation and the DJM equation are actually implied by two of the Virasoro conditions, but there are infinitely many other ones to derive Ward identities from.}, their Ward identities and recursion relations should manifest in the supersymmetric cases as well.

It is clear from open-closed string duality that the recursion relation between \eqref{eqn:Wrecursion} amongst the resolvents $W_{g,n}$ will maintain a geometric interpretation, insofar as we know that $W_{g,n}$ is intimately connected to the structure of worldsheets stretching between a set of $n$ branes. In particular, we can still imagine that the recursion relation is describing the number of ways to decompose a surface with $n$ geodesic boundaries and genus $g$ into constituent surfaces, either by pinching a handle of the initial surface or by cutting it open. As we will see, the close string operators $\sigma_k$ of the matrix model naturally interact with each other in such a way.

An important part of the Witten-Kontsevich theorem is that the matrix integral \eqref{eqn:WDmatrixintegral}, with $\beta = 2$, is related to the tau function of the KP hierarchy \cite{Witten:1990hr, cmp/1104250524}\footnote{In the double scaling limit, it is related to the tau function of the KdV hierarchy, which is of course our primary interest.}.  In particular, the tau function is $\tau = \sqrt{Z} = e^{-F/2}$, where $F$ is the free energy. By combining the KdV flow equations and the string equation, once can show that $\tau$ satisfies an infinite tower of partial differential equations called Virasoro conditions. The closed string Virasoro generators are
	\begin{equation}
	\begin{aligned}
		L_{-1} &= \sum_{k = 1}^\infty t_k \frac{\partial}{\partial t_{k -1}} + \frac{x^2}{4\hbar^2}, \\
		L_0 &= \sum_{k = 0}^\infty t_k \frac{\partial}{\partial t_{k }} + \frac{1}{16} ,\\
		L_n &= \sum_{k = 0}^\infty  t_k \frac{\partial}{\partial t_{k + n}}  + 4\hbar^2 \sum_{k = 1}^n \frac{\partial^2}{\partial t_{k-1} \partial t_{n - k}},
	\end{aligned}
	\end{equation}
and the constraints are $L_n \cdot \tau = 0$ \cite{Dijkgraaf:1990rs}. The key ingredients in deriving these equations can be generalized to models incorporating open strings and supersymmetry, and a version of the Virasoro conditions for such models was derived in \cite{Johnson:1993vk} and expanded on in \cite{Johnson:2004ut}.

By plugging in the coordinate transformation \eqref{eqn:shiftedcouplings}, we obtain 
	\begin{equation}
	\begin{aligned}
		L_{-1} &\to \sum_{k = 1}^\infty t_k \frac{\partial}{\partial t_{k -1}} +2\hbar\Gamma \sum_{k = 1}^\infty \zeta_k z^{-(2k+1)} \frac{\partial}{\partial t_{k-1}}+ \frac{(x+\hbar\Gamma z^{-1})^2}{4\hbar^2}, \\
		L_0 &\to \sum_{k = 0}^\infty  t_k \frac{\partial}{\partial t_{k }}+2\hbar\Gamma \sum_{k = 0}^\infty \zeta_k z^{-(2k+1)} \frac{\partial}{\partial t_{k}} + \frac{1}{16}, \\
		L_n &\to \sum_{k = 0}^\infty t_k \frac{\partial}{\partial t_{k + n}} +2\hbar\Gamma \sum_{k = 0}^\infty \zeta_k z^{-(2k+1)} \frac{\partial}{\partial t_{k+n}}  + 4\hbar^2 \sum_{k = 1}^n \frac{\partial^2}{\partial t_{k-1} \partial t_{n - k}}.
	\end{aligned}
	\end{equation}
The full form of the open string sector Virasoro operators is actually $\tilde{L}_n = L_n - (2n+2)\frac{\Gamma^2}{4}z^{2n} - z^{2n+2}\frac{\partial}{\partial z}$, where $L_n$ is denotes the shifted operators directly above \cite{Dalley:1992br, Johnson:1993vk}. Consider the correlation function $\langle \sigma_{k_I - 1} \rangle $ where $k_I$ denotes multi-index notation
	\begin{equation}
		 \sigma_{k_I - 1} \equiv \prod_{i \in I}\sigma_{k_i - 1},
	\end{equation}
for some set $I$. The open string sector Virasoro generators annihilate a new $\tau$-function $\tau = e^{-F/2}$, where $F$ is the Free energy of the open theory. The point-like operators $\sigma_k$ naturally act on the free energy via derivatives with respect to $t_k$. Notice that the Virasoro conditions can be rewritten to so that the operators act nonlinearly on $F$
	\begin{equation}
	\begin{aligned}
		L_n \cdot \tilde{\tau} &=  \Bigg[-\half \sum_{k = 0}^\infty \left(k + 1 \right) t_k \frac{\partial F}{\partial t_{k + n}}-\hbar\Gamma \sum_{k = 0}^\infty \zeta_k z^{-(2k+1)} \frac{\partial F}{\partial t_{k+n}}  \\
		&+4\hbar^2 \sum_{k = 1}^n \left(-\half \frac{\partial^2F}{\partial t_{k-1} \partial t_{n - k}} + \quarter \frac{\partial F}{\partial t_{k -1}} \frac{\partial F}{\partial t_{n-k}}\right)\Bigg]e^{-F/2} = 0.
	\end{aligned}
	\end{equation}
By acting on such correlation functions with the Virasoro generators, we arrive at a set of Ward identities. Performing the topological expansion on the correlation functions with a Virasoro generator inserted produces the recursion relation \cite{Johnson:1993vk}
	\begin{equation}
	\begin{aligned}
		\Bigg\langle &\Big(z^{2m+1} \omega_z + \sum_{k = 0}^\infty (k+1) t_k \sigma_{k+m-1} \Big)\sigma_{k_I-1} \Bigg\rangle_{g,h}  - \sum_{i \in I}(k_i+1)\langle \sigma_{k_i + m-1}\sigma_{k_{\hat{I}}-1}\rangle_{g,h} \\
		&= -\sum_{k = 1}^m\Bigg[4 \langle \sigma_{k-1}\sigma_{m-k}\sigma_{k_I-1} \rangle_{g-1,h} + 2\sum_{\substack{ g_1 + g_2 = g \\ Q \cup R = I }} \langle \sigma_{k-1}\sigma_{k_Q -1}\rangle_{g_1,h}\langle \sigma_{m - k} \sigma_{k_R-1} \rangle_{g_2,h} \Bigg], \label{eqn:Virasororecursion}
	\end{aligned}
	\end{equation}
for any $m \geq 1$. 

Structurally, the Ward identity \eqref{eqn:Virasororecursion} is already very close to the recursion relation for the matrix resolvents $W_{g,n}$ in \eqref{eqn:Wrecursion}. Notice that on both sides the multi-indices are so far unfixed, as is the set $I$. We can use the multi-indices on each side to introduce factors of the operator $\omega$ in the correlators on each side. Specifically, let $I = \{2, \dots,n\}$ and multiply by factors of $\zeta_{k_i} z^{-2k_{i}}$. When we sum over $k_I$ we get
	\begin{equation}
	\begin{aligned}
		\Bigg\langle \Big(z^{2m+1} \omega_z &+ \sum_{k = 0}^\infty (k+1) t_k \sigma_{k+m-1} \Big)\omega_{z_I} \Bigg\rangle_{g,h}  - \sum_{k_I = 1}^\infty \zeta_{k_I}z_I^{2k_I}\sum_{i \in I}(k_i+1)\langle \sigma_{k_i + m-1}\sigma_{k_{\hat{I}}-1}\rangle_{g,h} \\
		&= -\sum_{k = 1}^m\Bigg[4 \langle \sigma_{k-1}\sigma_{m-k}\omega_{z_I} \rangle_{g-1,h} + 2\sum_{\substack{ g_1 + g_2 = g \\ Q \cup R = I }} \langle \sigma_{k-1}\omega_{z_Q}\rangle_{g_1,h}\langle \sigma_{m - k} \omega_{z_R} \rangle_{g_2,h} \Bigg], 
	\end{aligned}
	\end{equation}
We still need two more factors of $\omega$ on the right hand side. Thinking of it as a two-index object $S_{k-1,m-k}$, notice that the sum over $k$ is a sum over the $m^\text{th}$ anti-diagonal. A sum over $m$ with a coefficient $C_m$, with $m$ going from 1 to $\infty$, will include each element of $S$ exactly once. Thus it can be rewritten
	\begin{equation*}
		\sum_{m = 1}^\infty C_m \sum_{k = 1}^\infty S_{k-1,m-k} = \sum_{i,j = 1}^\infty C_{i + j - 1} S_{i-1,j-1}
	\end{equation*}
Written in this manner, the indices of $S$ are decoupled, but it comes at the price of complicating the indexing of the coefficient $C$. Evidently we want $C_{i + j - 1} \propto \zeta_i \zeta_j z^{-2i-2j}$. A natural candidate for the numerical portion of $C_m$ comes from a polynomial $P$ built out of the Gelfand-Dikii polynomials via \cite{IMGelfand_1975}
	\begin{equation}
		\tilde{R}_i\tilde{R}_j' = \tilde{P}'_{i,j},
	\end{equation}
where we have used the Gelfand-Dikii normalization. Translated into the normalization used here, one can define a resolvent-esque generating function for $P_{i,j}$
	\begin{equation}
		P(x;z_1,z_2) = \sum_{i,j = 0}^\infty \zeta_i \zeta_j z^{-2(i + j + 1)}P_{i,j}
	\end{equation}
The function $P$ does not have any definite symmetry property under the exchange $z_1 \leftrightarrow z_2$, but its symmetric and anti-symmetric parts obey explicit relations determined by the resolvent $\hat{R}$. Since $P_{i,j}$ is a single polynomial, $P$ can be resummed to depend on just a single index $m$. The coefficients in the expansion of $P$ written this way should essentially be the coefficient $C_m$ that we desire.

By introducing the sum over $m$ with the coefficient just described, picking $|I| = n-1$, and including the other necessary sums to produce all insertions of $\omega$, the right hand side of \eqref{eqn:Virasororecursion} becomes
    \begin{equation}
    \begin{aligned}
        4 \langle \omega_z \omega_z \omega_I \rangle_{g-1,h} &+ 2 \sum_{\substack{ g_1 + g_2 = g \\ Q \cup R = I }} \langle \omega_z \omega_Q \rangle_{g_1,h}\langle \omega_z \omega_R \rangle_{g_2,h} = W_{g-1,n+1}(z,z,I) \\
        &+ \sum_{\substack{ g_1 + g_2 = g \\ Q \cup R = I }}W_{g_1,1+|Q|}(z,Q)W_{g_2,1 + |R|}(z,R),
    \end{aligned}
    \end{equation}
which is the essential content of the right hand side of the recursion relation \eqref{eqn:Wrecursion}. 

The left hand side of \eqref{eqn:Virasororecursion} needs to factorize in a very particular way after we introduce the sums. Change the explicit factor of $z$ to $z_1$, so that the all other explicit factors introduced in the extra sums are $z$. Then we desire the following outcome
	\begin{equation}
		\sum \text{Left hand side} \propto (z_1 + z)y(z)W_{g,n}(z,I).
	\end{equation}
If this happens, then we can divide both sides of the recursion relation by $(z_1^2 - z^2)y(z)$. On the left hand side this would leave $W_{g,n}(z,I)/(z_1-z)$, and the right hand side would match the argument of the residue in \eqref{eqn:Wrecursion}. But, taking the residue at $z = 0$ of both sides would yield $W_{g,n}(z_1,I)$ on the left hand side\footnote{We know that the resolvents will only have poles at $z = 0$, so taking the residue at $z = 0$ will still force the replacement $z \to z_1$.} , yielding the correct recursion relation.


\section{Concluding Remarks}

We have expanded upon two-dimensional open-closed string duality by exploring it in the context of FZZT branes and topological recursion, inspired by the recent developments in JT gravity and its supersymmetric generalizations. Using the general matrix models describing non-supersymmetric closed strings \cite{BREZIN1990144, DOUGLAS1990635, PhysRevLett.64.127, GROSS1990333, Kazakov:1989bc}, open strings \cite{KOSTOV1990181, Johnson:1992wr, Johnson:1993vk, Johnson:2004ut}, 0A open and closed strings \cite{Johnson:2003hy, Klebanov:2003wg}, and $\mathcal{N} = 2$ open and closed strings \cite{Johnson:2023ofr, turiaci2023mathcaln2}, we have presented an explicit equivalence between operator insertions in the closed string sector and derivatives acting on the free energy in the open string sector. This equivalence allowed us to compute perturbative contributions to correlation functions which have the interpretation of gravity path integrals in various matter backgrounds, on surfaces with both asymptotic and geodesic boundaries. The surfaces with geodesic boundaries are related to string worldsheets ending on FZZT branes. The results presented here highlight the immense power of the string equation formalism, in particular the large number of applications of the DJM equation. 

In section \ref{scn:stringequationperturbationtheory} we collected new explicit results for the perturbative solutions of various string equations, which generalize previously existing results to include parameters like the open string endpoint mass $-z^2$ and open-closed string coupling constant $\Gamma$. These perturbative solutions were utilized to compute the previously mentioned correlation functions in sections \ref{scn:macroscopicloopperturbationtheory} and \ref{scn:geodesicloopperturbationtheory}. Various examples were considered, including JT gravity and supergravity, as well as general $\mathcal{N} = 2$ matrix models. Matrix models describing systems with extended supersymmetry are relatively new, and any progress made in studying them is especially interesting. In particular, we have demonstrated that the open-closed string matrix model has the power to compute interesting objects, like the trumpet partition function, in theories with varying amounts of supersymmetry. In all cases, the result is independent of the coupling constants $t_k$ that define the matter background. This is in agreement with \cite{Okuyama:2021eju}. The trumpet partition function can be thought of as coming from the Schwarzian path integral on a hyperbolic surface with one asymptotic boundary and one geodesic boundary (see e.g. \cite{Saad:2019lba}). The universality of the result from the open-closed duality perspective agrees with the general relationship between the Schwarzian theory and two-dimensional CFTs presented in \cite{Mertens:2017mtv}. We also demonstrated the equivalence between two approaches to computing the FZZT brane partition function, one focused on macroscopic loops and the other on geodesic loops. It is possible to use the macroscopic loop formalism to compute correlation functions of determinant operators, but since this is not fully related to the main results of this paper we present these calculations in Appendix \ref{scn:extraFZZT}.

Geodesic boundaries are central to showing the equivalence between JT gravity and the corresponding matrix model. This dictionary between the two makes use of Mirzakhani's recursion relation for the Weil-Petersson volumes of the moduli spaces of hyperbolic Riemann surfaces with geodesic boundaries. The concept of a generalized Weil-Petersson volume, relating topological gravity in a general background to intersection theory, was explored in, e.g., \cite{Okuyama:2021eju}. In section \ref{scn:generalizedweilpeterssonvolumes} we showed that these generalized volumes appear naturally in the context of open-closed dualtiy, being the fixed length versions of certain brane-related correlation functions, which are naturally computed in terms of the branes' cosmological constants. One key strength of the string equation formalism is that it facilitates straightforward computations of the generalized volumes in supersymmetric theories.

In section \ref{scn:topologicalrecursion} we provided a heuristic argument connecting the Virasoro constraints of the open string sector of the matrix model with topological recursion. This connection was explored in \cite{mulase2006mirzakhani} starting from the recursion relation for the moduli space volumes, whereas here we have taken a matrix-model-first approach. The interpretation of the volumes in terms of correlation functions of closed string operators makes explicit the role that the KdV organization of observables in the theories plays in setting up the recursion relation via the Virasoro conditions.

To conclude, we consider possibilities for future work. It would be interesting to further explore matrix models with extended supersymmetry in the string equation formalism. The method proposed in section \ref{scn:stringequationperturbationtheory} for solving the string equation in $\mathcal{N} = 2$ theories is complicated by the rescaling of the 0A coupling constants. Compared to the better understood 0A models, the interpretation of $\hbar$ in the $\mathcal{N} = 2$ results here is shaky. We have noted that the results obtained here for the trumpet path integral and the moduli space volumes disagree with \cite{turiaci2023mathcaln2}, although in the end each respective gluing procedure produces the same macroscopic loop correlation functions, up to extra $E_0$-dependent terms which are subleading in the small $E_0$ limit. It is worth understanding why the two approaches agree in such a limit in the end despite the preliminary disagreements. Given the strong connection between the Schwarzian theory and 2D CFTs, it is natural to expect that $\mathcal{N} = 4$ theories have matrix model descriptions in the string equation formalism, although an interpretation in terms of minimal CFTs may be not be possible\footnote{The author thanks Nathan Benjamin for pointing this out.}. In particular, it would be interesting to determine how perturbation theory works for those string equations and how the theories' correlation functions differ from the $\mathcal{N} = 2$ ones.

Matrix models with $\mathcal{N} = 1$ supersymmetry require further attention as well. There is currently no string equation that describes the multi-brane open string sector in supersymmetric models. It is likely that this would require a generalization of the Gelfand-Dikii differential equation, and consequently the Gelfand-Dikii resolvent, to include multiple brane cosmological constants in analogy to the non-supersymmetric case. Some progress has been made here to construct the expected solutions, but it is possible that other contributions to the multi-brane sector could exist as well.

There is reason to believe that studying FZZT brane quantities may lead to some insight into the discrete structure of non-perturbative matrix model spectra (see e.g. \cite{Johnson:2021zuo, Johnson:2022wsr} and other recent works by Johnson). In \cite{Blommaert:2019wfy} the authors considered eigenbranes, which are essentially squared determinant operators that fix an eigenvalue in the double scaled spectrum. It is possible that by having an infinite number of these eigenbranes one could recover fully discrete spectrum. By interpreting this scenario in terms of branes, one might be able to extract a geometric (or lack thereof) description of non-perturbative quantum gravity microstates. The multi-brane formalism developed here is well-suited to studying this problem from a different angle. It would be interesting to apply the non-perturbative numerical techniques developed by Johnson to our multibrane solutions. 


\acknowledgments

The author thanks Clifford V. Johnson and Edward Mazenc for useful discussions. This work is supported in part by the US Department of Energy under grant DE-SC0011687.

\appendix

\section{Laplace Transform Equivalence} \label{apdx:laplacetransformequivalence}

Laplace transforms with explicit factors of the Laplace domain variable in the integrand are ubiquitous in the KdV formalism described above. This complicates performing the inverse Laplace transform. The remedy presented here is a formalized version of repeated integration by parts, and is included in this appendix for its potential use in simplifying more complicated inverse Laplace transforms that could arise in higher genus perturbative corrections to, say, macroscopic loop correlators. 

The goal is to write
    \begin{equation}
        \int_0^\infty dz\,e^{-\beta z} \beta^\mu z^\nu \simeq A(\mu,\nu) \int_0^\infty dz\,e^{-\beta z} z^{\gamma(\mu,\nu)},
    \end{equation}
for some coefficient $A$ and exponent $\gamma$. If the left hand side is finite, {\it i.e.} if $\nu$ is not a negative integer, it is equal to
    \begin{equation}
        \int_0^\infty dz\,e^{-\beta z} \beta^\mu z^\nu = \Gamma(\nu+1)\beta^{\mu-\nu-1}.
    \end{equation}
Similarly, if the right hand side is finite it is
    \begin{equation}
        A(\mu,\nu) \int_0^\infty dz\,e^{-\beta z} z^{\gamma(\mu,\nu)} = A(\mu,\nu)\Gamma(\gamma + 1)\beta^{-\gamma - 1}.
    \end{equation}
Clearly for the two to be equivalent we must take
    \begin{equation}
        \gamma(\mu,\nu) = \nu - \mu,\quad \& \quad A(\mu,\nu) = \frac{\Gamma(\nu+1)}{\Gamma(\nu-\mu+1)}.
    \end{equation}
In order for the transformation to be valid within the Laplace transform, we must also have $\gamma = \nu - \mu$ not be a negative integer.

We define a general transformation $\mathcal{T}_\mu$ on monomials using the coefficients $A$
    \begin{equation}
        \mathcal{T}_\mu \cdot z^\nu = A(\mu,\nu)z^{\nu - \mu},
    \end{equation}
where $A(\mu,\nu) \equiv 1$ when $\nu < \mu$, and $\nu$ is a negative integer or $\nu - \mu \in \mathbb{Z}$. This transformation extends to more general functions by linearity. Consider a function $f$ with Taylor series coefficients $f_n$. Then for $\mu \in \mathbb{N}$
    \begin{equation}
        \mathcal{T}_\mu \cdot f(z) = \sum_{0\leq n < \mu} f_n z^{n-\mu} + \sum_{n \geq \mu} \frac{n!}{(n-\mu)!}f_n z^{n-\mu}
    \end{equation}
In particular, for $\mu = 1$, 
    \begin{equation}
        \mathcal{T}_1 \cdot f(z) = \frac{f_0}{z} + \frac{df}{dz}.
    \end{equation}
Within the context of the Laplace transform, we define the equivalence $\beta^\mu f(z) \simeq \mathcal{T}_\mu \cdot f(z)$.

By using the transformation $\mathcal{T}$, $\mathcal{Z}_{1,1}(\beta)$ is equivalently written
    \begin{equation}
       \mathcal{Z}_1(\beta) = \frac{1}{2\sqrt{\pi}} \mathcal{L}_{u_0}\left[ \mathcal{T}_\half \cdot \left\{ \frac{u_2}{u_0'} + \frac{1}{12} \mathcal{T}_1 \cdot \left(\frac{2u_0''}{u_0'} - \mathcal{T}_1 \cdot u_0'  \right)  \right\}\right](\beta).
    \end{equation}
The proof of the formula for $\mathcal{Z}{1,1}(\beta)$ in \eqref{eqn:Z11} amounts to showing that 
    \begin{equation}
        \mathcal{T}_\half \cdot \left\{ \frac{u_2}{u_0'} + \frac{1}{12} \mathcal{T}_1 \cdot \left(\frac{2u_0''}{u_0'} - \mathcal{T}_1 \cdot u_0'  \right)  \right\} = \frac{1}{12}\mathcal{T}_{\frac{5}{2}} \cdot \left\{ \frac{1}{t_1} + \frac{2t_2u_0}{t_1^2} \right\},
    \end{equation}
irrespective of the choice of coupling constants.

Notice that for any values of the couplings $t_k$,
    \begin{equation}
        u_0' = -\frac{1}{t_0} + \frac{2t_2}{t_1^2}u_0 + \cdots,
    \end{equation}
which can be easily seen from the Taylor series expansion of $(f')^{-1}$. Therefore
    \begin{equation}
    \begin{aligned}
        \mathcal{T}_1 \cdot u_0' &= -\frac{u_0^{-1}}{t_1} + \frac{d}{du_0}u_0' \\
        &= -\frac{u_0^{-1}}{t_1} + \frac{u_0''}{u_0'}.
    \end{aligned}
    \end{equation}
Hence
    \begin{equation}
        \frac{2u_0''}{u_0'} - \mathcal{T}_1 \cdot u_0' = \mathcal{T}_1 \cdot u_0' + \frac{2u_0^{-1}}{t_1}.
    \end{equation}
Next,
    \begin{equation}
        \mathcal{T}_1\cdot \left\{\mathcal{T}_1 \cdot u_0' + \frac{2u_0^{-1}}{t_1}\right\} = \frac{u_0^{-2}}{t_1} + \frac{2t_2u_0^{-1}}{t_1^2} + \frac{d^2}{du_0^2}u_0'.
    \end{equation}
Using the fact that
    \begin{equation}
        \frac{d^2}{du_0^2}u_0' = -12 \frac{u_2}{u_0'},
    \end{equation}
we arrive at
    \begin{equation}
        \frac{u_2}{u_0'} + \frac{1}{12} \mathcal{T}_1 \cdot \left(\frac{2u_0''}{u_0'} - \mathcal{T}_1 \cdot u_0'  \right) = \frac{1}{12}\left( \frac{u_0^{-2}}{t_1} + \frac{2t_2u_0^{-1}}{t_1^2} \right).
    \end{equation}
Using the properties of $\mathcal{T}_\mu$, we conclude
    \begin{equation}
        \frac{1}{12}\left( \frac{u_0^{-2}}{t_1} + \frac{2t_2u_0^{-1}}{t_1^2} \right) = \frac{1}{12}\mathcal{T}_2 \cdot \left\{ \frac{1}{t_1} + \frac{2t_2u_0}{t_1^2} \right\},
    \end{equation}
which completes the argument.


\section{Microscopic Loop Calculations} \label{apdx:macroscopicloopcalculations}

\subsection{\texorpdfstring{$g = 1,n =2$}{g1n2}}

The two-point correlation function of the closed string operators $\sigma_k$ is given in general by
	\begin{equation}
		\langle \sigma_{k-1} \sigma_{l-1} \rangle = \hbar^{-2} \iint^\mu \xi_k \cdot R_l' \,(dx)^2.
	\end{equation}
The $g = 1$ contribution to the integrand is
	\begin{equation}
	\begin{aligned}
		\left[\xi_k \cdot R_l' \right]_1 &= - kl\frac{d^2}{dx^2} \Bigg[u_1 u_0^{k + l - 2} + \frac{(u_0')^2u_0^{k + l - 3}}{12}(-5 + 4 (k+l) - k^2- k l - l^2) \\
		&+ \frac{u_0''u_0^{k+l-3}}{6}(2 - (k +l))\Bigg].
	\end{aligned}
	\end{equation}
The second derivative cancels with the double integral, leaving all functions evaluated at $x = \mu$. In theories with $u_0(\mu) = 0$ this produces formally divergent results. Throwing away the diverging pieces yields the correct answer.

\subsection{\texorpdfstring{$g = 1, n = 3$}{g1n3}}

The three-point correlation function of the closed string operators is given in general by
	\begin{equation}
		\langle \sigma_{k-1} \sigma_{l-1} \sigma_{n-1} \rangle = \hbar^{-2} \iint^\mu \xi_k \cdot \xi_l \cdot R_l' \,(dx)^2
	\end{equation}
The $g = 1$ contribution to the integrand is
	\begin{equation}
	\begin{aligned}
		\left[\xi_k \cdot \xi_l \cdot R_l' \right]_1 &= kln \frac{d^3}{dx^3} \Bigg[u_1 u_0^{k + l + n - 3} - \left(\frac{k^2 + l^2 + n^2 + kl + kn + ln}{12}\right)(u_0')^2u_0^{k + l + n - 5} \\
		&- \left( \frac{k+ l + n}{6} \right)u_0''u_0^{k + l + n - 4} \Bigg] + kln \frac{d^2}{dx^2} \Bigg[
        \left(\frac{k l n}{12}\right)(u_0')^3 u_0^{k + l + n - 6}\Bigg].
	\end{aligned}
	\end{equation}
The double integral cancels with two derivatives in each term, leaving all functions evaluated at $x = \mu$. This contribution also has divergent terms that must be thrown away.


\section{FZZT Branes and Macroscopic Loops}\label{scn:extraFZZT}

The correlation function of an FZZT brane and a macroscopic loop has its own expansion in terms of macroscopic loops
    \begin{equation}
        \left\langle \Psi(E) e^{-\beta \mathcal{H}}\right\rangle = \sum_{n = 0}^\infty \frac{(-1)^n}{n!}\left\langle\prod_{i = 1}^n \int_0^\infty \frac{d\beta_i}{\beta_i}e^{\beta_i E}  e^{-\beta_i \mathcal{H}}e^{-\beta \mathcal{H}} \right\rangle. \label{eqn:brane+macloop}
    \end{equation}
A large number of contributions is captured by the factorization of the brane part from the macroscopic loop expectation value $\langle \Psi(E) \rangle \langle e^{-\beta \mathcal{H}}\rangle$.

One of the simpler contributions to eqn. (\ref{eqn:brane+macloop}) comes from breaking the odd-$n$ terms into products of two-point functions. For example, at $n = 1$ we have the integral
 \begin{equation}
        \mathcal{A} = \int_0^\infty \frac{d\beta_i}{\pi\beta_i} e^{\beta_i (E-u_0)} \frac{\sqrt{\beta_i\beta}}{\beta_i + \beta} = e^{-\beta (E-u_0)} \text{Erfc}(i\sqrt{\beta (E-u_0)}).
    \end{equation}
At $n = 3$ we have 3 terms that look like
    \begin{equation}
       \int_0^\infty \frac{d\beta_1}{\beta_1}\int_0^\infty \frac{d\beta_2}{\beta_2}\int_0^\infty \frac{d\beta_3}{\beta_3} e^{(\beta_1 + \beta_2 + \beta_3)E} \left\langle e^{-\beta_1 \mathcal{H}}e^{-\beta_2 \mathcal{H}}  \right\rangle \left\langle e^{-\beta_3\mathcal{H}}e^{-\beta \mathcal{H}}  \right\rangle \approx \mathcal{A}\mathcal{I}_1,
    \end{equation}
where
	\begin{equation}
       \mathcal{I}_1 = \frac{1}{\pi}\int_0^\infty \frac{d\beta_1d\beta_2}{\beta_1\beta_2} \frac{\sqrt{\beta_1\beta_2}}{\beta_1 + \beta_2}e^{-(\beta_1+\beta_2)(-E+u_0(\mu))} \cong -\frac{1}{2}\log(-E+u_0(\mu)),
    \end{equation}
The term with $k$ two-point functions will be $\mathcal{A} \mathcal{I}_1^{k-1}$ with a degeneracy of $\frac{(2k)!}{2^kk!}$. Since these appear when $n = 2k-1$ in eqn. (\ref{eqn:brane+macloop}), this contribution to the correlation function sums to be 
    \begin{equation}
         \left\langle \Psi(E) e^{-\beta \mathcal{H}}\right\rangle = \cdots - (E-u_0(\mu))^{-1/4} \text{Erfc}\left(i\sqrt{\beta (E-u_0(\mu))}\right)e^{-\beta (E-u_0(\mu))-\frac{i\pi}{4}} + \cdots.
    \end{equation}
This matches the result obtained for the Airy model in \cite{Okuyama:2021eju}.

The correlation function of two branes and one macroscopic loop is, at leading order
    \begin{equation}
        \left\langle \Psi(E)\Psi(E')e^{-\beta \mathcal{H}}\right\rangle \approx \sum_{n,m = 0}^\infty \frac{1}{n!m!} \prod_{i,j = 1}^{n,m} \sqrt{\frac{\beta}{\pi}}\int_0^\infty \frac{d\beta_i}{\pi \beta_i} \int_0^\infty \frac{d\beta'_j}{\pi \beta'_j} e^{-\beta_i E - \beta'_j E'} \sqrt{\beta_i\beta'_j}.
    \end{equation}
Using
    \begin{equation}
        \prod_{i,j = 1}^{n,m} \sqrt{\frac{\beta}{\pi}}\int_0^\infty \frac{d\beta_i}{\pi \beta_i} \int_0^\infty \frac{d\beta'_j}{\pi \beta'_j} e^{-\beta_i E - \beta'_j E'} \sqrt{\beta_i\beta'_j} = \left(\frac{\beta}{\pi}\right)^{\frac{n+m}{2}} E^{-n/2}E'^{-m/2},
    \end{equation}
we get
    \begin{equation}
        \left\langle \Psi(E)\Psi(E')e^{-\beta \mathcal{H}}\right\rangle \approx e^{-\sqrt{\frac{\beta}{\pi}} \left(\frac{1}{\sqrt{E}} + \frac{1}{\sqrt{E'}}\right)}.
    \end{equation}


\bibliographystyle{JHEP}
\bibliography{references.bib}

\end{document}